\documentclass[12pt]{article}
\usepackage[english]{babel}
\usepackage{amsthm}
\usepackage{amssymb}
\usepackage{amsmath}
\usepackage{float}
\usepackage{caption}
\usepackage{natbib}
\usepackage{bm}
\usepackage{multirow}
\usepackage{xcolor}
\usepackage{pdfpages}
\usepackage{titling}

\newtheorem{remark}{Remark}

\newtheorem{Theorem}{Theorem}
\usepackage{accents}

\usepackage[bf,SL,BF]{subfigure}
\usepackage{algorithm}
\usepackage{algpseudocode}
\usepackage{graphicx}
\usepackage{enumerate}
\usepackage{natbib}
\usepackage{url} 
\usepackage[colorlinks,linkcolor=blue,citecolor=blue,urlcolor=blue,filecolor=blue,backref=page]{hyperref}

\newcommand{\bbeta}{\mbox{\boldmath{$\beta$}}}
\newcommand{\Z}{\mathbf{Z}}

\newcommand{\T}{\text{T}}

\begin{document}

\def\spacingset#1{\renewcommand{\baselinestretch}%
{#1}\small\normalsize}
\spacingset{1.1}
\date{}

  \title{\bf \Large Bayesian prediction via nonparametric transformation models}
  \author{ Chong Zhong\thanks{
  \scriptsize
   The author is a PhD student of Department of Applied Mathematics, The Hong Kong Polytechnic University. },\hspace{.2cm}
     Jin Yang\thanks{  
    \scriptsize
    The author is a PostDoc of 
    Biostatistics and Bioinformatics Branch, \textit{Eunice Kennedy Shriver} National Institute of Child Health and Human Development, NIH, USA.},\hspace{.2cm}
 Junshan Shen\thanks{
    \scriptsize
    The author is an Associate Professor of School of Statistics, Capital University of Economics and Business, China.},\hspace{.2cm}
    Catherine C. Liu \thanks{
      \scriptsize
    The author is an Associate Professor of Department of Applied Mathematics, The Hong Kong Polytechnic University, HKSAR.} and \\
  Zhaohai Li\thanks{
    \scriptsize
    The author is a Professor of Department of Statistics, George Washington University, Washington, DC, USA.}
    }
  \maketitle

\vspace{-1.1cm}
\begin{abstract}

This article tackles the old problem of prediction via a nonparametric transformation model (NTM) in a new Bayesian way. Estimation of NTMs is known challenging due to model unidentifiability though appealing because of its robust prediction capability in survival analysis.
Inspired by the uniqueness of the posterior predictive distribution, we achieve efficient prediction via the NTM aforementioned under the Bayesian paradigm.
Our strategy is to assign weakly informative priors to nonparametric components rather than identify the model by adding complicated constraints in existing literature.
The Bayesian success pays tribute to i) a subtle cast of NTMs by an exponential transformation for the purpose of compressing  spaces of infinite-dimensional parameters to positive quadrants considering non-negativity of the failure time; ii) a newly constructed weakly informative quantile-knots I-splines prior for the recast transformation function together with the Dirichlet process mixture model assigned to the error distribution. 
In addition, we provide a convenient and precise estimator for the identified parameter component subject to the general unit-norm restriction through posterior modification, enabling effective relative risks.
Simulations and applications on real datasets reveal that our method is robust and outperforms the competing methods. 
An \texttt{R} package \texttt{BuLTM} is available to predict survival curves, estimate relative risks, and facilitate posterior checking.  

\end{abstract}
\vspace{-.1cm}
{

\hspace{1.45cm}Weakly informative priors. 
}

\newpage
\spacingset{1.8}
\section{Introduction}
\label{sec:intro}

The traditional linear transformation model raised by  Cheng, Wei, and Ying
is quite flexible,
covering whilst not limited to three commonly used survival models, proportional hazards (PH), proportional odds (PO), and accelerated failure time (AFT), and is formulated as
\begin{align}\label{basicLTM}
    h(T) = \bbeta^\T  \Z + \epsilon,
\end{align}
where 
 $T$ is the random censored failure time,  $\Z$ and $\bbeta$ are the $p$-dim predictor vector and the coupled vector of regression coefficients respectively,  
$h(\cdot)$ is a strictly increasing function that may be \textit{sign varying} on $\mathbb{R}^+$,
and $\epsilon$ is the model error with distribution function $F_\epsilon$ \citep{cheng1995analysis}.
Model \eqref{basicLTM} is called the \textit{nonparametric transformation model} (NTM) when both functional forms of $h$ and $F_\epsilon$ are unknown \citep{horowitz1996semiparametric, colling2019estimation}.

In predicting survival outcomes, the NTM is apparently preferable because of its model robustness compared to models of PH, PO, AFT, and other survival models assuming either or both of $h$, $F_\epsilon$ specified. 
However, it also poses the challenge to estimate infinite-dimensional parameters $h$, $S_\epsilon$ in the NTM owning to \textit{model unidentifiability}
in the sense that collections of triplet $(h, F_\epsilon, \bbeta)$ generate identical likelihood, called flat likelihood.
Nevertheless,  estimating such nonparametric components is essential for the prediction of failure times and conditional hazards \citetext{\citealp[pp.~207]{song2007semiparametric}; \citealp[pp.~980]{lin2017robust}}, to name a few.
This motivates us to overcome the challenging problem of prediction via the NTM. 

One may categorize existing approaches of predicting failure time via the NTM into two lines, 
\textbf{i)}, \textit{to make the model identifiable by adding constraints}. 
Econometricians impose scale normalization to the parametric component \citep{HardleStoker1989jasa}; and under NTM \eqref{basicLTM}, impose location normalization to either $h$  with specified root \citetext{\citealp{ gorgensHorowitz1999JOE, chen2002rank}; among others}, or $F_\epsilon$  with specified mean or median \citetext{\citealp{ye1997nonparametric, linton2008estimation, chiappori2015nonparametric}; among others}.
Such approaches focused on establishing theoretical results such as $\sqrt{n}$-convergence, while they did not \textit{touch upon computational feasibility in practice}. 
As a Bayesian counterpart, \cite{mallick2003bayesian} evidenced that imposing constrained priors such as the \textit{constrained} Polya tree prior for $F_\epsilon$ to identify the NTM is untractable, since an inappropriate center distribution of the Polya tree \textit{incurs slow convergence and poor mixing of posterior} \citep[pp.39]{Muller2015}. 
And \textbf{ii)}, \textit{to make strong \textit{priori} assumptions to circumvent the identifiability issue.} 
Frequentists either fixed $h$ such as the AFT model \citetext{\citealp{jin2003rank};  \citealp{ding2011sieve}; among others}, or made parametric assumptions on $F_\epsilon$ such as PH and PO models \citetext{\citealp{lu2004semiparametric}; \citealp{zeng2007efficient}; among others}. 
Alternatively, Bayesian used a two-step procedure to estimate all models and select the ``best"  \citetext{\citealp{zhao2009mixtures, de2014bayesian, zhou2018unified}}.
The \texttt{R} package \texttt{spBayesSurv} \citep{zhou2020spbayessurv} based on Zhou and Hanson, as far as we know, may be the optimal tool in prediction provided that it selected the correct model.  
Despite mathematical or computational convenience, \textit{designating $h$ or $F_\epsilon$ is at the risk of misspecification}, leading to inconsistent estimation, invalid statistical inference, and erroneous predictions.

In this article, we attempt to seek computationally tractable and robust \textit{Bayesian} prediction under the NTM \textit{without identifying the model}.  
The spirit of our methodology is based on two concerns. \\
\noindent{\textbf{i. The posterior predictive distribution (PPD) of a future observation is always unique regardless of model identifiability.}}
Although the parameters in triplet $(h, F_\epsilon, \bbeta)$ under NTM \eqref{basicLTM} are not separately identifiable, they are jointly estimable if their posterior distributions are \textit{proper}. 
Therefore, the unique PPD can be obtained by integrating all parameters out even though there are multiple solutions of triplet $(h, F_\epsilon, \bbeta)$ that provide the same likelihood; see subsection \ref{subsec:MCMC} for details. \\
\noindent{\textbf{ii. Weakly informative priors make Markov Chain Monte Carlo (MCMC) tractable.}} 
In Bayesian analysis, priors play a defining role, have a substantive impact on final model results \citetext{\citealp{depaoli2020importance, van2021bayesian}}, and are analog to constraints that make the model identifiable. 
Noninformative priors hinder posterior sampling under unidentified models since they cannot control posterior variance to be finite.
In contrast, the weakly informative prior is a kind of ``stronger" proper prior in the sense that it is able to \textit{control prior variance} moderately on the unconstrained support, and thus is able to \textit{dominate the posterior variance}. 
Consequently, it facilitates the convergence of posterior sampling by preventing the sampler from running to highly implausible values that are far away from its center \citep[pp.262]{mcelreath2020rethinking}.

The aforementioned two concerns stand by our methodology. 
We achieve PPDs of future observations computed from the posterior of $(h, F_\epsilon, \bbeta)$ by assigning two weakly informative priors to the infinite-dimensional parameters, a newly constructed quantile-knots I-splines prior for $h$, and a regular Dirichlet process mixture (DPM) model for $F_\epsilon$, together with a noninformative prior for the parametric component $\bbeta$.
In addition, we obtain an efficient Bayes estimator of identified $\bbeta$ through posterior projection so as to provide sound relative risks.

The contribution of this article is tri-folds.
Firstly, we solve the standing problem of prediction failure times via the NTM \eqref{basicLTM} efficiently and numerically conveniently.
This is realized by the joint strength of two weakly informative priors, quantile-knots I-splines prior for the transformation function, and the DPM model for model error distribution. 
It is based on I-spline basis functions \citep{ramsay1988monotone} and generates knots from the sample quantiles of censored and uncensored failure times directly.
Thus, \textit{a small size of knots enable us to capture the major shape of the transformation function well} rather than tuning the number of knots in traditional I-spline-based priors that select knots from a long series of equally spaced points \citetext{\citealp{CaiDunson2007JASA}; \citealp{wang2011Biometrika}; among others}. 
The proposed I-spline type prior is applicable to modeling monotone functions that are differentiable or nondifferentiable by adjusting the smoothness parameter.

Secondly, we provide a new and convenient Bayes estimator for the identified parameter $\bbeta$ through posterior projection. 
We impose a unit-norm normalization \citep{hardle2004nonparametric} rather than confining the first entry of the vector parameter to be $\pm 1$ \citetext{\citealp{gorgensHorowitz1999JOE, chen2002rank, song2007semiparametric}}
to avoid specifying the sign of a treatment effect associated with the survival outcomes. 
The presented posterior modification avoids extra sampling and thus is computationally expedient. 
In contrast, it is inapplicable to assign constrained priors for $\bbeta$ directly such as the Polar system prior \citep{Park2005JCGS} or \texttt{Stan}'s built-in prior since our prior elicitation has no constraints.

Finally, for practitioners, we supply the  \texttt{R} package \texttt{BuLTM}, which is computationally convenient and efficient to predict failure times and output estimates of predicted survival probability, conditional hazards, and relative risks.
For the prediction purpose, simulation studies demonstrate that \texttt{BuLTM} outperforms \texttt{spBayesSurv} under the PH, PO models, and model misspecification situations, and is comparable to \texttt{spBayeSurv} under the AFT model.
For the out-sample predictive capability, \texttt{BuLTM} is also competitive to \texttt{spBayesSurv} in application examples.

The remainder of this article is organized as follows. 
Section \ref{sec:mod} introduces the recast model of the NTM as the cornerstone of our Bayesian approach.
Section \ref{sec:prior} introduces weakly informative prior elicitation for infinite-dimensional parameters. 
Section \ref{sec:Posterior} introduces the posterior inference procedures including the PPD computation and the posterior projection procedure for $\bbeta$. 
Sections \ref{sec:sim} and \ref{sec:app}  assess and demonstrate our method compared with existing work by simulations and application examples, respectively.
Section \ref{sec:disc} concludes the article with a brief discussion. 
Related details are collected in the online supplementary materials. 
The \texttt{R} package \texttt{BuLTM} is available on GitHub \hyperlink{https://github.com/LazyLaker}{https://github.com/LazyLaker}.

\section{Recast: multiplicative relative risk model}
\label{sec:mod}
To resolve prediction via the NTM, we first impose the exponential transformation to NTM \eqref{basicLTM} and obtain a recast model 
\begin{align} \label{expmod1}
     H(T) = \xi\exp(\bbeta^\T  \Z),
\end{align}
where the recast transformation $H(\cdot) = \exp\{h(\cdot)\}$ and the model error $\xi = \exp(\epsilon)$ with distribution function $F_{\xi}$. 
The nonparametric transformation model \eqref{expmod1} with \textit{multiplicative relative risk } $\exp(\bbeta^T \Z)$ is abbreviated as MTM thereafter, 
where $H$ is \textit{positive} on $\mathbb{R}^+$ and the multiplicative random error $\xi$ is also positive.
Let $S_X = 1- F_X$, where the placeholder $X$ represents $\epsilon$ or $\xi$. 
MTM \eqref{expmod1} is equivalent to NTM \eqref{basicLTM} in the sense that they share common parametric component $\bbeta$,  and strictly $S_\epsilon(\cdot)= S_\xi\{\exp(\cdot)\}$ and $h(\cdot) = \log H(\cdot)$.

The above monotonic transformation step plays a critical role in establishing our Bayesian solution.  
In the Bayesian paradigm, prior elicitation and posterior sampling are two preliminary components of Bayesian inference.
Unfortunately, the infinite-dimensional parameter $h$ out of NTM is faced with unprecedented difficulties in both targets.

On one hand, most existing models for \textit{sign-varying} monotone functions are inapplicable to $h$ in that,  $h$ may not have an intercept such as the AFT model, preventing usage of approaches that  rely on an intercept term in modeling a counterpart of transformation $h$ \citetext{\citealp{neelon2004bayesian, shively2009bayesian, lenk2017bayesian}, among others}; 
it is also nontrivial to extend to censored observations
for those methods that impose a response-based monotonicity shape restriction to the model \citetext{\citealp{riihimaki2010gaussian, LinDunson2014Biometrika, wang2016estimating}, among others}.

 On the other hand, sampling for $h$ often gives rise to trouble if $h(0) \to -\infty$ and lifetimes are close to zero.
Take the logit transformed incomplete beta function in \cite{mallick2003bayesian} for instance. 
Sampling $h$  may be bothered by \textit{infinity gradient} caused by infinite $h$ under gradient-based samplers such as the Hamilton Monte Carlo and the No-U-Turn Sampler (NUTS) in \texttt{Stan} \citep{carpenter2017stan},
or by the \textit{poor proposal distributions whose center may disperse to infinity} under Metropolis-type samplers, leading to very slow convergence and a low acceptance rate. 

From the insight that it brings huge expedience if one is able to confine the transformation to be \textit{nonnegative}, we are driven to take the recasting as the foremost step to initiate our methodology.
Consequently, the exponential transformation compresses the space of infinite-dimensional parameters from $\mathcal{M}_{\mathbb{R}} \times \mathcal{S}_{\mathbb{R}}$ to a reasonable subset of $\mathcal{M}_{\mathbb{R}^+} \times \mathcal{S}_{\mathbb{R}^+}$, where $\mathcal{M}_{\mathcal{A}}$ denotes the space of monotone functions with range $\mathcal{A}$ and $\mathcal{S}_{\mathcal{A}}$ denotes the space of survival functions with support $\mathcal{A}$. 
Our spirit has allies in the literature about the transformation model where they rewrote their transformation as the logarithm of a cumulative hazard function \citetext{\citealp{scheike2006flexible}; \citealp{zeng2006efficient}; among others}.

 Besides its tractability and convenience, the recast MTM \eqref{expmod1} still maintains interpretability analogous to that of NTM \eqref{basicLTM}.
Let $\Lambda(\cdot)$ be the cumulative hazard function of a time-to-event. 
By some simple algebra, for MTM \eqref{expmod1}, the counterpart of expression (1.3)  of \cite{cheng1995analysis} that motivates NTM \eqref{basicLTM} can be represented as
\begin{align}\label{cumtrans}
 G\{\Lambda_{T|\Z}(t)\} = H(t)\exp(-\bbeta^\T  \Z),
\end{align}
where $G^{-1}(\cdot) = -\log\{1-F_\xi(\cdot)\}$ is the link working on the conditional cumulative hazard of the failure time.
Specifically, if the link functional forms of $G(s)$ are $s$ and $\{\exp(s)-1\}$,
or, $F_\xi(s)$ are $ \{1-\exp(-s)\}$ and $F_\xi(s) = (1+s)^{-1}$ in \eqref{expmod1}, 
or equivalently, the model error $\epsilon$ in \eqref{basicLTM} follows a standard extreme-value distribution and a standard logistic distribution, then
the model reduces to PH and PO models respectively.

\section{Likelihood and prior}
\label{sec:prior}
\vspace{-.5cm}
\subsection{Likelihood}
For the real survival time $T$ and the random censoring variable $C$, one denotes the observed time-to-event as $\widetilde{T} = \min(T, C)$. The censoring indicator  $\delta = I(T \le C)$.
Let $S_\xi$ and  $f_\xi$ be the survival probability and density function of $\xi$, respectively. 
In this section, we consider the following quite mild assumptions.\\
(A1) The exp-transformation $H$ is differentiable.\\
(A2) The multiplicative random error $\xi$ is continuous.\\
(A3) The covariate $\Z$ is independent of $\xi$.\\
(A4) The censoring variable $C$ is independent of $T$ given $\Z$.\\
(A1) is required since there is $H'$ functional in the likelihood representation below; (A2) is mild; (A3) is general; (A4) is the general noninformative censorship condition.

With independent triplets of observed data $\{(\widetilde{T}_i, \Z_i, \delta_i)\}_i^n$, one writes the complete data likelihood as
\begin{align}\label{likeli}
    \mathcal{L}(\bbeta, H, S_\xi, f_\xi|\widetilde{T}, \Z, \delta) = \prod_{i=1}^{n} [f_\xi\{H(\widetilde{T}_i)e^{-\bm{\beta}^\T  \Z_i}\} H'(\widetilde{T}_i)e^{-\bm{\beta}^\T  \Z_i}]^{\delta_i}[S_\xi\{H(\widetilde{T}_i)e^{-\bm{\beta}^\T  \Z_i}\}]^{1-\delta_i}.
\end{align}

Since $\xi$ is an arbitrary continuous positive random variable, the DPM model \citep{lo1984class} is a natural choice of the prior for $S_\xi$ and $f_\xi$. 
Then one specifies a nonnegative distribution family as the kernel in the DPM. 
Motivated by the commonly used Weibull mixture survival models \citetext{\citealp{kottas2006nonparametric, egleston2017latent, shi2019low}, among others}, we adopt the Weibull kernel so that
\begin{align}
\label{DPM}
    S_\xi = 1- \sum_{l=1}^{L}p_l F_w(\psi_l, \nu_l)&, f_\xi = \sum_{l=1}^{L}p_l f_w(\psi_l, \nu_l),
\end{align}
where $F_w$ and $f_w$ denote the CDF and density function of the Weibull distribution, respectively. 
An intuitive justification for the Weibull kernel is that the PH model is a special Weibull survival model. 
More details and justification about the DPM prior are deferred to \textit{Supplement S.2}. 

As shown in \eqref{likeli},  $H$ should be differentiable on $\mathbb{R}^{+}$, or its subset.
Most increments-based models \citetext{\citealp{kalbfleisch1978non, arjas1994nonparametric, Mckeague2000Biometrics}; among others} are not differentiable everywhere, and those differentiable ones may induce a complicated and intractable form of the derivative \citetext{\citealp{dykstra1981bayesian, hjort1990nonparametric}; among others}. 
In this article, we construct a quantile-knot I-splines type prior for $H$ and its derivative $H'$, which is computationally expedient. 
Details about the prior are discussed in the following subsection. 

\vspace{-.5cm}
\subsection{Quantile-knots I-splines prior}
\label{subsec:quantile_knots}
Suppose the survival outcome $T$ is observed on interval $D = (0, \tau]$, where $\tau$ is the largest survival time in the sample. 
Note that, $H$ is a nonnegative strictly increasing differentiable function on $D$ based on transformation model \eqref{expmod1} and likelihood function \eqref{likeli}.
It is natural to model $H$ and $H'$ by monotone splines,
\begin{align}\label{ispline}
    H(t) = \sum_{j=1}^{K}\alpha_j B_j(t),  H'(t) = \sum_{j=1}^{K}\alpha_j B_j'(t),
\end{align}
where $\{\alpha_j\}_{j=1}^{K}$ are positive coefficients to guarantee nondecreasing monotonicity, $\{B_j(t)\}_{j=1}^{K}$ are I-spline basis functions \citep{ramsay1988monotone} on $D$ and $\{B_j'(t)\}_{j=1}^K$ are corresponding derivatives. 
Unlike other I-splines approaches that include an unknown intercept, we simply set the intercept $H(0) = 0$ since it can be derived from assumption (A3) directly, referred to \textit{Supplement S.1}. 
The number of I-spline basis functions $K$ is the sum of the number of interior knots and the order of smoothness $r$ with $(r-1)$th order derivative existing. 
Empirically, $r$ may take value from $2$ to $4$ and we take the default value $r=3$ in \texttt{R} package \texttt{splines2}.
Interior knots cut the time interval $D$ into $(K-r+1)$ partitions. 
Then our concern lies in specifying the number and locations of interior knots for modeling the exp-transformation.

We construct an I-splines type prior based on representations \eqref{ispline} by selecting interior knots from empirical quantiles of survival times, namely quantile-knots I-splines prior.
First, we fix the initial number of interior knots $N_I$ which is much fewer than that in other typical I-splines type models coupled with the shrinkage prior. 
Our insight comes from the advantage of quantiles that a small number of quantiles quantify different ``locations" of a distribution and therefore they can be viewed as alternative measures of the shape of the predictive distribution of $T$. 
Meanwhile, the corresponding posterior is not sensitive within the range of a small number of knots, indicating that the proposed prior is free of tuning, referred to \textit{Supplement S.7.1}. 
It is expedient in implementation compared to those priors requiring tuning, referred to \textit{Supplement S.4.2}. 

Next, given the initial number of interior knots $N_I$, we propose a two-step data-driven procedure to specify their locations using the information of survival times and censoring states. 
Let $\hat{F}_{X}(t) = n^{-1}\sum_{i=1}^n I(X_i \le t)$ be the empirical CDF of $X$ and $\hat{Q}_{X}(p) = \hat{F}_{X}^{-1}(p) = \inf \{t: p\le \hat{F}_{X}(t)\}$ be the corresponding empirical quantile function, where $X$ is the placeholder for $T$ and $\tilde{T}$, uncensored and observed survival times, respectively. 
Let $j=0, \ldots, N_I-1$.  
\begin{enumerate}
    \item[Step 1:] Selects $N_I$ empirical quantiles of observed failure times as interior knots $0< t_0<\cdots<t_{N_I-1}\le\tau$, where $t_j = \hat{Q}_T\{j/(N_I-1)\}$.
    \item[Step 2:]  If $|\hat{F}_T(t_j) - \hat{F}_{\tilde{T}}(t_j)|>z_0 \ge 0.05$, then interpolate a new knot $t_j^* = \hat{Q}_{\tilde{T}}(j/(N_I-1))$. Output sorted series of $\{t_0, \ldots, t_j, t_j^*, \ldots, t_{N_I-1}\}$ as final interior knots. 
\end{enumerate}
In step 1, we choose equally spaced percentiles of uncensored survival times since information about $H'$ is provided by uncensored survival times only. 
In step 2, we make interpolation in case of high censoring of survival times and insufficient uncensored observations. 

\begin{figure}[!htp]
  \centering
   \includegraphics[width=3in]{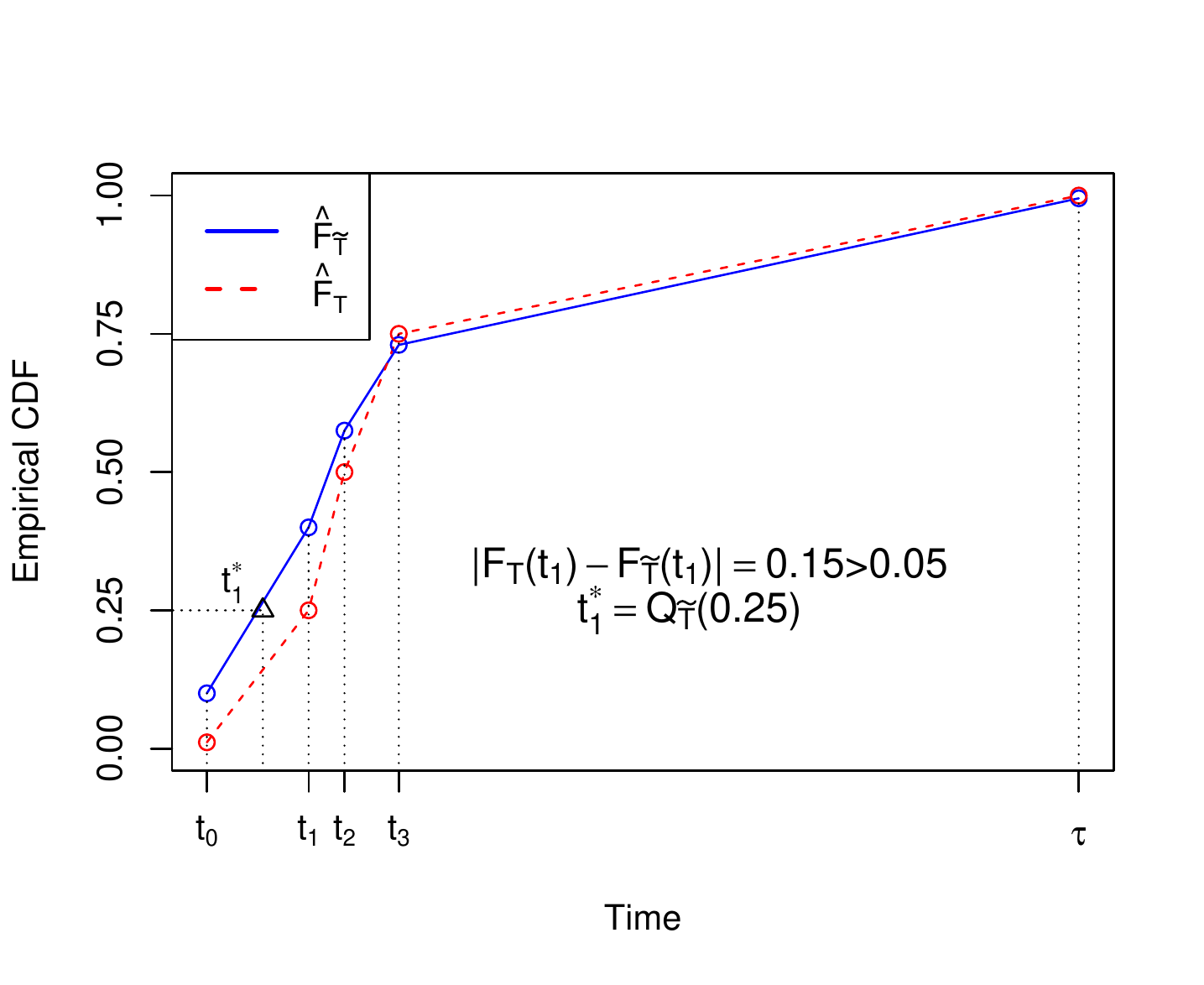}
  \caption{\label{QuantKnots}\footnotesize Example with $5$ initial knots.}
\end{figure}

Take $5$ initial knots for instance i.e. it contains $3$ quartiles and $2$ endpoints of uncensored survival times. 
In Figure \ref{QuantKnots}, there are apparent deviations between uncensored and observed curves on the first three interior knots.
Therefore, we interpolate by three new knots $t_j^{*} = Q_{\tilde{T}}(j/4)$, for $j=0, 1, 2$.  
Finally, we obtain $(t_0^*, t_0, t_1^*, t_1, t_2^*, t_2, t_3, t_4=\tau)$ as our interior knots.
By the above operation, I-spline basis functions $\{B_j(t)\}_{j=1}^{K}$ are specified.
We further assign an exponential prior for $\{\alpha_j\}_{j=1}^K$. 
Consequently, we have built our quantile-knots I-splines prior for $H$, which is weakly informative by the fact that, 
given  $\alpha_j \sim \exp(\eta)$, $E\{H(t)\} = \eta^{-1}\sum_{j-1}^{K}B_j(t)<\infty$ and $\text{Var}\{H(t)\} = \sum_{j=1}^{K} \eta^{-2} B_j^2(t)< \infty$ for any $\eta>0$ and $t<\infty$.


\begin{remark}\label{rmk:spline2}
The quantile-knots I-splines prior can also be applied to model nondifferentiable functions. 
The proposed prior can be viewed as a combination of NII processes, referred to \textit{Supplement S.3}. 
Particularly, when $r=1$, the I-spline function reduces to a straight line on each partition, and the proposed prior reduces to the piecewise exponential prior. 
\end{remark}

\vspace{-.5cm}
\section{Posterior inference}
\label{sec:Posterior}
\subsection{MCMC and posterior prediction}
\label{subsec:MCMC}
According to above prior settings, nonparametric parameters $H$ and  $S_\xi$ in MTM \eqref{expmod1} are encapsulated in elements of $\bm{\alpha}$ and $(\bm{p}, \bm{\psi}, \bm{\nu})$, respectively, where $\bm{\alpha} = \{\alpha_j\}_{j=1}^K$, $\bm{p} = \{p_l\}_{l=1}^L, \bm{\psi} = \{\psi_l\}_{l=1}^L,$, and $\bm{\nu}=\{\nu_l\}_{l=1}^L$. 
Consequently, the nonparametric components $(h, S_\epsilon)$ in the original NTM \eqref{basicLTM} are expressed as
$$
h(t) =  \log \{\sum_{j=1}^K \alpha_j B_j(t)\}, ~ S_\epsilon(x) = 1- \sum_{l=1}^L p_l F_w\{\exp(x) | \psi_l, \nu_l\}. 
$$
Then the estimators of triplet parameters $(h, S_\epsilon, \bbeta)$ can be obtained from the posterior distribution of parameters $\Theta = (\bm{\alpha}, \bbeta, \bm{p}, \bm{\psi}, \bm{\nu})$. 
The posterior density of $\Theta$ is 
\begin{align}
\label{posterior}
\pi(\Theta|\widetilde{T}, \Z,\delta)&\propto \mathcal{L}(\Theta|\widetilde{T}, Z,\delta)p(\bm\alpha)p(\bbeta)p(\bm{p})
\prod_{l=1}^{L}G_0(\psi_l, \nu_l),
\end{align}
where $\mathcal{L}$ is the likelihood for $\Theta$  defined by \eqref{likeli} and $p(\cdot)$ represents a prior density. 
For each parameter in the posterior density, we set their priors as
\begin{eqnarray}\label{priors}
 \begin{aligned}
    &\alpha_j \sim  \exp(\eta), p(\bbeta) \propto 1, G_0(\psi_l, \nu_l) = \text{Gamma}(a, b) \times \text{Gamma}(a, b), \\
    & p_{_l} =q_{_l}\prod_{L=1}^{l-1}(1-q_{_l}),
    q_{_l} \sim \text{Beta}(1, c),
    l=1, \dots, L-1; p_{_L} = 1-\sum_{l=1}^{L-1}p_{_l}.
 \end{aligned}
\end{eqnarray}
Here $\eta$ is the hyper-parameter of the prior for $\bm \alpha$. 
The prior for $\bbeta$ is an improper uniform prior, which is ``purely" noninformative. 
One may either assign a hyperprior for $\eta$ or fix it to a constant, referred to \textit{Supplement S.7.2} for sensitivity analysis of $\eta$.  
Parameters $\{q_l\}_{l=1}^{L}$ are stick-breaking weights of the DPM. 
We fix $c=1$ as the default total mass parameter in \texttt{BuLTM}. 
For the base measure $G_0$, we recommend fixing it as that in \eqref{priors} rather than assigning it another hyperprior, referred to \textit{Supplement S.2} for justification.

We note that choices of the prior for $\bbeta$ are flexible, either weakly informative or noninformative. 
We suggest a pure noninformative improper prior for $\bbeta$ since it simplifies the form of the posterior and its gradient so as to speed up the MCMC sampler. 
Even though the prior for $\bbeta$ is improper, the following theorem tells very mild conditions such that the posterior in \eqref{posterior} is still proper. 
\begin{Theorem}
\label{theo:proper}
With the improper uniform prior for $\bbeta$, the posterior distribution in \eqref{posterior} is proper under the following conditions: (i) $0<\widetilde{T_i}<\infty$, for $i=1, \ldots, n$, (ii) priors for $\{\psi_l, \nu_l\}_{l=1}^{L}$, $\{p_l\}_{l=1}^L$ in model \eqref{DPM} and $\{\alpha_j\}_{j=1}^{K}$ in model \eqref{ispline} are proper, (iii) $0 < K, L < \infty$ in models \eqref{DPM} and \eqref{ispline}, (iv) the kernel $f_w$ in model  \eqref{DPM} satisfies that $xf_w(x) < \infty$ for all $x>0$, (v) let $\Z^*$ be the $n_1 \times p$ matrix of the covariates of uncensored observations, where $n_1 = \sum_{i=1}^{n}\delta_i$, and $\Z^*$ is of full rank $p$. 
\end{Theorem} 
\noindent {This theorem indicates that the impact of the prior for $\bbeta$ on the prediction is inferior to that of priors for nonparametric components. 
The proof and justifications for the above conditions are deferred to \textit{Supplement S.5}. 
}

We implement the NUTS in \texttt{Stan} as our MCMC sampler since the domain of $\Theta$ is continuous. 
NUTS is a tuning-free extension of Hamilton Monte Carlo, which is robust and efficient for continuous-variable models.
\texttt{Stan} has become popular and appealing in recent years since it provides clear automatic posterior sampling procedures. 
Therefore, users are released from complicated probabilistic deriving and implementation.
Our \texttt{R} package \texttt{BuLTM} is developed based on \texttt{Stan}.  
We approximate the improper uniform prior for $\bbeta$ through $N(0, 10^6 I_p)$ to avoid possible computational issues caused by improper priors in \texttt{Stan}. 

For prediction purposes, the \textit{posterior predictive survival probability} of a future observation $T_0$ given covariates $\Z_0$, denoted by $S_{T_0|\Z_0}(t)$, is an average of
conditional predictions over the posterior distribution of $\Theta$ \citep[pp.7]{gelman2013bayesian}. 
Mathematically, ${S}_{T_0|\Z_0}(t)$ is the integral of product of conditional survival probability given $\Theta$ and $\pi(\Theta|\widetilde{T}, \Z,\delta)$,
\begin{align}\label{PPD}
    {S}_{T_0|\Z_0}(t) = \int S_{T_0|\Z_0}(t|\Theta) \pi(\Theta|\widetilde{T}, \Z,\delta)d\Theta = \int [S_\xi\{H(t) \exp(-{\bbeta}^\T  \Z_0)\}] \pi(\Theta|\widetilde{T}, \Z,\delta)d\Theta,
\end{align}
where $S_\xi$ and $H$ are expressed by elements of $\Theta$ as in \eqref{DPM} and \eqref{ispline}, respectively. 
Note that, alternatively, \eqref{PPD} can also be expressed by $(h, S_\epsilon, \bbeta)$. 
By definition, unidentified MTM \eqref{expmod1} means that collections of triplets $(\bbeta, H, S_\xi)$ generate unique likelihood \eqref{likeli}, which has the same form as $S_{T_0|\Z_0}(t|\Theta)$.  
The uniqueness of $S_{T_0|\Z_0}(t|\Theta)$
determines the uniqueness of $S_{T_0|\Z_0}(t)$ if the posterior $\pi(\Theta|\widetilde{T}, \Z,\delta)$ is proper. 
Numerically, this integral is approximated by averaging all posterior samples. 
Suppose that we have drawn $M$ samples of  $\bbeta$ and sample paths of $H$ and $S_\xi$, denoted by $\bbeta^{(i)}$, ${H}^{(i)}$ and ${S}_\xi^{(i)}$ respectively, for $i=1, \ldots, M$. 
Then the conditional survival probability $S_{T_0|\Z_0}$ and the conditional cumulative hazard $\Lambda_{T|\Z_0}$ are estimated by
\begin{align}
\label{pred}
   \hat{S}_{T_0|\Z_0}(t) = N^{-1}\sum_{i=1}^{M} {S}_\xi^{(i)}\{{H}^{(i)}(t)\exp({\bbeta^{(i)}}^\T  Z)\}, ~\hat{\Lambda}_{T_0|\Z_0}(t) = -\log(\hat{S}_{T_0|\Z_0}(t)). 
\end{align}

\subsection{Posterior projection for parametric estimation}

Note that without any constraints, we assign two weakly informative priors to nonparametric components  $(H, S_\xi)$ or $(h, S_\epsilon)$ and a noninformative prior to $\bbeta$. 
Then the joint posterior \eqref{posterior} of triplet $(h, S_\epsilon, \bbeta)$ is obtained under prior settings in \eqref{priors}. 
Although the posterior of the full set of parameters $(h, S_\epsilon, \bbeta)$ is jointly estimable, the marginal posterior of each component is  meaningless. 
Nonetheless, it is essential for practitioners to have the marginal estimator of the parametric component $\bbeta$  and related quantities such as relative risks $\exp(-\hat{\bbeta}^T \Z)$. 
To this end, let $\bbeta$ be restricted to $||\bbeta||=1$, where $||\cdot||$ is the $L_2$ norm in the Euclidean space. 
Our interest focuses on marginal posterior inference and estimation of the \textit{identified unit vector} $\bbeta/||\bbeta||$, denoted by $\bbeta^*$, hereafter. 

We obtain a Bayes estimator of $\bbeta^*$ through posterior modification. 
This is inspired by a state-of-the-art posterior projection technique. 
In essence, it is to project the marginal posterior of unconstrained $\bbeta$ to the constrained parameter space of $\bbeta^*$. 
Note that the parameter space of $\bbeta^*$, the unit hyper-sphere $||\bbeta^*|| = 1$, is exactly the Stiefel manifold $\text{St}(1, p)$ in $\mathbb{R}^p$. 
Define a metric projection operator into a set $\mathcal{A}$ as the mapping $m_{\mathcal{A}}: \mathbb{R}^p \to \mathcal{P}(\mathcal{A})$, where $\mathcal{P}(\mathcal{A})$ is the power set of $\mathcal{A}$. 
Let $\text{dist}(\bm{x}, \mathcal{A}) = \inf \{||\bm{x}- x^*||, x^*\in \mathcal{A}\}$ be the distance between $\bm{x} \in \mathbb{R}^p$ and $\mathcal{A}$.
The metric projection operator $m_\mathcal{A}$ is determined by
\begin{align*}
    m_{\mathcal{A}}(\bm{x}) = \{x^*\in \mathcal{A}: ||\bm{x}-x^*||=\text{dist}(\bm{x}, \mathcal{A})\}.
\end{align*}

\noindent{Then, the metric projection of any vector $\bm{\beta} \in \mathbb{R}^p$ into $\text{St}(1, p)$ is uniquely determiened as} $m_{\text{St}(1, p)}(\bm{\beta}) = \bm{\beta}/||\bm{\beta}||$ \citep[Proposition 7]{absil2012projection}.
Consequently, the projected posterior distribution of $\bbeta^*$ is always proper by proposition 3 in \cite{sen2022constrained} since
the posterior of $\bbeta$ in \eqref{posterior} is proper and absolutely continuous. 
Note that one only samples the posterior of unconstrained $\bbeta$ and obtains the posterior of $\bbeta^*$ by projection. 
Then the point estimate of $\bbeta^*$ is given by mean or median of the projected posterior.
Numerical studies reveal that our estimator of $\bbeta^*$ enjoys excellent frequentist performance in the sense of low bias and credible intervals that reach the nominal rate, reconciling the frequentist and Bayesian measures of uncertainty quantification. 

In summary, our whole posterior inference procedure takes the following steps,
\begin{enumerate}
    \item[1. ] \textbf{Initialization}. Initialize the MCMC procedure with initial values of $\bm{\alpha}, \bm{p}, \bm{\phi}$ and $\bm{\nu}$ sampled from their priors. Randomly generate an initial for $\bbeta$ so that $||\bbeta|| > 0$. 
    \item[2. ] \textbf{MCMC}. Draw $M$ posterior samples of $\Theta = (\bm{\alpha}, \bbeta, \bm{p}, \bm{\phi}, \bm{\nu})$  from the posterior \eqref{posterior} by NUTS. 

    \item[3. ]  \textbf{Prediction}.  Compute posterior predictive survival functions given $\bm{z}_0$ following \eqref{pred}.

    \item[4. ]  \textbf{Estimation of $\bbeta^*$}. Generate the $i$th posterior sample of  parameter $\bbeta^*$ as $\bbeta^{(i)}/||\bbeta^{(i)}||$, where $\bbeta^{(i)}$ is the $i$th posterior sample of $\bbeta$ drawn in Step 2, for $i=1, \ldots, M$. 
\end{enumerate}


\vspace{-.5cm}

\section{Simulations}
\label{sec:sim}
Extensive simulations are conducted to evaluate the robustness of prediction of failure times by the proposed \texttt{BuLTM} method and performance of the parametric estimation under the nonparametric transformation model setting. 
We compare \texttt{BuLTM} with \texttt{spBayesSurv} by \cite{zhou2018unified}, which provides a unified two-step Bayesian route for fitting and selecting mainstream transformation models of PH, PO, and AFT.
Details about reproducibility and simulation results in highly-censored cases are put into \textit{Supplements S.6.1} and \textit{S.6.2}. 

Simulated failure times are generated following model \eqref{basicLTM}. Under each setting, we generate $300$ Monte Carlo replicates, each with sample size $n = 200$.
The vector of regression coefficients is $\mathbf{\bbeta} = (\beta_1, \beta_2, \beta_3)^\T =(\sqrt{3}/3, \sqrt{3}/3, \sqrt{3}/3)^\T $ such that $||\bbeta||=1$. 
Therefore, the identified $\bbeta^*$ estimated by \texttt{BuLTM} is coincided with the true $\bbeta$ in data generation, leading to the interpretation. 
For covariates $\Z=(z_1, z_2, z_3)$,  we set $z_1 \sim \text{Bin}(0.5)$ indicating a discrete/categorical variable, $z_2, z_3 \sim N(0, 1)$ as continuous variables with correlation coefficient $0.2$, and $z_1$ is independent of  $(z_2, z_3)$. 

We assess the performance of \texttt{BuLTM} under four true model cases including PH, PO, AFT models, and a case where none of them is the true model. 
\begin{eqnarray*}
  \begin{aligned}
    &\textbf{Case 1.} \text{ Non-PH/PO/AFT}: \epsilon \sim  0.5N(0.5, 0.5^2)+0.5EV(\log(1.5), 1),\\
    & h(t) = \log[(0.6t+0.78t^{1/2}+0.745)\{0.5\Phi_{0.5, 1}(t) + 0.5\Phi_{4, 0.5}(t) - c_1\}], C \sim \text{U}(4.5, 5.5);\\
   &\textbf{Case 2.} \text{ PH model}: \epsilon \sim \text{EV}(0,1), \\
   & h(t) = \log[(t+1.213t^{1/2}+1.5)\{0.5\Phi_{0.5, 1}(t) + 0.5\Phi_{3.5, 0.3}(t) - c_2\}],  C \sim \text{U}(1, 5);\\
   &\textbf{Case 3:} \text{ PO model}: \epsilon \sim \text{Logistic}(0, 1^2), \\
   & h(t) = \log[(t+1.213t^{1/2}+1.5)\{0.5\Phi_{1, 0.5}(t) + 0.5\Phi_{4.5, 0.3}(t) - c_3\}],  
   C \sim \text{U}(3.5, 5);\\
   &\textbf{Case 4:} \text{ AFT model}: \epsilon \sim \text{Normal}(0, 1), 
    h(t) = \log(t),  
   C \sim \text{U}(2.5, 5).
  \end{aligned}
\end{eqnarray*}
Here $\Phi_{\mu, \sigma}$ denotes the CDF of $N(\mu, \sigma^2)$, $\text{EV}$ denotes the extreme value distribution  such that its exponential follows $\text{Weibull}\{\exp(a), 1/b\}$, and $c_k$ is the constant such that $\exp\{h(0)\}=H(0)=0$, for $k=1, 2, 3$.
The censoring variable $C$ is generated independent of $\Z$, leading to approximately $27\%, 29\%$, $24\%$, and $25\%$ censoring rates respectively. 

Case 1 can neither be expressed by any of PH, PO, and AFT models nor be incorporated by the Box-Cox transformation models in \cite{de2014bayesian}. 
In Case 2, 
$
S_{T|\Z}(t) = \exp\{-\exp[h(t)]\exp(-\bbeta^T \Z)\}
$. 
Therefore, the conditional hazard function is
$$
\lambda_{T|\Z}(t) =  \exp[h(t)]h'(t)\exp(-\bbeta^T \Z), 
$$
which is exactly a PH model. 
In Case 3, 
$
S_{T|\Z}(t) = \{1+\exp[H(t)] \exp(-\bbeta^T \Z)\}^{-1}. 
$
Then, the conditional odds function is 
$$
\frac{1-S_{T|\Z}(t)}{S_{T|\Z}(t)} = \exp[h(t)]\exp(-\bbeta^T\Z), 
$$
which is exactly a PO model. 
Case 4 is exactly exactly an AFT model. 
Therefore, in these three cases the regression coefficients $\bbeta$ have the same interpretation as that of PH, PO, and AFT models, respectively. 

\vspace{-.5cm}
\subsection{Prediction of conditional survival probability}
We assess the accuracy of the prediction of failure times and visualize predictive survival probability and cumulative hazard functions. 
Following \eqref{pred}, \texttt{BuLTM} computes the PPD by posterior samples of triplet $(H, S_\xi, \bbeta)$. 
The accuracy of prediction is assessed by the  $L_2$ distance between real conditional survival curves and the PPD.
Numerically, the $L_2$ distance is approximated by root integrated square error (RISE) on the observed time interval. 
The smaller RISE, the better the prediction. 
For each prediction scenario, we compare PPDs of three new observations with different sets of covariates: $\Z_1 = (0, 0, 0)^\T , \Z_2= (1, 1, 1)^\T $ and $\Z_3 = (0, 1, 1)^\T $, respectively.

Table \ref{TPreNon} shows that, under these three sets of new observations, \texttt{BuLTM} overwhelmingly outruns \texttt{spBayesSurv} in performance of predicting conditional survival probability under non-PH/PO/AFT, PH, and PO models, and is comparable with \texttt{spBayesSurv} under the AFT model.
It is reasonable that \texttt{BuLTM} is superior to \texttt{spBayesSurv} in Case 1 since the non-PH/PO/AFT model is beyond the application scope of \texttt{spBayesSurv};
\texttt{BuLTM} still outperforms \texttt{spBayesSurv} with smaller $L_2$ distance under Cases 2 and 3, where the true model is PH and PO, respectively.
In addition, for the first three cases, even for estimating baseline survival probability that determines the approach of \texttt{spBayesSurv}, which corresponds to the prediction case where all covariates are zero, \texttt{BuLTM} still surprisingly outplays \texttt{spBayesSurv}. 
Once the underlying model is the AFT model, \texttt{spBayesSurv} outperforms. 
\begin{table}[H]
  \centering
  \small
  \def~{\hphantom{0}}
  \tabcolsep 8pt
  \caption{\label{TPreNon}\footnotesize{The RISEs between the conditional survival curves and true curves predicted by \texttt{BuLTM} (MTM) and \texttt{spBayesSurv} under Cases 1 to 4. Data size $n=200$. } }
  \footnotesize
    \begin{tabular}{c|cccc|cc|cc|cc}
      \hline
      \multicolumn{1}{c}{}& \multicolumn{4}{c}{Case 1: Non- PH/PO/AFT} & \multicolumn{2}{c}{Case 2: PH} &
      \multicolumn{2}{c}{Case 3: PO} &  \multicolumn{2}{c}{Case 4: AFT}\\
      \hline
       $\Z$ & MTM  &PH  & PO &AFT & MTM &  PH  & MTM & PO  & MTM & AFT\\
     $\Z_1$ & \textbf{0.063} & 0.068 & 0.071 & 0.142 & \textbf{0.067} & 0.073 & \textbf{0.078} & 0.083 & 0.074  & \textbf{0.060}\\ 
      $\Z_2$ & \textbf{0.086} & 0.148 & 0.097 & 0.122 & \textbf{0.140} & 0.229
      & \textbf{0.118} & 0.122 & 0.104   & \textbf{0.090} \\ 
      $\Z_3$ & \textbf{0.136} & 0.245 & 0.163 & 0.221 & \textbf{0.130} & 0.220
      & \textbf{0.120} & 0.128 & 0.113  & \textbf{0.095}\\ 
      \hline
    \end{tabular}
\end{table}
\noindent It is not surprising since the AFT model is an ``ideal" linear regression with the \textit{log} transformation on the time-to-event. 
On the other hand, it might be evidence that the transformation function plays a dominating role in the NTM. 
\begin{figure}[H]
    \centering
\includegraphics[width=1.5in]{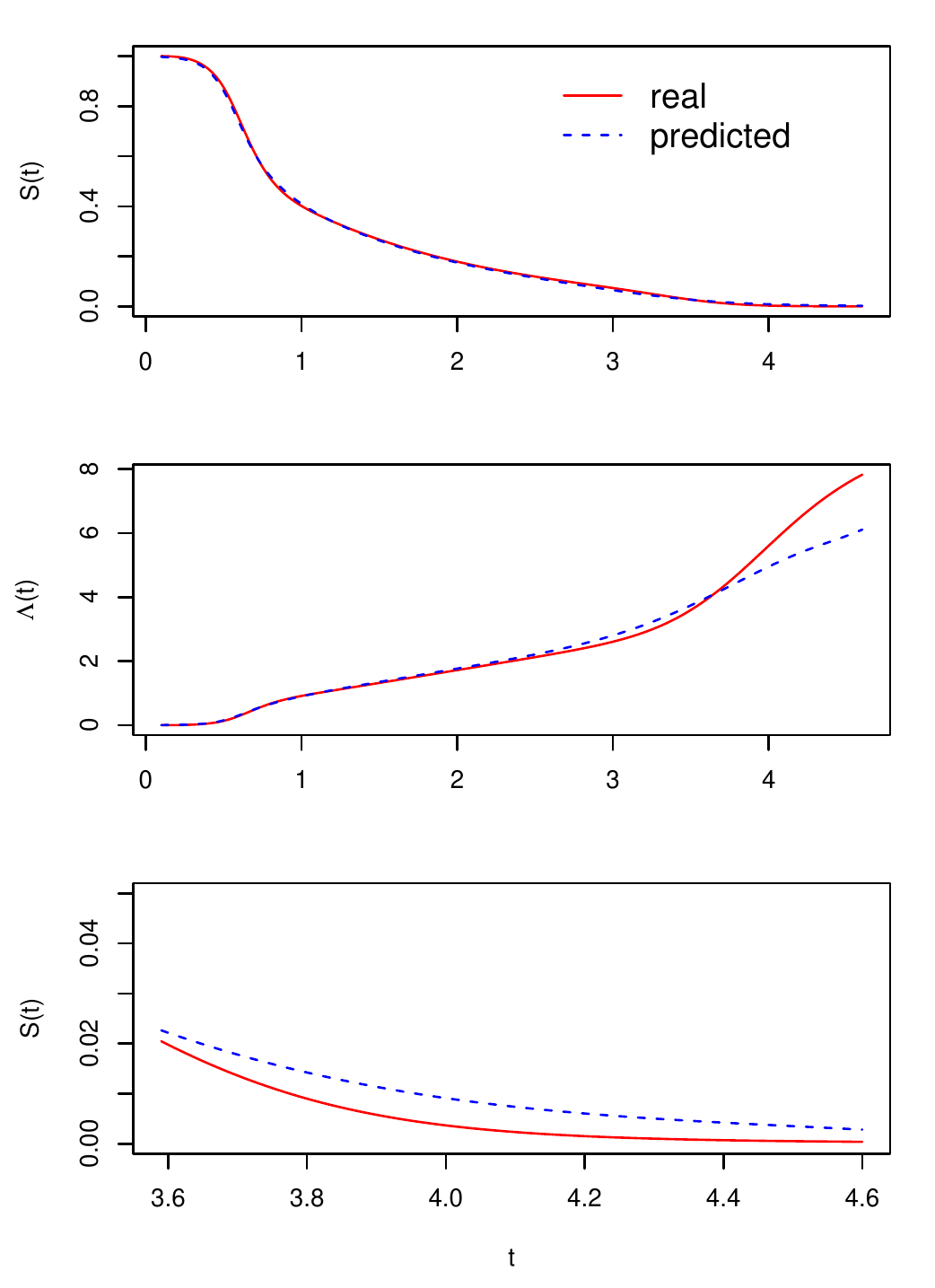}
\includegraphics[width=1.5in]{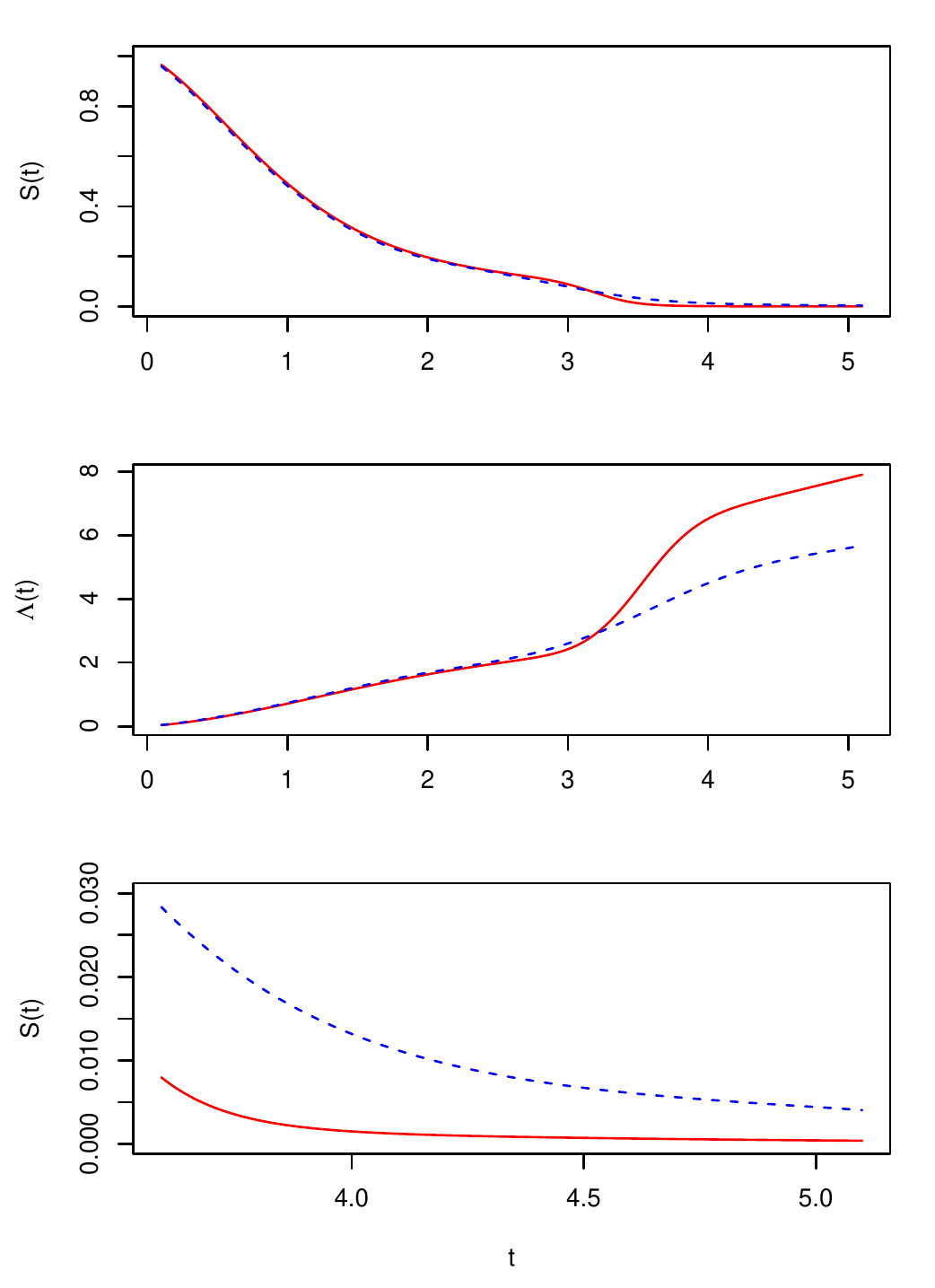}
\includegraphics[width=1.5in]{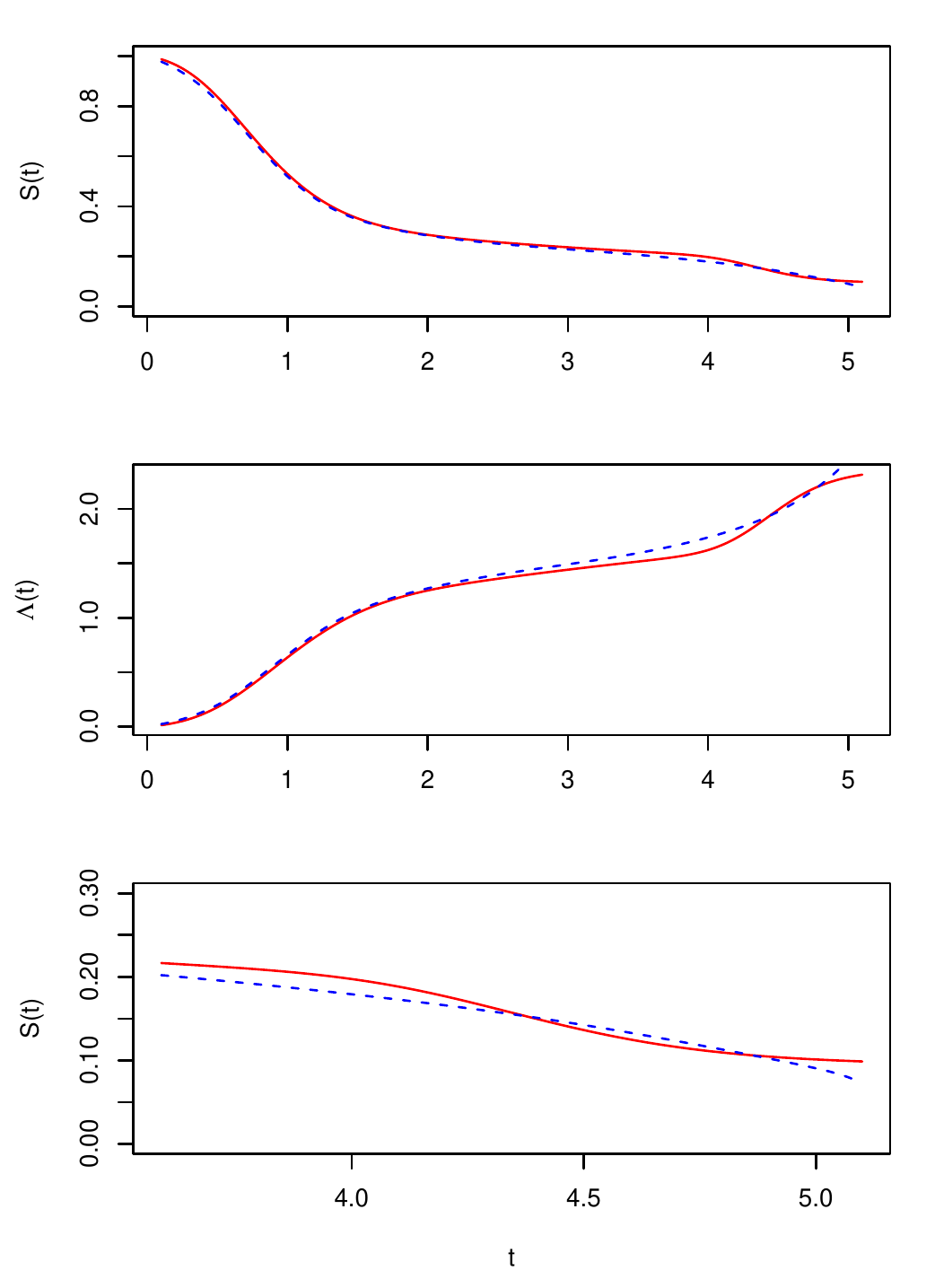}
\includegraphics[width=1.5in]{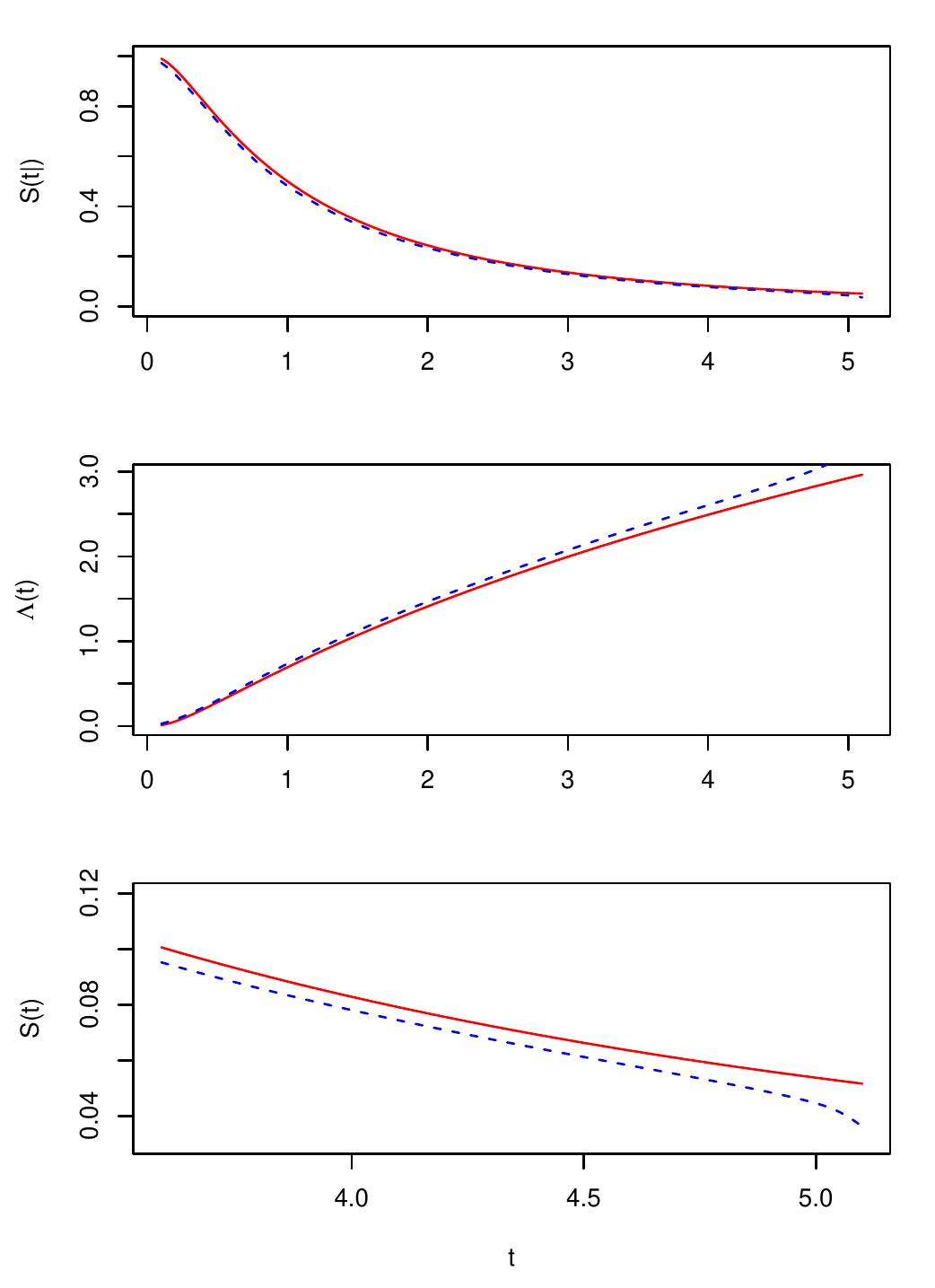}
\caption{\footnotesize The predicted conditional survival probability curve ($S(t)$) and the conditional cumulative hazard function ($\Lambda(t)$) for $Z=(0, 0, 0)^\T $; row 1: survival probability; row 2: cumulative hazard functions; row 3: tails of survival probability; column 1: Case 1; column 2: Case 2; column 3: Case 3; column 4: Case 4. }
\label{FigNon}
\end{figure}

Next, Figure \ref{FigNon} displays the predicted baseline survival probability curves and corresponding baseline cumulative hazard curves for Cases 1-4. As shown in the first row of Figure \ref{FigNon}, \texttt{BuLTM} fits baseline survival probability curves pretty well. 
In terms of baseline cumulative hazard curves shown in the second row, we find some deviation at tails in Cases 1 and 2.
This is reasonable by the zoomed-in tail analysis of survival probability curves shown in the third row of Figure \ref{FigNon}.
Despite negligible bias in the tail of the survival probability, the corresponding cumulative hazard function deviates significantly as its log transformation.

\vspace{-.3cm}
\subsection{Parametric estimation}
We evaluate the performance of \texttt{BuLTM} in estimating the identified parameter $\bbeta^*$, which has the same interpretation as the true unit vector $\bbeta$ in all simulation settings. 
We consider the following frequentist operating characteristics for evaluation, the average bias of estimates (BIAS), the square root of the mean squared error of the estimator (RMSE), the average posterior standard error (PSD), the standard error of the estimated values (SDE), and the coverage probability of the $95\%$ credible interval (CP), as usual.
The pointwise bias of \texttt{BuLTM} should be computed in a different way from \texttt{spBayesSurv}. 
Among all simulations, we re-scale the mean vector of estimated $\hat{\bbeta}^*$ into a unit vector and then compute the pointwise bias.
Otherwise, the result is surely biased no matter what kind of unit-norm estimator is used. 
The reason is that \texttt{BuLTM} provides an estimate of a unit vector in each replication of simulations, while the element-wise mean of a series of unit vectors is not a unit vector anymore since for unit vectors $v_1, \ldots, v_n \in \text{St}(1, p)$, $||n^{-1}\sum_{i=1}^n v_i||\le 1$ by triangle inequality. 

Results of parametric estimation are summarized in Table \ref{ParaPHPO} for Cases 1-3. Results under the AFT model are put into \textit{Supplement S.6.3}. 
It is worth noting that the interpretation of the true $\bbeta$ in Case 1 is different from that of any PH, PO, and AFT models fitted by \texttt{spBayesSurv}. 
Therefore, none of PH, PO, and AFT models provides reasonable parametric estimation in Case 1, and we leave the place of their assessment results blank. 
In contrast, the parametric estimation given by \texttt{BuLTM} has little bias, the PSD is quite close to the SDE, and the CP is close to the nominal level in this case. 
In Cases 2 and 3, where the true model is one of PH and PO models, \texttt{BuLTM} has a lower bias for almost all parameters and has lower RMSE for all parameters than \texttt{spBayesSurv}. 
These results demonstrate that \texttt{BuLTM} estimates the fully identified parameter $\bbeta^*$ well.


\begin{table}[H]
  \centering
  \small
  \def~{\hphantom{0}}
  \tabcolsep 3pt
  \caption{\label{ParaPHPO}\footnotesize{The performance of parametric estimation of \texttt{BuLTM} and \texttt{spBayesSurv} under Cases 1-3. }}
  \footnotesize
    \begin{tabular}{cc|ccccc|ccccc}
     \hline

     \hline
     \multicolumn{2}{c}{Case 1: Non-PH/PO/AFT}& \multicolumn{5}{c}{BuLTM}& \multicolumn{5}{c}{spBayesSurv}\\
     
     \hline
     & Parameter & BIAS   &  RMSE &  PSD  &  SDE  &  CP  &   &   &   &   &  \\ 
      & $\beta_1 $ &  0.005 & 0.065 & 0.066 & 0.065 & 94.3 &  &  &  &  & \\
    & $\beta_2 $ &  -0.008 & 0.051 & 0.050 & 0.050 & 95.3 &  & &  &  & \\
    & $\beta_3 $ & -0.013 & 0.052 & 0.050 & 0.051 & 91.7 &  & &  &  & \\
      \hline
      \multicolumn{2}{c}{}& \multicolumn{5}{c}{Case 2: PH} & \multicolumn{5}{c}{Case 3: PO} \\
      \hline
      Method      & Parameter & BIAS   &  RMSE &  PSD  &  SDE  &  CP  &  BIAS  & RMSE  &  PSD  &  SDE  & CP   \\ 
      BuLTM      & $\beta_1$  & \textbf{-0.013} & 0.123 & 0.123 & 0.121 & 94.7 & 0.012 & 0.169 & 0.174 & 0.167 & 94.0 \\ 
                  & $\beta_2$  & \textbf{0.006} & 0.083 & 0.087 & 0.081 & 95.0 & \textbf{-0.008} & 0.130 & 0.123 & 0.123 & 92.3 \\ 
                  & $\beta_3$  & \textbf{0.006} & 0.086 & 0.088 & 0.085 & 95.0 & \textbf{-0.005} & 0.130 & 0.122 & 0.124 & 94.0 \\ 
     spBayesSurv & $\beta_1$  & -0.032 & 0.172 & 0.175 & 0.170 & 95.0 &  \textbf{0.002} & 0.258 & 0.256 & 0.259 & 94.7 \\ 
                  & $\beta_2$  & -0.026 & 0.088 & 0.095 & 0.084 & 95.3 &  0.010 & 0.142 & 0.136 & 0.142 & 94.7 \\ 
                  & $\beta_3$. & -0.027 & 0.102 & 0.095 & 0.098 & 93.0 &  0.013 & 0.135 & 0.136 & 0.135 & 95.0 \\
      \hline
   \end{tabular}
\end{table}

\begin{figure}[H]
  \centering
  \subfigcapskip=-18pt
  \subfigure[]{
    \begin{minipage}[t]{0.3\linewidth}
      \centering
      \includegraphics[width=2in]{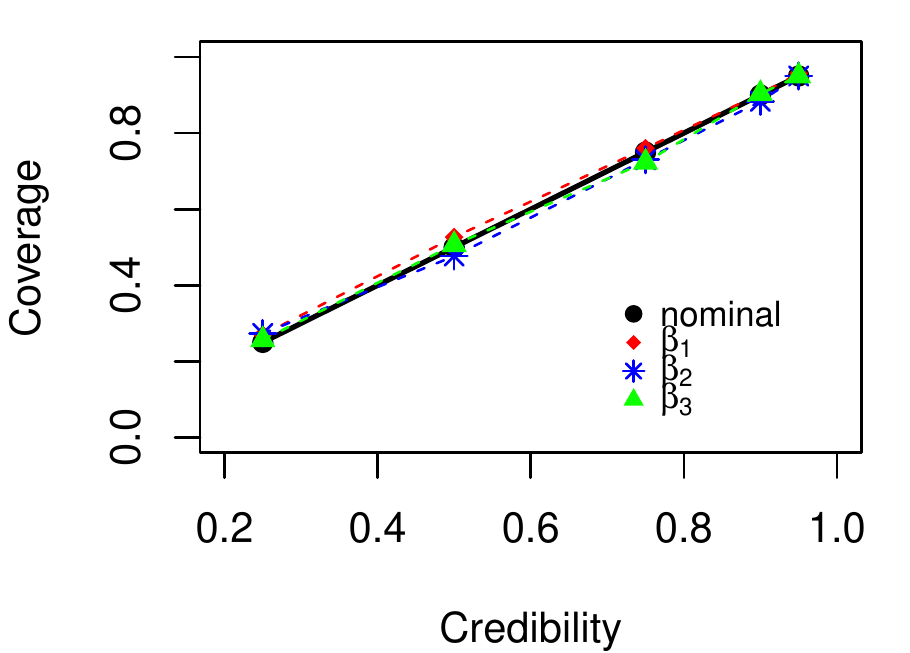}
      \label{CoverNonLow}
    \end{minipage}
}
  \subfigure[]{
    \begin{minipage}[t]{0.3\linewidth}
      \centering
      \includegraphics[width=2in]{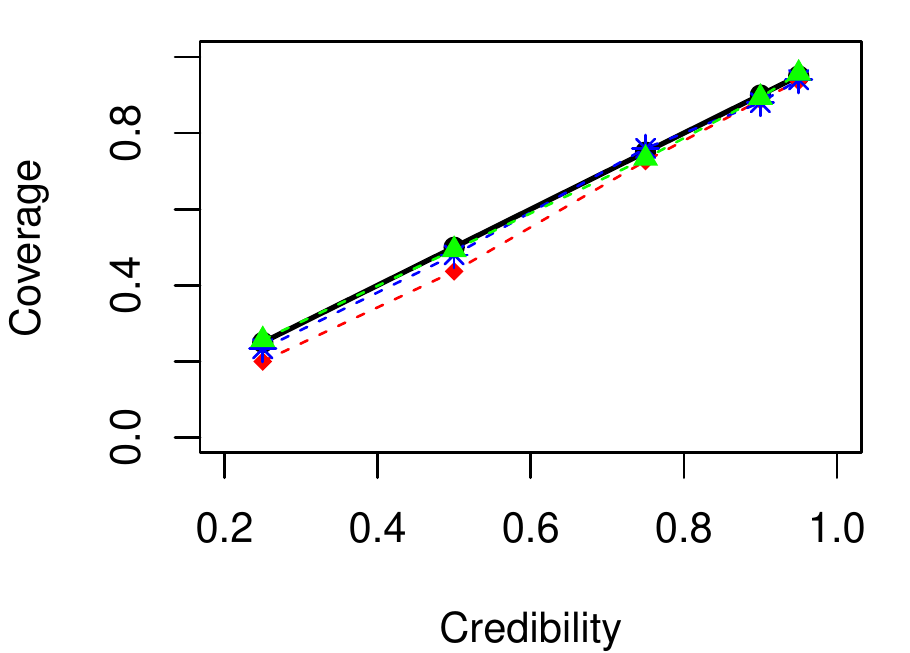}
      \label{CoverPHLow}
    \end{minipage}
}
\subfigure[]{
    \begin{minipage}[t]{0.3\linewidth}
      \centering
      \includegraphics[width=2in]{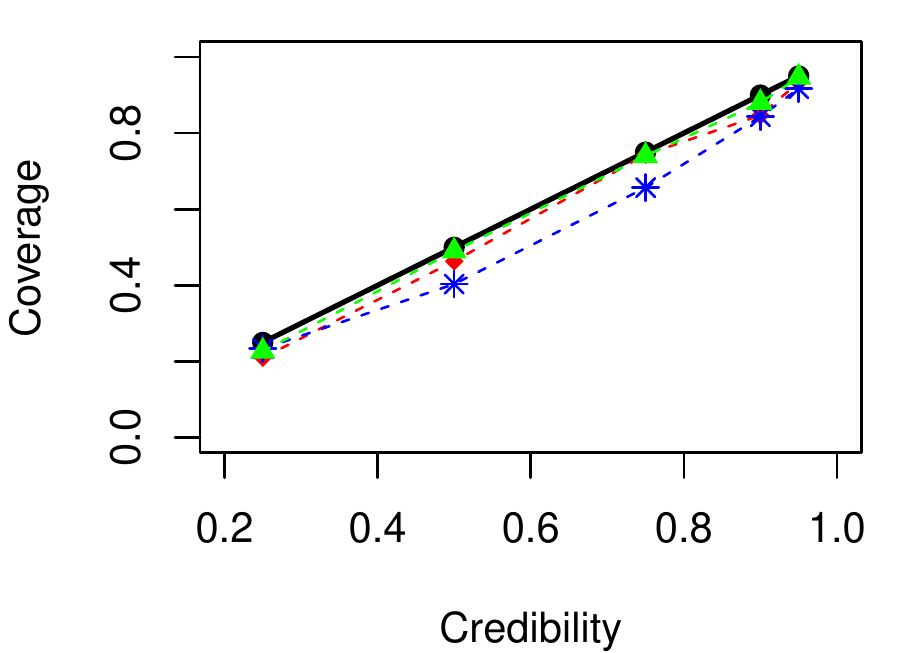}
      \label{CoverPOLow}
    \end{minipage}
}
\vspace{-.6cm}

\caption{\footnotesize  Coverage of credible intervals given different credibility; (a), Case 1; (b), Case 2; (c), Case 3.}
\label{FigCov}
\end{figure}
We visualize obtained coverage of credible intervals given credibility at 25\%, 50\%, 75\%, 90\%, and 95\%.
Figure \ref{FigCov} shows plots of obtained coverage in Cases 1-3. 
We find that for Non-PH/PO/AFT and PH models, coverage of all parameters is close to nominal rates given different credibility levels; while
for the PO model, coverage of $\beta_2$ has deviation under credibility levels $50\%$ and $75\%$.
It implies that the posterior sampled by \texttt{BuLTM} describes the true posterior well and thus interval estimation is precise under almost all credibility levels.

\section{Applications}
\label{sec:app}
\subsection{PO case: veterans lung cancer data}
The first example is the veterans lung cancer dataset from \texttt{R} package \texttt{survival} \citep{Terry2022}. 
It contains $137$ patients from a randomized trial receiving either a standard or a test form of chemotherapy. 
In the study, the failure time is one of the primary endpoints for the trial and $128$ patients were followed to death. 
We include six covariates, the first five of which are $Z_1 = \text{karno}/10$ (karnofsky score), $Z_2 = \text{prior}/10$ (prior treatment, with 0 for no therapy and 10 otherwise), $Z_3 = \text{age}/100$ (years), $Z_4 = \text{diagtime}/100$ (time in months from diagnosis to randomization), and $Z_5 = I (\text{treatment} = \text{test form of chemotherapy})$. The remaining is the covariate of the cell type which has four categories, adeno, squamous, small cell, and large cell.
Thus we include indicator variables to associate with  time-to-death, that is, $Z_6 = I(\text{cell type} = \text{squamous}), Z_7 = I (\text{celltype} = \text{small}), ~\text{and} ~Z_8 = I (\text{celltype} = \text{large})$.  

The proposed method is implemented by \texttt{R} package \texttt{BuLTM}. 
Recall that \texttt{spBayesSurv} fits three survival models first and then selects one, and it selects the PO model in this case.

\begin{figure}[H]
  \subfigcapskip=-14pt
  \centering
  \subfigure[]{
    \begin{minipage}[t]{0.45\linewidth}
      \centering
      \includegraphics[width=2.5in]{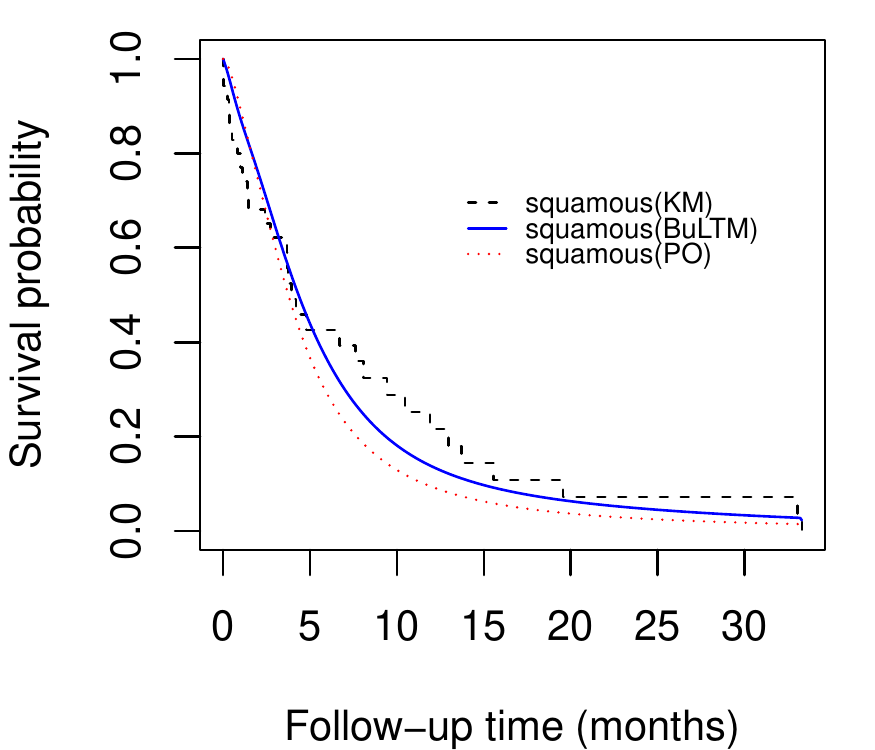}
      \label{Squamous}
    \end{minipage}
}
  \vspace{-.5cm}
  \subfigure[]{
    \begin{minipage}[t]{0.45\linewidth}
      \centering
      \includegraphics[width=2.5in]{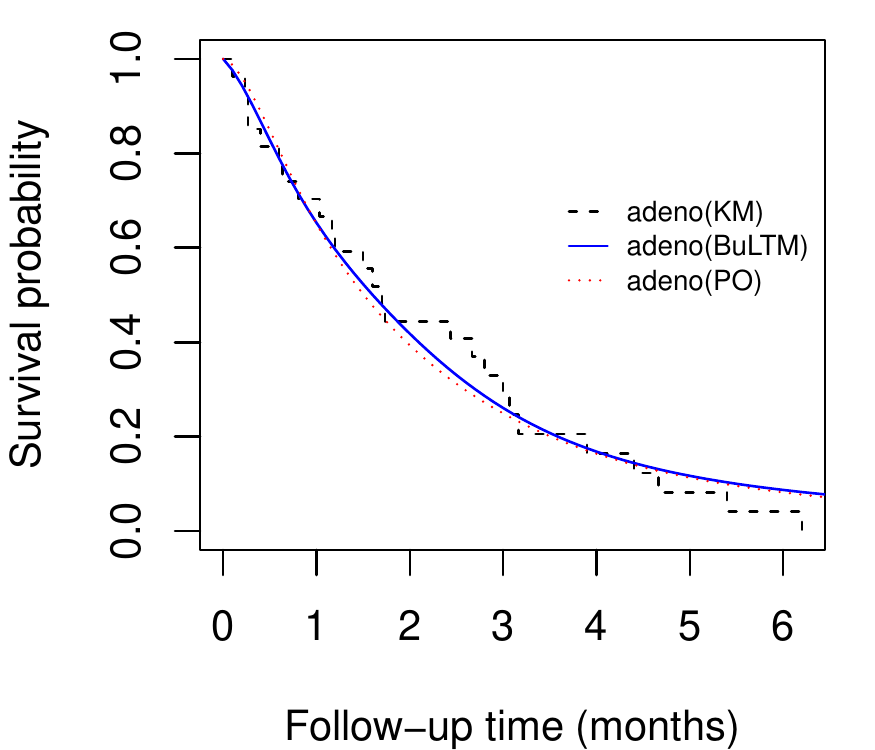}
      \label{Adeno}
    \end{minipage}
}
\caption{\footnotesize Estimated curves of survival probability given by BuLTM, spBayesSurv, and the K-M estimator under strata categorized by celltypes; (a) the stratum of squamous; (b) the stratum of large cell.   }
\label{Lung}
\end{figure}

\noindent \textbf{Prediction} 
We compare the curves of estimated survival probability given by \texttt{BuLTM} with that of  \texttt{spBayesSurv} first. 
We divide the dataset into four strata based on their cell types. 
For each stratum, the survival curves given by \texttt{BuLTM} and \texttt{spBayesSurv} are estimated through the predicted survival probability conditional on the mean values of covariates of all individuals within the stratum. 
For comparison, we use the Kaplan-Meier (K-M) estimator of that stratum as the baseline result.
Figure \ref{Squamous} and \ref{Adeno} display the results of estimated survival curves. 
For the squamous stratum, the survival curve given by \texttt{BuLTM} is significantly closer to the K-M estimator than that of \texttt{spBayesSurv};
for the adeno stratum, the survival curve given by \texttt{BuLTM} is slightly closer to the K-M estimator in the middle range of the following-up period.  
Since \texttt{BuLTM} and \texttt{spBayesSurv} perform similarly to each other on the remaining two strata, we simply omit their results here. 
The comparison with the K-M estimator supports the nonparametric transformation model. 

To further compare their predictive capability, we randomly split the full dataset into the training and testing sets with proportions $90\%$ and $10\%$, respectively. 
We repeat this procedure 10 times. 
We fit survival models based on the training data first and then predict survival outcomes on the testing set. 
The prediction capability is assessed by the commonly used Concordance index \citep[C index,][] {harrell1982evaluating}, which is an extension of the area under the curve (AUC) as a measure of concordance between a predictive biomarker and the right-censored survival time. 
A higher C index implies better prediction capability of a model. 
In this article, the C index is computed by \texttt{R} package \texttt{SurvMetrics} \citep{Zhou2022RpkgSurvMetriocs} following the procedure in \cite{Ishwaran2008random}. 
Details about metrics for prediction evaluation of survival models in this article are deferred to \textit{Supplement S.10}. 
Since most observations in the example are uncensored, a natural prediction of the survival time of a future observation is the median based on its PPD. 
And then we use this predicted survival time as the diagnostic marker to compute C index. 
We also compute the mean of absolute error (MAE) between the predicted survival times and the true survival times of uncensored observations. 

\begin{figure}[H]
  \centering
    \subfigcapskip=-15pt
  \subfigure[]{
    \begin{minipage}[t]{0.45\linewidth}
      \centering
      \includegraphics[width=2.5in]{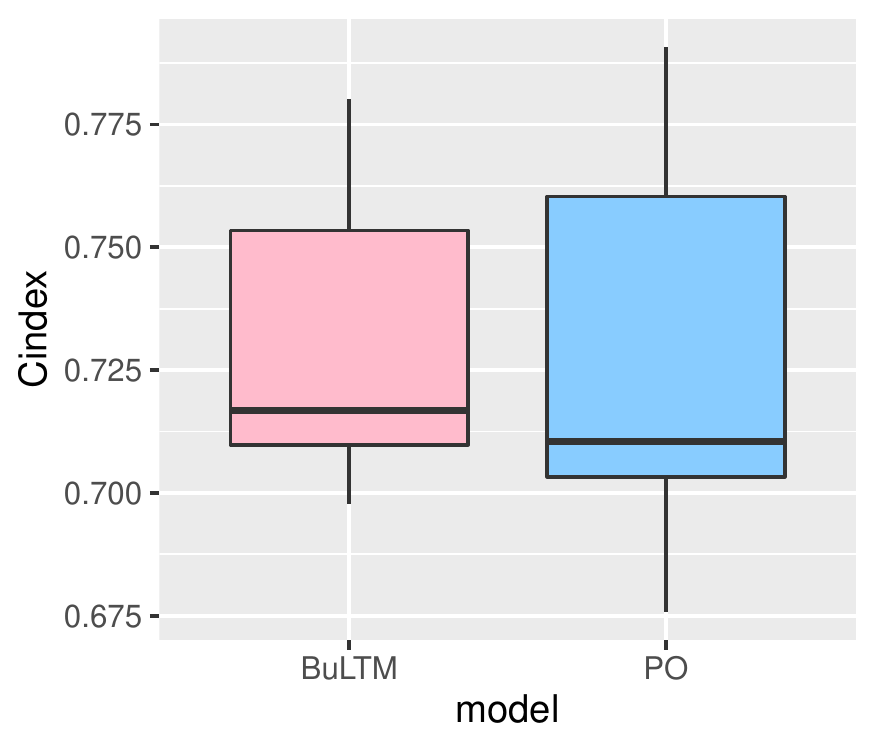}
      \label{Cindex}
    \end{minipage}
}
  \vspace{-.5cm}
  \subfigure[]{
    \begin{minipage}[t]{0.45\linewidth}
      \centering
      \includegraphics[width=2.5in]{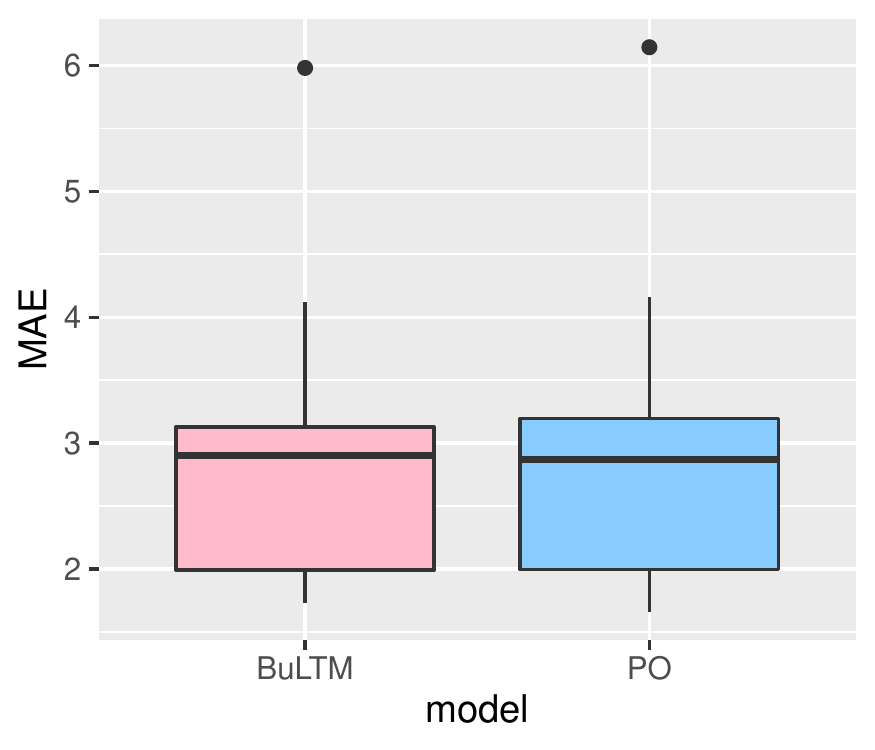}
      \label{MAE}
    \end{minipage}
}
\caption{\footnotesize (a) The box plot of the C index computed on 10 testing sets; 
(b) the box plot of MAE between predicted and true survival times of uncensored observations on 10 testing sets.}
\label{KMplots}
\end{figure}

Figure \ref{Cindex} shows that among $10$ testing sets, the median C index of \texttt{BuLTM} is higher than that of \texttt{spBayesSurv}. 
Although \texttt{spBayesSurv} provides a relatively higher C-index in the best case, it is worse than \texttt{BuLTM} in the worst case. 
The average C index given by \texttt{BuLTM} (0.729) is also slightly higher than that of \texttt{spBayesSurv} (0.725). 
This is consistent with the result of the MAE assessment. 
As shown by Figure \ref{MAE}, the MAE of predicted survival times given \texttt{BuLTM} and \texttt{spBayesSurv} have almost the same median and $25\%$ quantile, while \texttt{BuLTM} has lower  $75\%$ quantile and the maximum MAE in the worst case. 
These two results demonstrate that \texttt{BuLTM} has better predictive capability on this dataset. 

\noindent \textbf{Estimation of relative risks} 
In terms of estimation of relative risks, we add the smoothed partial rank (SPR) estimator \citep{song2007semiparametric} into our comparison. 
Although quantitative interpretations of $\bbeta$ ($\bbeta^*$ in \texttt{BuLTM}) under different models are different,  their qualitative interpretations such as the relative importance of the predictors such as relative importance of treatment effects are relatively stable \citep{solomon1984effect}. 
Our analysis demonstrates this point of view since results of parametric estimation given by different methods are consistent, referred to \textit{Supplement S.8.1}. 

\begin{figure}[H]
  \centering
    \subfigcapskip=-15pt
  \subfigure[]{
    \begin{minipage}[t]{0.45\linewidth}
      \centering
      \includegraphics[width=2.7in]{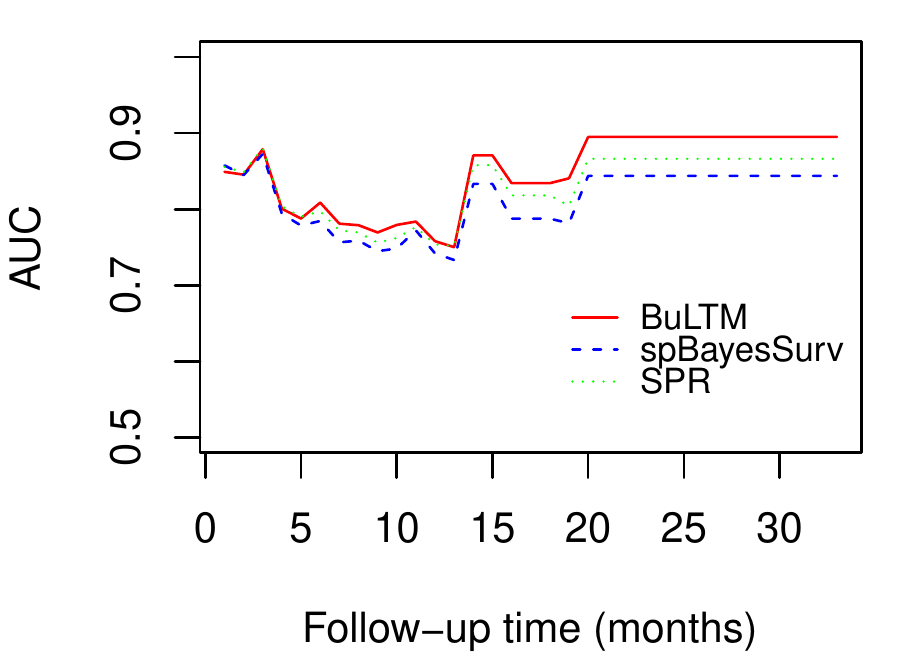}
      \label{LungAUC_KM}
    \end{minipage}
}
  \vspace{-.5cm}
  \subfigure[]{
    \begin{minipage}[t]{0.45\linewidth}
      \centering
      \includegraphics[width=2.7in]{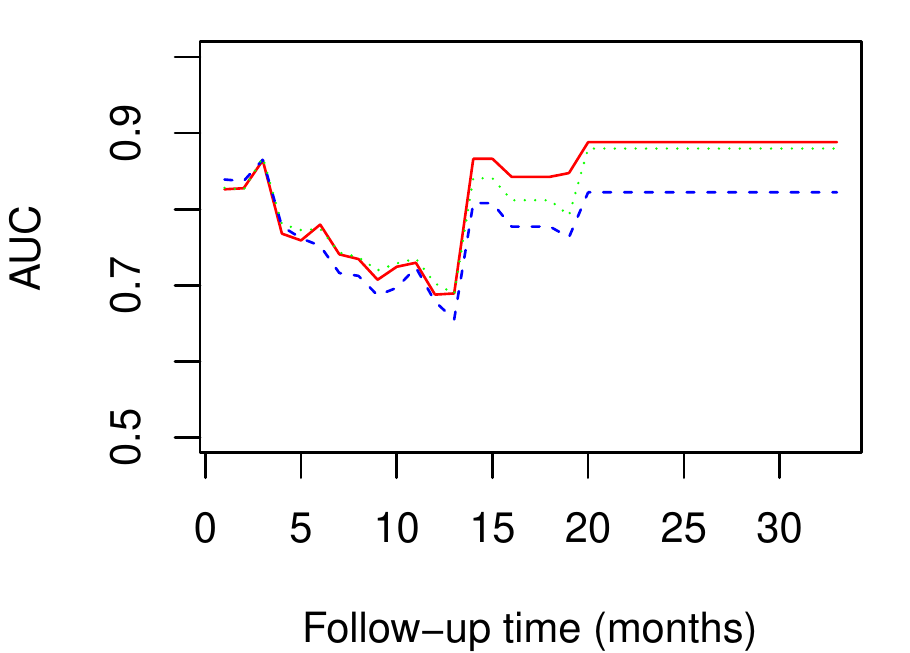}
      \label{LungAUC_NNE}
    \end{minipage}
}
\caption{\footnotesize  Time dependent survival $\text{AUC}(t)$ computed by estimated relative risks. (a), method ``K-M"; (b), method ``NNE".}
\label{LungScoreAUC}
\end{figure}

According to the model selection result by \texttt{spBayesSurv}, the underlying survival model of this dataset is more possible to be the PO model. 
Under the PO model, the odds given covariates $\Z$ are proportional to the relative risk $\exp(-\bbeta^T\Z)$ at any time $t$. 
Hence, it is important to evaluate the estimated relative risk $\exp(-\hat{\bbeta}^T \Z)$ given by the above three methods ($\exp(-\hat{\bbeta}^{*T} \Z)$ by \texttt{BuLTM}). 
Naturally, we assess the estimated relative risk through the area under the time-dependent $\text{ROC(t)}$ curve (AUC) for censored failure time by treating the survival status as a binary response.  
Figure \ref{LungScoreAUC} displays the dynamic AUCs using the estimated relative risks given by \texttt{BuLTM}, \texttt{spBayesSurv}, and SPR as diagnostics. 
We find \texttt{BuLTM} is superior to the other two methods evaluated by both survival AUCs computed by the K-M and the nearest neighbor estimator (NNE) methods presented by \cite{heagerty2000time}. 

\subsection{PH case: heart failure clinical records data}
We apply \texttt{BuLTM} to analyze the heart failure clinical records data first published by \cite{chicco2020machine}.
The dataset records 299 heart failure patients collected at the Faisalabad Institute of Cardiology and at the Allied Hospital in Faisalabad, from April to December in 2015 \citep{ahmad2017survival}. 
The dataset consists of 105 women and 194 men, with a range of ages between 40 and 95 years old. 
In the dataset, 96 observations are recorded as death and the remaining 203 are censored, leading to a censoring rate of 67.9\%, which is relatively high. 
The dataset contains 11 covariates reflecting one's clinical, body, and lifestyle information. 
Among the 11 covariates, 5 of them are binary variables: anaemia, high blood pressure, diabetes, sex, and smoking. 
The dataset considers a patient having anaemia if haematocrit levels were lower than 36\%, while 
the criterion for high blood pressure is unclear in the study. 
Other continuous covariates are age (year), creatinine phosphokinas (level of the creatinine phosphokinas enzyme in the blood, mcg$/$L), ejection fraction (percentage of blood leaving the heart at each contraction), platelets (platelets in blood, kiloplatelets$/$mL), serum creatinine (level of creatinine in blood, mg$/$dL), and serum sodium (level of sodium in blood, mEq$/$L). 
The survival times are recorded in days. 
In our data pre-processing, we transfer the survival time to months by days$/30$.  
We report the results of prediction here compared with \texttt{spBayesSurv}. 
Parametric estimation results and estimation of relative risks given by the two methods are similar and deferred to \textit{Supplment S.8.2}. 

\noindent \textbf{Prediction} Likewise, we compare the curves of estimated survival probability given by \texttt{BuLTM} with that of \texttt{spBatesSurv} first. 
In this case, \texttt{spBayesSurv} selects the PH model. 
We consider two strata of observations: the high-risk (HR) stratum where observations have both anaemia and high blood pressure, and the low-risk (LR) stratum where observations have neither anaemia nor high blood pressure. 
For each stratum, the survival curves given by \texttt{BuLTM} and \texttt{spBayesSurv} are estimated through the predicted survival probability conditional on the mean values of covariates
of all individuals within the stratum. 
We also use the K-M estimator as the baseline result for comparison.

\begin{figure}[H]
    \centering
    \includegraphics[width=2.8in]{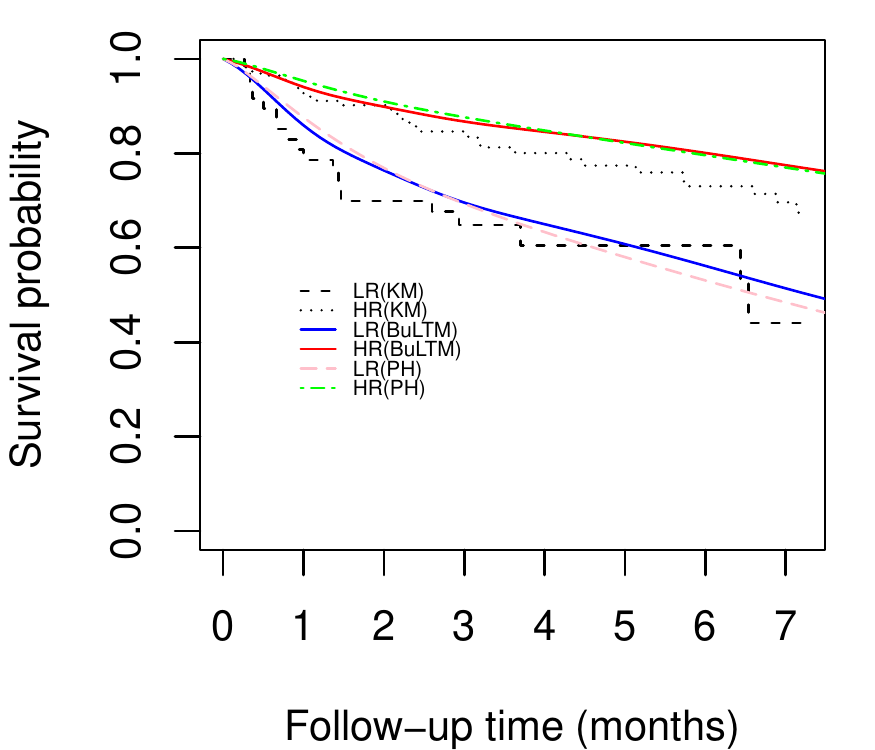}
    \caption{\label{FigHRLR}\footnotesize{Estimated curves of survival probability given by BuLTM, spBayesSurv, and the K-M estimator under high-risk and low-risk strata. }}
\end{figure}

As shown by Figure \ref{FigHRLR}, for the LR stratum, the survival curve estimated by \texttt{BuLTM} is closer to the K-M estimator than that of \texttt{spBayesSurv} both at the beginning follow-up time period and months from 5 to 6, and \texttt{spBayesSurv} is closer to the K-M estimator at other times. 
For the HR stratum, \texttt{BuLTM} performs slightly better at the beginning and provides almost the same result as \texttt{spBayesSurv} at the tail. 
It is reasonable that \texttt{BuLTM} performs better at the beginning time period on this highly-censored dataset since most quantiles of failure times are distributed at the beginning period and the quantile-knots I-splines prior generates more knots at the beginning. 
For comparison of their predictive capability on this dataset, we still randomly split the full dataset into the training and testing sets with proportions 90\% and 10\%, respectively, and repeat this procedure 10 times.
Again, we evaluate the predictive capability by the C index.
According to the censoring rate ($68.9\%$), we select the $70\%$ quantiles of PPDs to compute the C index.  
Besides, we consider the Brier score \citep[BS,][]{graf1999assessment} to assess the prediction curve error i.e. expected value of the square of the difference between the true survival state of a sample and its predicted survival probability at some specific time points. 
To evaluate the BS on all follow-up time intervals, we consider the integral of BS functions (IBS) on a given interval as another assessment. 
As a kind of square error, the lower the IBS, the better the prediction. 
We don't consider the MAE as an assessment in this case since most observations are censored and hence, the MAE loss is meaningless.

\begin{figure}[!htp]
  \centering
      \subfigcapskip=-15pt
  \subfigure[]{
    \begin{minipage}[t]{0.45\linewidth}
      \centering
      \includegraphics[width=2.3in]{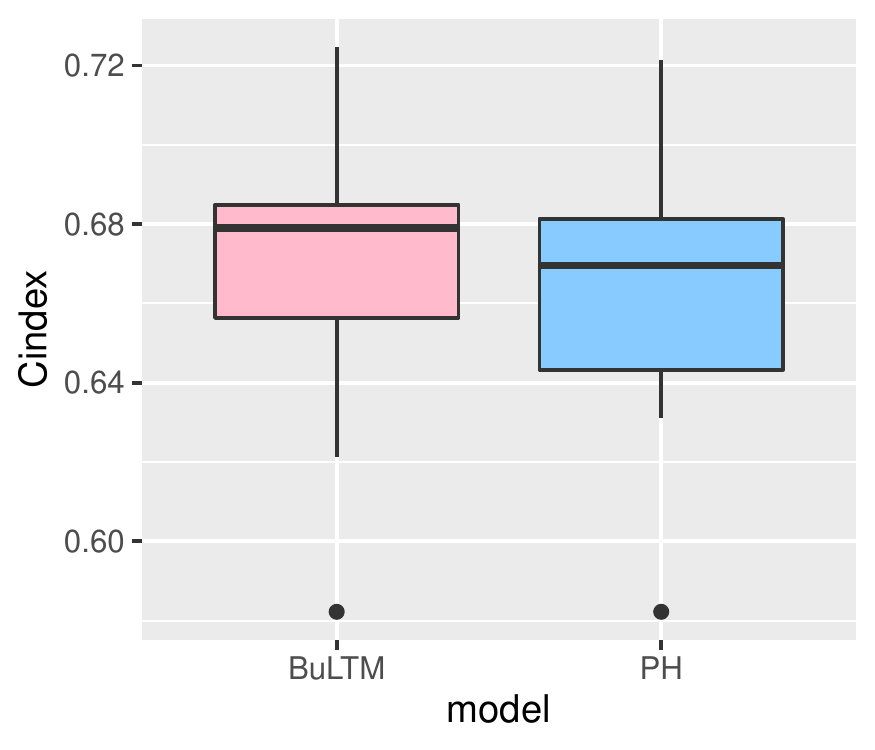}
      \label{CindexHeart}
    \end{minipage}
}
  \vspace{-.5cm}
  \subfigure[]{
    \begin{minipage}[t]{0.45\linewidth}
      \centering
      \includegraphics[width=2.3in]{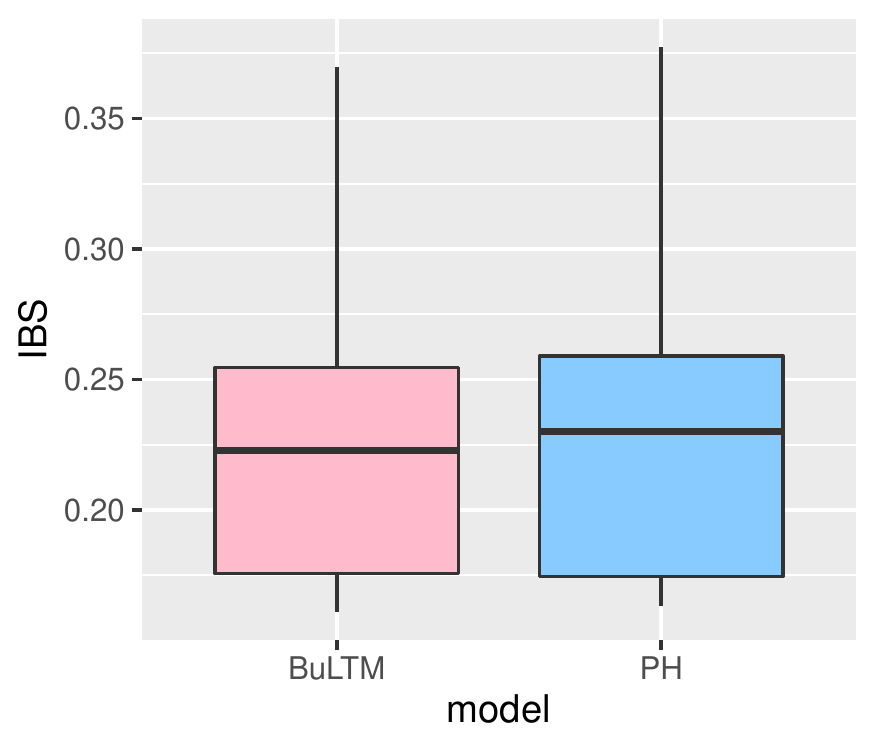}
      \label{IBSHeart}
    \end{minipage}
}

  \label{Transplant}
\caption{\footnotesize{Prediction comparison between BuLTM and spBayesSurv; (a), C index; (b), Integrated Brier score. } }
\end{figure}

As shown by Figure \ref{CindexHeart}, among the 10 testing sets, \texttt{BuLTM} enjoys a higher median and a higher $75\%$ quantile of C indices. 
Meanwhile, the average C index of \texttt{BuLTM} (0.669) is again slightly higher than that of the PH model (0.664). 
In terms of the IBS, as shown by Figure \ref{IBSHeart}, \texttt{BuLTM} enjoys a lower median, $75\%$ quantile, and the maximum value than the PH model among the 10 testing sets. 
The average IBS of \texttt{BuLTM} (0.233) is lower than the average value of the PH model (0.238) too. 
These results support that \texttt{BuLTM} has better out-sample predictive capability on this dataset.

\section{Discussion}
\label{sec:disc}
In this article, instead of imposing strong restrictions to make the NTM identified, 
we assign two weakly informative priors for the nonparametric components,  
the quantile-knots I-splines prior to the transformation function and
a Weibull kernel DPM model to the error distribution, 
and employ a noninformative prior to the parametric component, to achieve prediction through computing PPDs under NTM \eqref{basicLTM}. 
We are not the daredevils to do so since existing literature has had a few explorations in other environments, where weakly informative priors were modeled to avoid burdensome computation caused by constraints for model identification \citetext{\citealp{MccullochRossi1994JOE, Branscum2008SIM, Burgette2021BA, berchuck2021bayesian}; among others}.

We explored the use of constrained priors for $H$ while the posterior on the constrained support is too difficult to sample; see \textit{Supplement S.4.1} for details. 
For posterior inference in \texttt{BuLTM}, although we admit that a few inner points of the posterior surface (percentage less than $0.5\%$) may exceed the maximum tree depth of NUTS \citep{hoffman2014no} in MCMC sampling,
our method enjoys fast convergence and well mixing of MCMC chains with high effective sample size (ESS) in MCMC diagnosis; see \textit{Supplement S.6.4}; 
 the posterior is neither sensitive to subjective choices of hyperparameters in the weakly informative priors nor similar to priors, referred to \textit{Supplements S.7} and \textit{S.9}, respectively.

The coupled  \texttt{R} package \texttt{BuLTM} is a good competitor of  \texttt{spBayesSurv}, which is possibly the best  Bayesian method in existing literature designed to cover the main specials of NTMs, PH, PO, and AFT models. 
In finite sample situations (data size $<600$), \texttt{BuLTM} outperforms \texttt{spBayesSurv} 
in prediction in most cases and is expedient to output results of conditional survival probability, cumulative hazards, and relative risks, not restricting applications to the aforementioned three mainstay transformation models; 
while the two-step \texttt{spBayesSurv} method is remarkable in handling extremely large-scale survival data with a sample size of more than $45000$. 
The superiority of \texttt{BuLTM} might be explained from three aspects. 
The first is the influence of nonparametric priors used in these two methods.
\texttt{BuLTM} combines two weakly informative priors, the newly proposed quantile-knots I-splines prior for $H$ and the DPM prior for $S_\xi$ to compute PPDs;
and \texttt{spBayesSurv} employs the transformed Bernstein polynomial (TBP) prior for baseline survival probability and transfers it to PPDs. 
However, the quantile-knots I-splines prior may be more effective in catching the \textit{majority shape information} of the PPD than the TBP prior, since it \textit{aggregates more knots on the majority of time intervals} (referred to Figure \ref{QuantKnots}) rather than the spirit of \textit{equally spaced knots} in the TBP prior. 
The more knots, the more information.
Secondly, for estimation of the fully identified parametric component, \texttt{BuLTM} enjoys lower RMSE than \texttt{spBayesSurv} since it incorporates the information of $||\bbeta^*||=1$, which significantly reduces the posterior variance after posterior projection.
Finally, the excellent performance of \texttt{BuLTM} may benefit from the  use of NUTS, which is purely designed for sampling of continuous parameters;
whilst \texttt{spBayesSurv} has to design adaptive Metropolis samplers for discrete parameters to incorporate spatially referenced data.

A natural next step work may use the spirit of solving estimation of the NTM to estimation of single index models from the Bayesian perspective; another natural extension is to study random effects models where the nonparametric transformation acts as the functional random effect. 

\bigskip
\begin{center}
{\large\bf SUPPLEMENTARY MATERIAL}
\end{center}

\begin{description}

\item[Title:] Supplementary material for ``Bayesian prediction via nonparametric transformation models". (PDF file)

\item[R-package:] The \texttt{R} package \texttt{BuLTM} is available in GitHub \hyperlink{https://github.com/LazyLaker}{https://github.com/LazyLaker}.
The code and data for real-world data analysis are uploaded.

\end{description}

\begin{center}
    {\LARGE \bf Supplementary materials for ``Bayesian prediction via nonparametric transformation models" }
\end{center}
\renewcommand*{\thesection}{S.\arabic{section}}
\renewcommand*{\thesubsection}{S.\arabic{section}.\arabic{subsection}}
\setcounter{section}{0}

\spacingset{1.7} 

\section{Deriving $H(0)=0$ from assumption (A3)}
\begin{proof}
 Suppose $H(0) = a$ , where $a$ is a positive constant. 
It is natural that $Pr\{T>0\}=1$. 
Then we have
$$
Pr\{T>0\} = \int_{D}  Pr\{T>0|\Z=\bm{z}\}f_\Z(\bm{z})d\bm{z} = 1,
$$
where $D$ denotes the support of covariate $\Z$ and $f_\Z$ denotes the density of $\Z$.  
According to the transformation model, $Pr\{T>0|\Z=\bm{z}\} = Pr\{H(T)>a|\Z=\bm{z}\}=Pr\{\xi\exp(\bbeta^T\bm{z})>a\} = Pr\{\xi>a\exp(-\bbeta^T \bm{z})\}$. 
As a counterexample, we suppose the covariate $\Z = Z \sim N(0, 1)$ is univariate, the model error $\xi \sim \exp(1)$, and $\bbeta = \beta_1=-1$. 
 Since $\xi$ and $Z$ are independent, we have 
 $
Pr\{T>0\} = \int_{\mathbb{R}}\int_{a\exp(z)}^{+\infty} \exp(-t) \phi(z;0, 1)dtdz <1, 
 $
 where $\phi(\cdot; 0, 1)$ denotes the density of $N(0,1)$. 
This contradicts the fact that $Pr\{T>0\}=1$. 
 Therefore, $H(0)=0$. 
\end{proof}

\section{The DPM model for $S_\xi$}
\label{PriorLambda}
A regular Dirichlet process mixture (DPM) model \citep{lo1984class} is assigned for $S_\xi$, the survival probability function of the positive random variable $\xi$. 
The DPM is a kernel convolution to the Dirichlet process (DP).
We use the stick breaking representation for $G \sim \text{DP}(c, G_0)$ \citep{sethuraman1994constructive}
$$
G(\cdot) = \sum_{l=1}^{\infty} p_{_l} \delta_{\theta_{_l}}(\cdot), \theta_{_l} \sim G_0, p_{_l} \sim \text{SB}(1, c)
$$
where $\delta(\cdot)$ is the point mass function, and $\text{SB}$ is the stick-breaking representation. 
We call $G_0$ as the base measure and $c$ as the total mass parameter, acting as the center and precision of the DP, respectively. 

Following the above stick-breaking representation, we construct the truncated DPM priors for $S_\xi$ and $f_\xi$ with the Weibull kernel such that
\begin{align*}
S_\xi = 1- \sum_{l=1}^{L}p_{_l} F_w(\psi_{_l}, \nu_{_l}), f_\xi = \sum_{l=1}^{L}p_{_l} f_w(\psi_{_l}, \nu_{_l}),
p_{_l} \sim \text{SB}(1, c), (\psi_{_l}, \nu_{_l}) \sim G_0, 
\end{align*}
where $L$ is the truncation number, and $F_w$ and $f_w$ denote CDF and density of Weibull distribution, respectively. 
We fix the truncation number $L$ rather than sampling it to simplify computation as a common strategy \citep{rodriguez2008nested}.  
Let $S_\xi^{(\infty)}$ denote the limit of the DPM model, and $S_\xi ^{(L)}$ denote the truncated form.
The truncation number $L$ is generally selected such that the $L_1$ error between the limit form and the truncated form, denoted as $\int_{0}^{+\infty} |S_\xi^{(\infty)}(s) - S_\xi ^{(L)}(s)|ds$, is as small as possible. 
As shown by \cite{ishwaran2002approximate}, this $L_1$ error is bounded by $4n\exp\{-(L-1)/c\}$, where $n$ denotes the sample size. 
In practice, an error bound of $0.01$ is considered to be sufficiently small \citep{ohlssen2007flexible}. 
Since we fix the total mass parameter $c=1$ as a common practice, for sample size $n<600$, $L=12$ is a suitable choice of truncation number. 
In our numerical studies, we find that an $L$ in the range of $10-15$ is appropriate to approximate the DPM model well. 
Users of \texttt{BuLTM} are free to adjust the truncation number according to the data size. 

Let $G_0$ be the base measure for $(\psi_{_l}, v_{_l})$. 
We recommend choosing $G_0 = \text{Gamma}(1, 1) \times \text{Gamma}(1, 1)$ as the specified base measure without any hyperprior for it. 
The setting of $G_0$ in our approach implies that $E\{F_\xi(t)\} = 1 - \exp(-t)$ i.e the nonparametric transformation model is centering around the PH model.
Such elicitation of the DPM model is a weakly informative prior for $S_\xi$ since the variance of the DP is finite \citep{nieto2012time}. 
Note that it is nontrivial to select the hyperprior for $G_0$. 
For the base measure in the DPM with Weibull kernel, \cite{kottas2006nonparametric} proposed a Uniform-Pareto (Upar) prior, and \cite{shi2019low} proposed a low information omnibus (LIO) prior, while neither of them is applicable to our method. 
The Upar prior is not applicable to our unidentified models since the Upar prior is noninformative to $(\psi, \nu)$;
otherwise, the MCMC algorithm can hardly converge. 
The LIO prior is a kind of hierarchical specification, which is too complicated to be incorporated into our method with a heavy computation burden.

\section{Relationship between the quantile-knots I-splines prior and the NII process}
\label{propSpline}
We summarize the relationship between the quantile-knots I-splines prior and the nonnegative independent increment process here. Let $s_0=0 < s_1 < s_2 < \cdots < s_J=\tau$ and we get $J$ disjoint partitions  $[0, s_1], (s_1, s_2], \cdots, (s_{J-1}, s_J]$ of $D$. 
Note that each I-spline function starts at $0$ in an initial flat region, increases in
the mid region, and then reaches $1$ a the end \citep{wang2011semiparametric}.
Therefore, the range of all I-spline functions is $[0, 1]$. 
Then we determine the I-spline basis functions with knots $s_0=0 < s_1 < s_2 < \cdots < s_J=\tau$ and smoothness order $r>1$ as $\{B_j(t)\}_{j=1}^{K=J+r}$. 
We call two I-spline functions $B_{j_1}(t)$ and $B_{j_2}(t)$ are ``\textit{joint}" on a certain interval $D_i$ for $i=1, \cdots, J$, if $\exists t' \in D_i$ such that $B_{j_1}(t'), B_{j_2}(t') \in (0, 1)$. 
Otherwise, they are ``\textit{disjoint}" on $D_i$. 
We also call an I-spline function $B_j(t)$ ``\textit{crosses}" an interval $D_i$ if $\exists t' \in D_i$ such that $0 <B_j(t_0)<1$.

We divide all $K$ I-spline basis functions into $r$ groups. 
Among the $r$ groups, for $\iota = 1, \ldots, r$, the $\iota$th group consists of $B_\iota, B_{\iota+r}, B_{\iota+2r}, \ldots$ such that all I-spline functions in this group are disjoint. 
That is, for any $D_i$, only one of the I-spline functions within the $\iota$th group crosses the interval $D_i$.  
We define the combination of I-spline functions within the $\iota$th group as
$$
H_\iota(t) = \sum_{k\ge1} \alpha_{\iota+kr}B_{\iota+kr}(t). 
$$
Then $H_\iota(t)$ has independent increments among all knots $s_0=0 < s_1 < s_2 < \cdots < s_J=\tau$, if the coefficients $\{\alpha_{\iota+kr}\}_{k\ge1}$ are independent positive variables. 
Therefore, $H\iota$, the combination of I-splines functions within the $\iota$th group is an NII process with independent increment on fixed locations \citep[pp.129]{phadia2015prior}. 
Then we rewrite the equation (6) in the manuscript, the I-splines model into the sum of $H_\iota$
$$
H(t) = \sum_{j=1}^{K=J+r}\alpha_j B_j(t) = \sum_{\iota=1}^r H_\iota(t).
$$
This equation clearly shows that the quantile-knots I-splines prior is a combination of $r$ groups of NII processes. 
Specifically, when $r=1$, all I-spline functions are disjoint and therefore, the combination of them reduces to the piecewise exponential model if $\alpha_j \sim \exp(\eta)$ independently. 
Actually, the first step of determining the initial knots in the quantile-knots I-splines prior is similar to the construction of the piecewise exponential prior in survival models, where partitions of time axis are often taken on empirical quantiles of observed failure times \citep{de2014bayesian}. 

\section{Alternative I-splines priors for $H$}
One may consider other alternative choices of parametric and nonparametric priors for the triplet $(\bbeta, H, S_\xi)$. 
Here we introduce some alternative choices of priors. It includes how to construct constrained priors to make the MTM identified. 
Another construction of I-splines prior with shrinkage prior for $H$ is also given here. 
\subsection{Fully identified priors}
In this subsection, we discuss the construction of identified priors. Our spirit is from Horowitz's normalization conditions. 
Like the manuscript, we use the unit scale condition that $||\bbeta||=1$ as an equivalent condition of Horowitz's scale normalization. 
Rather than applying posterior projection, we assign the uniform distribution on the $p$-dim unit hypersphere as the prior for the fully identified $\bbeta$. 
It is conducted by the following transformation
$$
 \bbeta_* \sim N(0, I), \bbeta = \bbeta_* / ||\bbeta_*||^{1/2}.
$$
Still, we need the location normalization, which assumes that the $H(t_0)=1$ or $h(t_0)=0$  for some finite $t_0$ \citep{horowitz1996semiparametric}.
We adopt the I-spline priors as our initial. 
We formulate $H$ by
$$
H(t) = \sum_{j=1}^{K} \alpha_j B_j(t),
$$
where $K=J+r$ is the number of I-spline functions (see \textit{Section \ref{propSpline}}).  
By the characteristic of I-spline functions on interval $D = (0, \tau]$, if $\sum_{j=1}^{K}\alpha_j = 1$,  $H$ will surely pass the point $(\tau, 1)$. 
That is, for $h=\log H$, we have $h(\tau) = 0$. 
Therefore, the location normalization condition is transferred to a sum-to-one restriction, that is, $(\alpha_1, \ldots, \alpha_K)$ is a $K$-dim simplex. 
We consider two choices of priors for the $p$-dim simplex.
The first one is the Dirichlet prior
$$
(\alpha_1, \ldots, \alpha_K) \sim \text{Dir}(a_1, \ldots, a_K),
$$
where $\{a_j\}_{j=1}^K$ are hyperparameters of Dirichlet distribution. 
Alternatively, we may consider a kind of transformed prior. For $j=1, \ldots, K$, 
$$
\alpha_j^* \sim \exp(\eta), \alpha_j = \alpha_j^* \bigg/ \sum_{j=1}^K \alpha_j^*.
$$
Both these two priors normalize the location of $H$ and therefore, fully identify the transformation function. 

The above priors make the transformation model fully identified. 
However, with these priors, we find that the MCMC procedure by NUTS converges very slowly and suffers from poor mixing.
What's worse, the prediction accuracy is poor. 
These two drawbacks force us not to work on a fully identified model.

\vspace{-.3cm}
\subsection{The shrinkage prior and comparison}
We here introduce the commonly used shrinkage priors for I-spline functions as an alternative to the proposed quantile-knots I-splines prior for $H$. 
All I-splines variant priors for $H$ and $H'$ have the same shell
\vspace{-.3cm}
\begin{align*}\label{ispline}
    H(t) = \sum_{j=1}^{K}\alpha_j B_j(t),  H'(t) = \sum_{j=1}^{K}\alpha_j B_j'(t).
\end{align*}
However, unlike the proposed prior which selects knots from empirical quantiles of observed failure times, the traditional I-splines prior selects sufficiently many (usually from $10$ to $30$) equally spaced knots from the observed time interval 
\citetext{\citealp{CaiDunson2007JASA}; \citealp{wang2011Biometrika}; among others}.
Then, to avoid overfitting due to using too many knots, one has to incorporate a shrinkage prior for the coefficients $\alpha_j$ to select appropriate I-spline functions. 
We here consider the truncated generalized double Pareto prior :
\vspace{-.2cm}
\begin{align*}
    \alpha_j \sim N^+(0, \sigma_j^2), \sigma_j \sim \exp(\eta_j), \eta_j \sim \text{Ga}(\theta, \zeta),
\end{align*}
where $N^+$ denotes the truncated Gaussian distribution such that $\alpha_j>0$. 
This is a truncated form of the widely used generalized double Pareto prior as shrinkage prior for coefficients of basis functions \citep{gelman2013bayesian}. 
In general,  $\theta =\zeta=1$ are typical default hyperparameters. 
In \texttt{BuLTM}, we further simplify this prior as $\sigma_j \sim \exp(1)$. 
The use of shrinkage prior for I-splines functions may be sensitive to the number of knots \citep{perperoglou2019review}. 
In our experience, as the number of knots increases, the computation burden of the shrinkage prior becomes heavier while it may not improve the accuracy of final model results.
Therefore, the use of shrinkage priors may be accompanied by a time-consuming tuning procedure to determine the best number of equally spaced knots. 
We compare the shrinkage prior using $15$ equally spaced knots and the proposed quantile-knots I-splines prior under model setting Case 1 in the manuscript. 
Table \ref{PriorComp} shows the parametric estimation and root integrated square error (RISE) of estimated baseline survival probability functions using these two nonparametric priors in $100$ Monte Carlo replications. 
We find that both priors provide similar estimation results whereas the proposed quantile-knots I-splines prior perform slightly better. 
\begin{table}[H]
\centering
\small
   \caption{\label{PriorComp}\footnotesize{Parametric estimation results employing two nonparametric priors for $H$ (standard deviation in bracket) and RISE of estimated baseline survival probability functions. }}\vspace{.3cm}
   \tabcolsep 20pt
   \begin{tabular}{c|cc}
    \hline
  
    & Quantile-knots & Shrinkage \\ 
    \hline
    $\beta_1=0.577$ & 0.579(0.070) & 0.581(0.069)  \\ 
    $\beta_2=0.577$ & 0.578(0.050) & 0.576(0.049)  \\ 
    $\beta_3=0.577$ & 0.575(0.050) & 0.574(0.050)  \\ 
       RISE        & 0.063 & 0.064  \\ 
    \hline
\end{tabular}
\end{table}

\section{Proof of Theorem 1}
Conditions required for Theorem 1 are quite mild. 
Conditions $(i)$ is a general setting for right censored data. 
Conditions $(ii)$ and $(iii)$ are general settings for Bayesian analysis. 
Condition $(iv)$ is satisfied in \texttt{BuLTM} by using the Weibull kernel in the DPM model. 
It is easy to show that both log-logistic and log-normal kernels fulfill this condition. 
Condition $(v)$ is similar to condition $(ii)$ in \cite{de2014bayesian}, which is a common condition within the survival context. 
In right censoring case, condition $(v)$ is also required by \cite{zhou2018unified} as their algorithm employs a Cholesky decomposition to the covariate matrix. 

\begin{proof}
 Let $\Theta = (\bm{\alpha}, \bbeta, \bm{p}, \bm{\psi}, \bm{\nu})$ and $p(\Theta)$ be the product of priors of elements in $\Theta$. 
 To show the posterior $\pi(\Theta)$ is proper is equivalent to show that $\int_{D_\Theta} \pi(\Theta)d\Theta < \infty$, where $D_\Theta$ is the domain of $\Theta$.

 Let $B_j$ be the I-splines functions, for $j=1, \ldots, K$. 
 Let $f_w\{\cdot; \psi_{_l}, \nu_{_l}\}$ be the Weibull PDFs with parameters $\psi_{_l}$ and $\nu_{_l}$, for $l = 1, \ldots, L$. 
 By condition $(v)$, let $n_1$ be the number of uncensored observations and $n_0$ be the number of censored observations such that $n=n_1+n_0$, and then we have 
 \begin{align*}
     \mathcal{L}(\Theta) < {\mathcal{L}}^*(\Theta) &\equiv \prod_{i=1}^{n_1} f_\xi\{H(T_i) \exp(-{\bbeta}^T \Z_i)\}H'(T_i)\exp(-{\bbeta}^T \Z_i)\\
     &= \prod_{i=1}^{n_1}  \sum_{j=1}^{K}\alpha_j B_j'(T_i) \exp(-{\bbeta}^T \Z_i)\sum_{l=1}^{L} p_{_l} f_w\{\exp(-{\bbeta}^T \Z_i)\sum_{j=1}^{K}\alpha_j B_j(T_i); \psi_{_l}, \nu_{_l}\} .
 \end{align*}
 
By condition $(ii)$, we first integrate out all $p_l$ and it remains to show that 

\begin{align*}
\mathcal{A}_{_l}=&\int_{D_{\Theta|-p_{_l}}} \Bigg\{\prod_{i=1}^{n_1}  [\exp(-{\bbeta}^T \Z_i)f_w\{\exp(-{\bbeta}^T \Z_i)\sum_{j=1}^{K}\alpha_j B_j(T_i); \psi_{_l}, \nu_{_l}\}\sum_{j=1}^{K}\alpha_j B_j'(T_i)]\\
                &\times p(\Theta|-p_l)d(\Theta|-p_l)\Bigg\} < \infty, 
\end{align*}
for all $l$, where $\Theta|-p_{_l}$ denotes all parameters except $p_{_l}$s and $D_{\Theta|-p_{_l}}$ denotes corresponding domains. 

Let $\bm{\alpha} = (\alpha_1, \cdots, \alpha_K)^T, \bm{\phi}_i = (B_1'(T_i), \cdots, B_K'(T_i))^T$ and $\bm{\Phi}_i = (B_1(T_i), \cdots, B_K(T_i))^T$.
For any $0<T_i<\infty$, by the definition of I-splines function, we have 
$0<\bm{\alpha^T} \phi_i<\infty$ and $0<\bm{\alpha^T} \Phi_i<\infty$. 
Therefore, we have $0<\bm{\alpha^T}\phi_i/ \bm{\alpha^T}\Phi_i<\infty$. 
Let $M_0 = \max(\bm{\alpha^T}\phi_1/ \bm{\alpha^T}\Phi_1, \ldots, \bm{\alpha^T}\phi_{n_1}/ \bm{\alpha^T}\Phi_{n_1})$. 
Then by condition $(iv)$, 
$$
\exp(x)f_w\{\exp(x)\bm{\alpha}^T\bm{\phi}_i; \psi_{_l}, \nu_{_l}\}\bm{\alpha}^T\bm{\Phi}_i
\le M_0 \{\exp(x)\bm{\alpha}^T\bm{\phi}_i\} f_w\{\exp(x)\bm{\alpha}^T\bm{\phi}_i; \psi_{_l}, \nu_{_l}\} < \infty
$$
for all $x\in \mathbb{R}$. 

By condition $(v)$, we can find $p$ uncensored observations such that the $p \times p$ matrix of their covariates, with each row being the vector of covariates of one observation, is full rank. Let $Z^*$ denote that full rank $p$ matrix and let $\bm{\gamma} = -Z^*\bbeta = (\gamma_1, \cdots, \gamma_p)^T$. 
Thus, any $-\bbeta^T \Z_i$ can be expressed as a linear combination of $(\gamma_1, \ldots, \gamma_p)$ i.e $-\bbeta^T \Z_i = \sum_{h=1}^p c_{ih}\gamma_h$. 
That is, for $i =1, \ldots, n_1$
$$
f(\gamma_1, \ldots, \gamma_p) = \exp(\sum_{h=1}^p c_{ih}\gamma_h)f_w\{\exp(\sum_{h=1}^p c_{ih}\gamma_h)\bm{\alpha}^T\bm{\phi}_i; \psi_{_l}, \nu_{_l}\}\bm{\alpha}^T\bm{\Phi}_i<\infty. 
$$
Meanwhile, since $Z^*$ is a one-on-one linear operation of $\bbeta$, the integrand $\bbeta$ can be transferred to $\bm{\gamma} = (\gamma_1, \ldots, \gamma_p)$. 
Let $T^* = (T_1^*, \ldots, T_p^*)$ denote the survival outcomes of the $p$ subjects with covariates $(Z_1^*, \ldots, Z_p^*)^T = Z^*$. 
By simple algebra, we have
\begin{align*}
    \mathcal{A}_{_l} \le & 
     ~ M_1\int_{D_{(\Theta|-p_{_l}, \beta^*)}} p(\Theta|-p_{_l}) [ \int_{\mathbb{R}^p} \prod_{h=1}^{p} \exp(\gamma_h) f_w\{\exp(\gamma_h)\bm{\alpha}^T \bm{\Phi}_h; \psi_{_l}, \nu_{_l}\}\\
     & \times \bm{\alpha}^T\bm{\phi}_h d\gamma_1\cdots d\gamma_p ] d(\Theta|-p_{_l})\\
     \le& ~M_1\int_{D_{(\Theta|-p_{_l}, \beta^*)}} p(\Theta|-p_{_l}) d(\Theta|-p_{_l}) \prod_{h=1}^{p} \int_{-\infty}^{+\infty}  \exp(\gamma_h) f_w\{\exp(\gamma_h)\sum_{j=1}^{K}
     \bm{\alpha}^T\bm{\Phi}_h; \psi_{_l}, \nu_{_l}\}\\
     & \times \bm{\alpha}^T \bm{\phi}_h d\gamma_h \equiv \mathcal{B}_{_l},
\end{align*}
where $M_1$ is a constant. The first inequality can be derived directly from previous results and the second inequality is the Cauchy–Schwarz inequality. 

Finally, we have
\begin{align*}
 \mathcal{B}_{_l} &\le M_1 M_0^p \int_{D_{(\Theta|-p_{l})}} p(\Theta|-p_{_l}) d(\Theta|-p_{_l}) \prod_{h=1}^p \int_{\infty}^{+\infty} \exp (\gamma_h) f_w\{\exp(\gamma_i); \psi_{_l}, \nu_{_l}\}d\gamma_h \\
 &= M_1M_0^p \int_{D_{(\Theta|-p_{l})}} p(\Theta|-p_{_l}) d(\Theta|-p_{_l}) \prod_{h=1}^p \int_0^{+\infty} f_w\{\exp(\gamma_h); \psi_{_l}, \nu_{_l}\}d\{\exp(\gamma_h)\} \\
 &= M_1M_0^p < \infty. 
\end{align*}
The first equation includes product of (p+1) integrals of PDFs  $p(\Theta|-p_{_l})$ and $f_w\{ \exp(\gamma_h); \psi_{_l}, \nu_{_l}\}$, $l = 1, \ldots, p$. 
Therefore, the posterior is proper. 
\end{proof}

\section{Additional simulation results}
\label{sec:addsim}
We report additional simulations here. 
We first introduce the reproducibility of all simulations, and report the results of simulations in highly-censored cases, results of parametric estimation under AFT models, results of effective sample size (ESS) given by \texttt{BuLTM} in simulations, and results of prediction and estimation on data sets with size $100$. 
\subsection{Reproducibility of simulations}
\label{subsec:reprod}
This subsection is about details for the reproducibility of our simulation results. 
In all simulations, we run four independent parallel chains in \texttt{BuLTM} as the default setting in \texttt{Stan}. 
The length of each chain is $2500$ with the first $500$ iterations burn-in and we aggregate four chains to obtain total $8000$ posterior samples without any thinning. 
The MCMC procedure in {\texttt{spBayesSurv}} draws the same number of samples as ours.
In all simulations, we set $L=12$ for the truncation number of DPM $v=1$ for the total mass parameter, and $r=3$ for the order of smoothness of I-spline functions. 
In case the censoring rate is higher than $50\%$, we use $5$ initial knots; when the censoring rate is less than $50\%$ we use $6$ initial knots in constructing the quantile-knots I-splines prior. 
The coefficients $\{\alpha_j\}_{j=1}^K$ are assigned exponential prior with parameter $1$. 
The credible interval of estimates given by \texttt{BuLTM} is the default central posterior interval in \texttt{Stan}; 
the credible interval of estimates given by \texttt{spBayesSurv} is the highest posterior density interval computed by \texttt{R} package \texttt{HDInterval}. 
All numerical studies are realized in \texttt{R} version 4.1.0 with \texttt{rstan} version 2.26.4.  

\subsection{Highly censored cases}
\label{subsec:highcensor}
We assess \texttt{BuLTM} under four cases with high censoring rates. These model settings are similar to the model settings used in the manuscript while the censoring rates are all higher $50\%$. 
\begin{eqnarray*}
  \begin{aligned}
    &\textbf{HCase 1.} \text{ Non- PH/PO/AFT}: \epsilon \sim  0.5N(-0.5, 0.5^2)+0.5N(1.5, 1^2),\\
    & h(t) = \log[(0.8t+t^{1/2}+0.825)(0.5\Phi_{1, 0.3}(t) + 0.5\Phi_{3, 0.3}(t) - c_1)], C \sim \text{U}(1.5, 3);\\
   &\textbf{HCase 2.} \text{ PH model}: \epsilon \sim \text{EV}(0,1), \\
   & h(t) = \log[(0.8t+t^{1/2}+0.825)(0.5\Phi_{0.5, 0.2}(t) + 0.5\Phi_{2.5, 0.3}(t) - c_2)], C \sim \min(\exp(1), 2.5);\\
   &\textbf{HCase 3:} \text{ PO model}: \epsilon \sim \text{logistic}(0, 1), \\
   & H(t) = \log[(0.8t+t^{1/2}+0.825)(0.5\Phi_{0.5, 0.2}(t) + 0.5\Phi_{2.5, 0.3}(t) - c_3)],
    C \sim \min(\exp(3/4), 3.5);\\
   &\textbf{HCase 4:} \text{ AFT model}: \epsilon \sim N(0, 1^2), 
    h(t) = \log(t),  
   C \sim \min(\exp(3/4), 5).
  \end{aligned}
\end{eqnarray*}
Here $\Phi_{\mu, \sigma}$ denotes the CDF of $N(\mu, \sigma^2)$, and $c_k$ is the constant such that $H(0)=0$, for $k=1, 2, 3$.
The censoring variable $C$ is generated independent of $\Z$, leading to approximately 57\%, 58\%, 59\%, and 61\% censoring rates, respectively. 
For each prediction scenario, we compare the PPDs of three new observations with sets of covariates: $\Z_1 = (0, 0, 0)^T, \Z_2= (1, 1, 1)^T$ and $\Z_3 = (0, 1, 1)^T$, respectively.

\begin{table}[ht]
  \centering
  \small
  \def~{\hphantom{0}}
  \tabcolsep 6pt
  \caption{\label{PreHigh}\footnotesize{The RISE between true conditional survival functions and functions predicted by \texttt{BuLTM} and \texttt{spBayesSurv} under HCases 1 to 4.} }
  \vspace{.3cm}
  \footnotesize
    \begin{tabular}{c|cccc|cc|cc|cc}
      \hline
      \multicolumn{1}{c}{}& \multicolumn{4}{c}{HCase 1: Non- PH/PO/AFT} & \multicolumn{2}{c}{HCase 2: PH} &
      \multicolumn{2}{c}{HCase 3: PO} & \multicolumn{2}{c}{HCase 4: AFT}\\
      \hline
       $\Z$  & BuLTM  &PH  & PO & AFT & BuLTM &  PH  & BuLTM & PO & BuLTM & AFT \\
      $\Z_1$ & \textbf{0.068} & 0.103 & 0.126 & 0.118 & \textbf{0.074} & 0.080
       & 0.010 & \textbf{0.098} & 0.100 & \textbf{0.079}\\ 
      $\Z_2$ & \textbf{0.060} & 0.146 & 0.083 & 0.224 & \textbf{0.077} & 0.084
      & \textbf{0.125} & 0.126 & 0.158 &  \textbf{0.125}\\ 
      $\Z_3$ & \textbf{0.074} & 0.121 & 0.091 & 0.131 & \textbf{0.100} & 0.110
      & 0.139 & \textbf{0.135} & 0.178 & \textbf{0.132}\\ 
      \hline
    \end{tabular}
\end{table}

Table \ref{PreHigh} shows that \texttt{BuLTM} still works well when the censoring rate goes high. 
We find that when the censoring rate is higher than $50\%$, \texttt{BuLTM} outperforms \texttt{spBayesSurv} under Non-PH/PO/AFT and PH models, is comparable under the PO model, and is slightly worse than \texttt{spBayesSurv} under the AFT model. 
This result is in line with the results we report in the manuscript. 
Readers may wonder why \texttt{BuLTM} does not work well under the PO model. 
We conjecture a possible reason is that the log-logistic distribution is heavy-tailed and may be hard to be approximated by the DPM. 
One future work is to model the distribution of model error $\xi$ through a more complicated nonparametric prior. 

Results of parametric estimation are summarized in Table \ref{ParaHigh} for HCases 1-3. 
 Case 1 is out of the application scope of the \texttt{spBayesSurv} where none of PH, PO, and AFT models provides reasonable parametric estimation, and hence we omit results of \texttt{spBayesSurv}. 
We find that parametric estimation given by \texttt{BuLTM} has little bias and the PSD is quite close to the SDE, and the CP is close to the nominal level in all cases. 
When the true model is one of PH and PO models, \texttt{BuLTM} has a lower bias for almost all parameters and has lower RMSE for all parameters than \texttt{spBayesSurv}. 
These results demonstrate that \texttt{BuLTM} estimates the fully identified $\bbeta$ quite well.

\begin{table}[!htp]
  \centering
  \footnotesize
\tabcolsep 3pt
  \caption{\label{ParaHigh}\footnotesize{Results of estimation of $\bbeta$ by \texttt{BuLTM} and \texttt{spBayeSurv} in HCases 1 to 3.  }}\vspace{.3cm}
  \footnotesize
    \begin{tabular}{cc|ccccc|ccccc}
    \hline
     \multicolumn{2}{c}{Case 1: Non-PH/PO/AFT}& \multicolumn{5}{c}{BuLTM}& \multicolumn{5}{c}{spBayesSurv}\\
     \hline
& Parameter & BIAS   &  RMSE &  PSD  &  SDE  &  CP  &   &   &   &   &  \\ 
  \hline
& $\beta_1$& -0.003 & 0.098 & 0.092 & 0.097 & 94.0 &  &  &  &  &  \\ 
 & $\beta_2$& -0.006 & 0.072 & 0.067 & 0.071 & 92.0 &  &  &  &  &  \\ 
& $\beta_3$&  0.009 & 0.072 & 0.067 & 0.068 & 94.0 &  &  &  &  &  \\ 
   \hline        
      \hline
      \multicolumn{2}{c}{}& \multicolumn{5}{c}{HCase2: PH} & \multicolumn{5}{c}{HCase3: PO} \\
      \hline
      Method      & Parameters & BIAS   &  RMSE &  PSD  &  SDE  &  CP  &  BIAS  & RMSE  &  PSD  &  SDE  & CP   \\ 
      BuLTM       & $\beta_1$ & \textbf{0.005} & 0.159 & 0.152 & 0.158 & 93.7 & 0.011 & 0.218 & 0.211 & 0.214 & 92.7  \\ 
             & $\beta_2$  & \textbf{-0.002} & 0.122 & 0.107 & 0.118 & 93.3 & \textbf{-0.000} & 0.148 & 0.146 & 0.138 & 95.3  \\ 
             & $\beta_3$  & \textbf{-0.003} & 0.109 & 0.108 & 0.105 & 93.3& \textbf{-0.011} & 0.149 & 0.146 & 0.135 & 95.3 \\ 
      spBayesSurv & $\beta_1$  & 0.018 & 0.240 & 0.227 & 0.240 & 92.0 & \textbf{0.000} & 0.335 & 0.315 & 0.335 & 94.7 \\ 
                  & $\beta_2$  & 0.025 & 0.137 & 0.122 & 0.135 & 92.7 &  0.021 & 0.172 & 0.167 & 0.171 & 94.7  \\ 
                  & $\beta_3$. & 0.023 & 0.128 & 0.122 & 0.126 & 93.7  &  0.014 & 0.164 & 0.168 & 0.164 & 95.0 \\
      \hline
   \end{tabular}
\end{table}

\subsection{Parametric estimation under AFT models}
Results of parametric estimation are given by Table \ref{TabparaAFT}, where we find \texttt{BuLTM} has lower RMSE than \texttt{spBayesSurv} for all parameters. 
In terms of BIAS, \texttt{BuLTM} outperforms \texttt{spBayesSurv} in the highly-censored case and is comparable in the case with the lower censoring rate. 
This result as well as results of prediction demonstrate that \texttt{BuLTM} performs robustly under the AFT model.

\begin{table}[!htp]
\centering
\tabcolsep 3pt
\caption{\label{TabparaAFT} Results of estimation of $\bbeta $ under AFT models. }
\small
\begin{tabular}{cc|ccccc|ccccc}
\hline
 \multicolumn{2}{c}{}& \multicolumn{5}{c}{Case4: AFT1, 25\% Censored} & \multicolumn{5}{c}{HCase4: AFT2, 61\% Censored} \\
  \hline
Method& Parameter& BIAS & RMSE & PSD & SDE & CP & BIAS & RMSE & PSD & SDE & CP \\ 
  \hline
BuLTM & $\beta_1$ &0.017 & 0.107 & 0.102 & 0.107 & 92.3 & \textbf{0.011} & 0.138 & 0.130 & 0.138 & 94.0 \\ 
 & $\beta_2$& -0.009 & 0.079 & 0.076 & 0.076 & 92.3 & \textbf{-0.004} & 0.101 & 0.095 & 0.098 & 93.0 \\ 
 & $\beta_3$& -0.008 & 0.079 & 0.077 & 0.081 & 92.7 & \textbf{-0.007} & 0.101 & 0.094 & 0.093 & 95.0\\ 
spBayesSurv& $\beta_1$& \textbf{0.000} & 0.159 & 0.150 & 0.159 & 90.3 & 0.016 & 0.207 & 0.194 & 0.206 & 92.0 \\ 
 & $\beta_2$& \textbf{0.002} & 0.078 & 0.079 & 0.078 & 92.3 & 0.016 & 0.105 & 0.103 & 0.104 & 91.0 \\ 
  & $\beta_3$& \textbf{0.003} & 0.084 & 0.079 & 0.084 & 92.0 & 0.014 & 0.101 & 0.103 & 0.100 & 93.7 \\ 
   \hline
\end{tabular}
\end{table}

\subsection{Effective sample size of $\bbeta$}
The effective sample size (ESS) is useful as a first-level check when analyzing the reliability of inference.
It measures how many independent draws contain the same amount of information as the dependent posterior samples obtained by the MCMC procedure. 
ESS is usually accompanied by $\hat{R}$, the diagnostics of convergence of MCMC. 
In an MCMC procedure, especially the case where multiple chains are used, very low ESS may be caused by divergent chains or poor mixing and hence, large $\hat{R}$. 
If one obtains sufficient ESS (ESS that is greater than $400$ is considered to be sufficient by \cite{vehtari2021rank}) after sampling, it is highly possible that all chains are converged and well mixed. 
Therefore, we report ESS of $\bbeta$ in our simulation studies here as the diagnosis of MCMC.

Results of the average estimated ESS of $\bbeta$ in all the simulation studies in the manuscript are given by Table \ref{samplesize}, from which we find in each simulation the ESS of $\bbeta$ is sufficiently large.  
This is owed to the NUTS used by \texttt{Stan}, which is more possible to sample nearly independent draws \citep{hoffman2014no}. 
In terms of other parameters, only a few parameters suffer from low ESS in sporadic Monte Carlo replications as a drawback of the analysis of unidentified models. 
Even so, the MCMC algorithm is still well converged and mixed examined by $\hat{R}$ in \texttt{Stan} and thus the final model results are reasonable. 
Therefore, when using \texttt{BuLTM}, one can simply increase the length of MCMC chains to obtain sufficient ESS for all parameters in all situations regardless of the lack of identifiability.
Particularly, if one's interest falls on estimating $\bbeta$, the vector of regression parameters, the length of chains needed is quite small, and the required computation burden is mild. 
\begin{table}[H]
\centering
\small
\tabcolsep 7pt
\caption{\label{samplesize}\footnotesize{The average estimated ESS of $\bbeta = (\beta_1, \beta_2, \beta_3)^T$ in simulation studies.}}\vspace{.3cm}
\begin{tabular}{c|cccccccc}
\hline
\multicolumn{1}{c}{}& Case 1 & HCase 1 & Case 2 & HCase 2 & Case 3 & HCase 3 & Case 4 & HCase 4 \\ 
\hline
$\beta_1$ & 5935.29 & 6201.22 & 6697.80 & 6187.15 & 6026.63 & 5744.58 & 7014.90 & 6431.38 \\ 
$\beta_2$ & 5573.79 & 6243.93 & 7305.17 & 6800.33 & 6697.69 & 6302.88 & 7497.02 & 6900.31 \\ 
 $\beta_3$ & 5591.05 & 6193.69 & 7307.38 & 6757.07 & 6689.56 & 6263.12 & 7487.56 & 7053.08 \\ 
\hline
\end{tabular}
\end{table}

\subsection{Simulations on data sets with size 100}
Additional simulations on data sets with size $100$ are conducted to evaluate the performance of \texttt{BuLTM} on moderate data size.  
We consider the same settings of Case 1-4 in the manuscript for true models. 

\noindent \textbf{Prediction} Table \ref{STPreNon} summaries results of prediction performance, where we find \texttt{BuLTM} outperforms \texttt{spBayesSurv} in prediction when the true model is one of non-PH/PO/AFT, PH, and PO models with small data size, and is comparable with \texttt{spBayesSurv} when the true model is the AFT model. 
These results are consistent with that on data size of $200$. 
It demonstrates that \texttt{BuLTM} still performs well on small data sets in prediction.

\begin{table}[H]
  \centering
  \small
  \def~{\hphantom{0}}
  \tabcolsep 7pt
  \caption{\label{STPreNon}\footnotesize{The RISE between the true conditional survival functions and functions predicted by \texttt{BuLTM} (MTM) and \texttt{spBayesSurv} under Cases 1 to 4. Data size $n=100$. } }
  \vspace{.1cm}
  \small
    \begin{tabular}{c|cccc|cc|cc|cc}
      \hline
      \multicolumn{1}{c}{}& \multicolumn{4}{c}{Case 1: Non- PH/PO/AFT} & \multicolumn{2}{c}{Case 2: PH} &
      \multicolumn{2}{c}{Case 3: PO} &  \multicolumn{2}{c}{Case 4: AFT}\\
      \hline
       $\Z$ & MTM  &PH  & PO &AFT & MTM &  PH  & MTM & PO  & MTM & AFT\\
      $\Z_1$ & \textbf{0.068} & 0.068 & 0.069 & 0.120 & \textbf{0.096} & 0.098 & 0.106 & \textbf{0.104} & 0.104 & \textbf{0.088}\\
      $\Z_2$ & \textbf{0.102} & 0.148 & 0.118 & 0.199 & \textbf{0.140} & 0.229
      & \textbf{0.147} & 0.157 & 0.133 &\textbf{0.111} \\ 
      $\Z_3$ & \textbf{0.149} & 0.245 & 0.164 & 0.240 & \textbf{0.130} & 0.220
      & \textbf{0.155} & 0.161 & 0.149  & \textbf{0.129}\\ 
      \hline
    \end{tabular}
\end{table}

\begin{table}[!htp]
\centering
\tabcolsep 4pt
 \small
  \def~{\hphantom{0}}
  \caption{\label{SParaPHPO}\footnotesize{The performance of parametric estimation of \texttt{BuLTM} and \texttt{spBayesSurv} under Cases 1-3 with size 100. spBayesSurv cannot provide reasonable estimation in Case 1 and we omit it.}}
\footnotesize
\begin{tabular}{cc|ccccc|ccccc}
    \hline
    \multicolumn{2}{c}{Case 1: Non-PH/PO/AFT}& \multicolumn{5}{c}{BuLTM}& \multicolumn{5}{c}{spBayesSurv}\\
    \hline
& Parameter & BIAS   &  RMSE &  PSD  &  SDE  &  CP  &   &   &   &   &  \\ 
  \hline
& $\beta_1$&-0.006 & 0.098 & 0.095 & 0.097 & 93.7 &  &  &  &  &  \\ 
 & $\beta_2$& -0.010 & 0.074 & 0.069 & 0.071 & 94.3 &  &  &  &  &  \\ 
& $\beta_3$&  0.016 & 0.074 & 0.068 & 0.069 & 93.7 &  &  &  &  &  \\ 
   \hline    
   \hline
  \multicolumn{2}{c}{}& \multicolumn{5}{c}{Case 2: PH} & \multicolumn{5}{c}{Case 3: PO} \\
  \hline
 Method & Parameter &BIAS & RMSE & PSD & SDE & CP & BIAS & RMSE & PSD & SDE & CP \\ 
  \hline
BuLTM & $\beta_1$ &\textbf{0.010} & 0.156 & 0.165 & 0.155 & 96.3 & 0.060 & 0.209 & 0.218 & 0.209 & 93.3 \\ 
  & $\beta_2$ & -0.015 & 0.123 & 0.119 & 0.114 & 91.3 & \textbf{-0.028} & 0.177 & 0.159 & 0.155 & 93.0 \\
 & $\beta_3$& \textbf{0.005} & 0.123 & 0.118 & 0.112 & 93.3 & -0.037 & 0.177 & 0.159 & 0.160 & 93.0 \\ 
 spBayesSurv & $\beta_1$& 0.022 & 0.238 & 0.250 & 0.237 & 97.3 & \textbf{0.043} & 0.407 & 0.368 & 0.364 & 89.3 \\ 
 & $\beta_2$& \textbf{0.008} & 0.135 & 0.136 & 0.135 & 95.7 & \textbf{0.028} & 0.223 & 0.194 & 0.195 & 89.7 \\ 
 &$\beta_3$& 0.011 & 0.143 & 0.138 & 0.143 & 95.7 & \textbf{0.026} & 0.240 & 0.195 & 0.213 & 91.0 \\ 
   \hline
\end{tabular}
\end{table}
\noindent \textbf{Parametric estimation} Table \ref{SParaPHPO} summarizes results of parametric estimation on data sets with size $n=100$ under model settings of non-PH/PO/AFT, PH, and PO, where we find the BIAS and RMSE are low, the PSD is close to SDE, and the CP is close to the nominal rate of credibility. 
Compared to \texttt{spBayesSurv}, \texttt{BuLTM} has lower BIAS under the PH model and is comparable under the PO model.

\section{Sensitivity analysis}
We analyze the sensitivity of the proposed quantile-knots I-splines prior for $H$ in this section. 
There are two pre-specified hyperparameters in the prior, the hyperparameter $\eta$ for the exponential prior, and the number of initial knots. 
Here we show that the final prediction results are not sensitive to either the initial number of initial knots or the hyperparameter $\eta$. 

\subsection{Sensitivity of number of initial knots}
\begin{figure}[!ht]
    \centering
    \includegraphics[width = 0.5\textwidth]{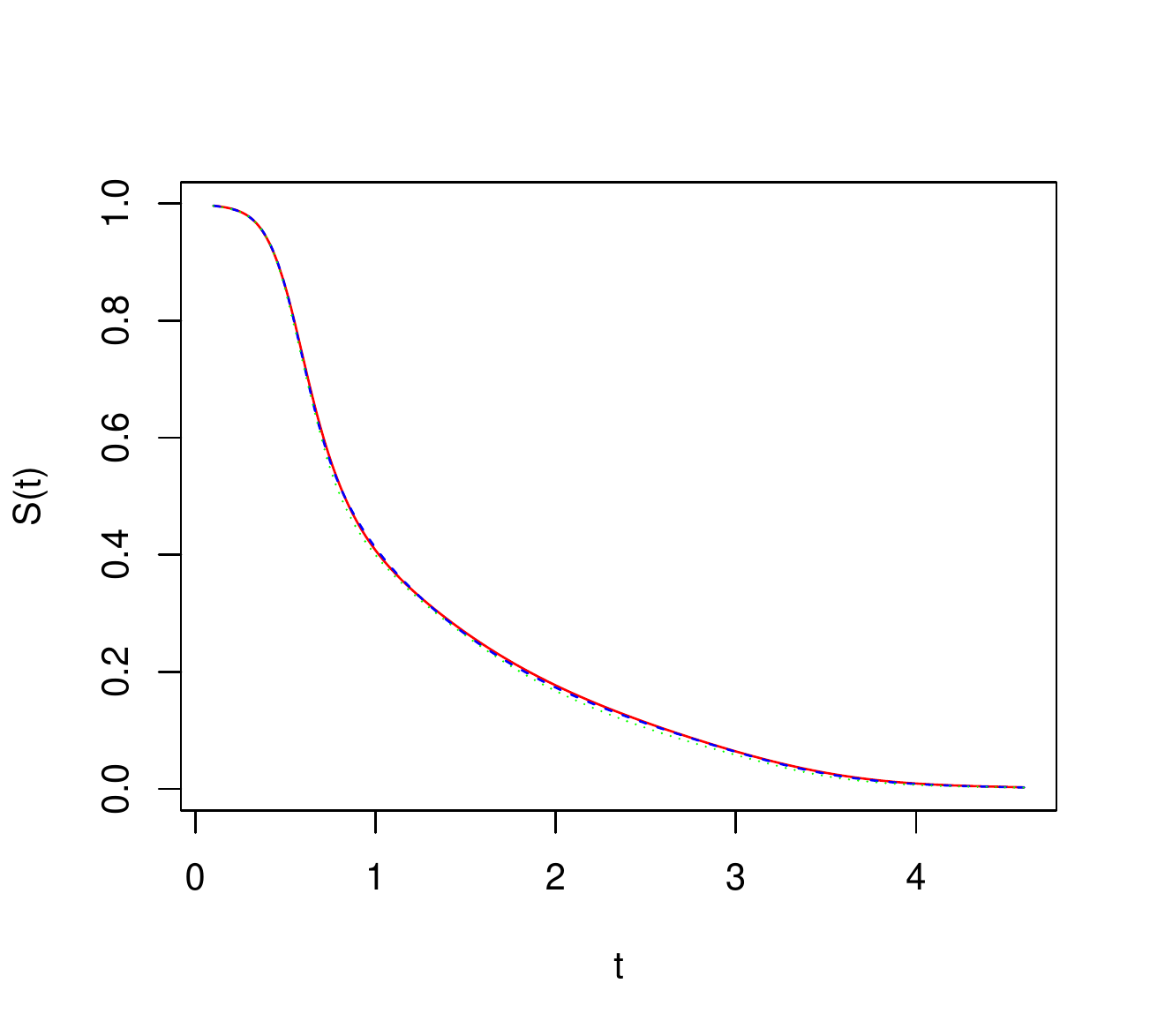}
    \caption{\label{FigSensKnots}\footnotesize{Pointwise mean estimated baseline survival probability curves under $100$ replications. Real line, $N_I=5$; dash line, $N_I=6$; dotted line, $N_I=11$. }}
\end{figure}

Sensitivity analysis of the choice of the initial number of basic knots ($N_I$) in the quantile-knots I-splines prior is conducted by $100$ Monte Carlo studies under Case 1 setting in the manuscript. 
Candidates for the number of initial knots are taken from the range $5$ to $11$, where we display results of using $5, 6$, and $11$ initial knots here for comparison.
Results of parametric estimation and the RISE of estimated baseline survival probability curves among different numbers of initial knots are shown in Table \ref{SensKnots}, where we find with different choices of $N_I$, both results of parametric estimation and RISE of estimated survival probability curves have very mild variation. 
Figure \ref{FigSensKnots} displays plots of average estimated baseline survival probability curves under three choices of the number of initial knots, where we find they are close to each other. 
This sensitivity analysis numerically demonstrates that the quantile-knots I-splines prior is not sensitive to its choice of the number of knots. And therefore, it is generally tuning-free and computationally expedient. 
\begin{table}
\centering
\footnotesize
   \caption{\label{SensKnots}\footnotesize{Parametric estimation results (standard deviation in bracket) and RISE of estimated baseline survival probability functions under different choices of $\eta$ .}}\vspace{.3cm}
   \tabcolsep 12pt
   \begin{tabular}{c|ccc}
    \hline
  
    & $N_I=5$ & $N_I=6$ & $N_I=11$ \\ 
    \hline
    $\beta_1=0.577$ & 0.578(0.070) & 0.580(0.070) & 0.586(0.069) \\ 
    $\beta_2=0.577$ & 0.575(0.051) & 0.575(0.051) & 0.572(0.052) \\ 
    $\beta_3=0.577$ & 0.561(0.058) & 0.560(0.058) & 0.557(0.058) \\ 
       RISE        & 0.063 & 0.063 & 0.066 \\ 
    \hline
\end{tabular}
\end{table}

\subsection{Sensitivity of $\eta$}
Let $\eta$ be the hyperparameter of exponential prior for coefficients of the quantile-knots I-splines prior in equation (8) in the manuscript. 
Sensitivity analysis of $\eta$ is conducted under the setting Case1 in the manuscript. For the sensitivity of $\eta$, among $100$ Monte Carlo replications, we choose $\eta$ from three candidates of $\eta = 1, 5,$ and $0.2$, corresponding to three levels of informative priors. Notice that we should avoid using too small $\eta$ since it implies too large prior variance, then the prior is not sufficiently informative anymore. Similarly, too large $\eta$ induces too small variance, which is too informative to provide sufficient uncertainty. 

\begin{figure}[!htp]
    \centering
    \includegraphics[width = 0.5\textwidth]{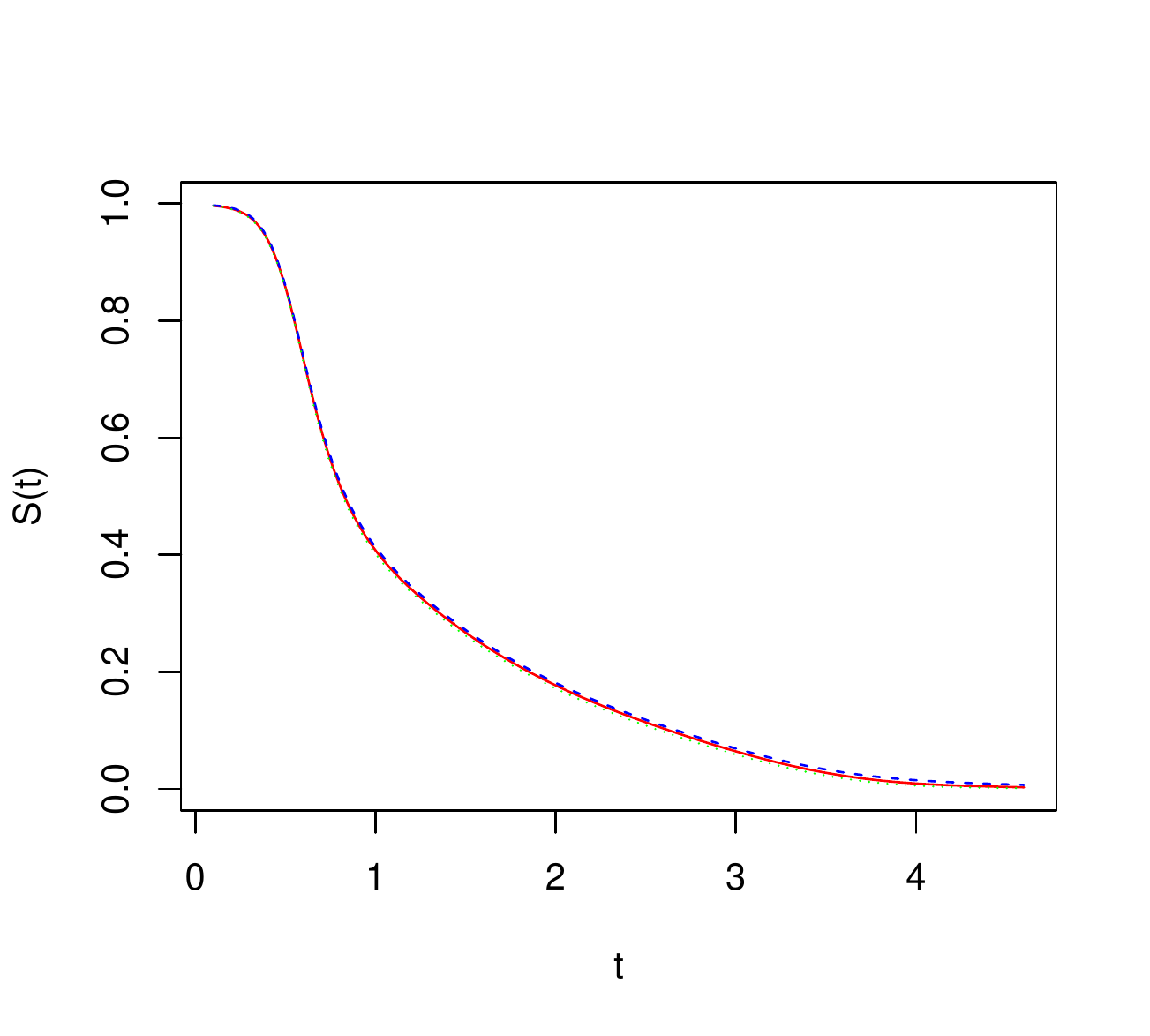}
    \caption{\label{FigSensEta}\footnotesize{Pointwise mean estimated baseline survival probability curves in $100$ replications. Real line, $\eta = 1$; dash line, $\eta = 5$; dotted line, $\eta = 0.2$. }}
\end{figure}

Results of parametric estimation and RISE of estimated baseline survival curves are given in Table \ref{Sens_eta}. From the table, we find that estimation of the parametric component varies quite little among all choices of $\eta$ and the RISE of estimated baseline survival curves is almost the same with different values of $\eta$. 
For visualization, plots of estimated survival curves given different values of $\eta$ are shown in Fig \ref{FigSensEta}, where we find all estimated curves are close to each other. 
This sensitivity analysis numerically demonstrates that the quantile-knots I-splines prior are not sensitive to the choice of $\eta$ within the range $0.2$ to $5$. 
Therefore, it is safe to fix $\eta$ rather than to assign a hyperprior for it. 

\begin{table}[H]
\centering
\tabcolsep 6pt
\footnotesize
   \caption{\label{Sens_eta}\footnotesize{Parametric estimation results (standard deviation in bracket) and RISE of estimated baseline survival probability functions under different choices of $\eta$. }}\vspace{.3cm}
   \tabcolsep 10pt
   \begin{tabular}{c|ccc}
    \hline
    \multicolumn{1}{c}{} &   $\eta=1$   &  $\eta = 5$  & $\eta = 0.2$ \\ 
    \hline
    $\beta_1=0.577$ & 0.580(0.070) & 0.571(0.071) & 0.592(0.068) \\ 
    $\beta_2=0.577$ & 0.575(0.052) & 0.578(0.051) & 0.569(0.052) \\ 
    $\beta_3=0.577$ & 0.560(0.058) & 0.564(0.057) & 0.553(0.059) \\ 
       RISE        & 0.063 & 0.064 & 0.065 \\ 
    \hline
\end{tabular}
\end{table}
\vspace{-.5cm}

\section{Results of parametric estimation on real datasets }

\subsection{Veterans lung cancer data}
Results of parametric estimation for the veterans lung cancer data given by \texttt{BuLTM}, \texttt{spBayesSurv}, and the smooth partial rank estimator are displayed in Table \ref{tabLung}. 
The three methods provide similar significance levels for all coefficients. 
Although some signs of estimated coefficients are different, say $\beta_3$ and $\beta_7$, they are not significant since their credible/confidence intervals cover zero. 
That implies qualitative interpretations of the estimates of the regression parameter under the three models are stable. 
\begin{table}[H]
  \centering
  \small
  \def~{\hphantom{0}}
  \caption{\label{tabLung}\footnotesize{Results of estimated $\bbeta$ for veterans administration lung cancer data. Credible intervals are given on $95\%$ credibility. The confidence interval of SPR is a 95\% Wald-type confidence level. }}
  \tabcolsep 3pt
  \footnotesize
    \begin{tabular}{c|cc|cc|cc}
      \hline
      \multicolumn{1}{c}{} &  \multicolumn{2}{c}{BuLTM} & \multicolumn{2}{c}{spBayesSurv (PO)} & \multicolumn{2}{c}{SPR} \\
      \hline
      Covariate   & Estimate &     Credible interval     & Estimate &   Credible interval    & Estimate &  Confidence interval   \\
      $Z_1 $ &  0.119  & (0.045, 0.246)  &  0.617   & (0.449, 0.800)  & 1  & -\\
      $Z_2 $ &  -0.302  & (-0.951, 0.897) &  -1.391 & (-8.597, 6.028) & -0.626 &  (-1.831, 0.579)\\
      $Z_3 $ &{-0.006} & (-0.700, 0.671) &  {1.426}   & (-1.643, 4.477) &  {0.827} & (-3.071, 4.725)\\  
      $Z_4$ & 0.081 & (-0.693, 0.730) & 0.033 & (-3.533, 3.469)& -0.752& (-7.320, 5.826) \\ 
      $Z_5 $ &  -0.044  & (-0.227, 0.117) & -0.147   & (-0.739, 0.487) & -0.839 &  (-1.878, 0.200)\\
      $Z_6 $ &  0.350  & (0.093, 0.694) &  1.387   & (0.396, 2.334)  & 2.764 & (0.469, 5.060)\\
      $Z_7 $ &  {-0.005}  & (-0.242, 0.205) &  {0.058}   & (-0.739, 0.916) & {0.082} & (-0.771, 0.935)\\
      $Z_8 $ & 0.274 & (0.053, 0.571)  &  1.367   & (0.444, 2.308) & 2.642 & (1.446, 3.838) \\
      \hline
    \end{tabular}
\end{table}

\subsection{Heart failure clinical records data}
Results of parametric estimation for the veterans lung cancer data given by \texttt{BuLTM} and \texttt{spBayesSurv} are displayed in Table \ref{tabHeart}. 
We find that \texttt{BuLTM} is consistent with \texttt{spBayesSurv} in the detection of significance. 
Meanwhile, all regression coefficients, their estimates given by \texttt{BuLTM} have the same signs as that given by \texttt{spBayesSurv}.
That implies the qualitative interpretations of estimates of regression coefficients given by \texttt{BuLTM} and \texttt{spBayesSurv} are consistent on this dataset. 

\begin{figure}[!htp]
  \centering
    \subfigcapskip=-15pt
  \subfigure[]{
    \begin{minipage}[t]{0.45\linewidth}
      \centering
      \includegraphics[width=2in]{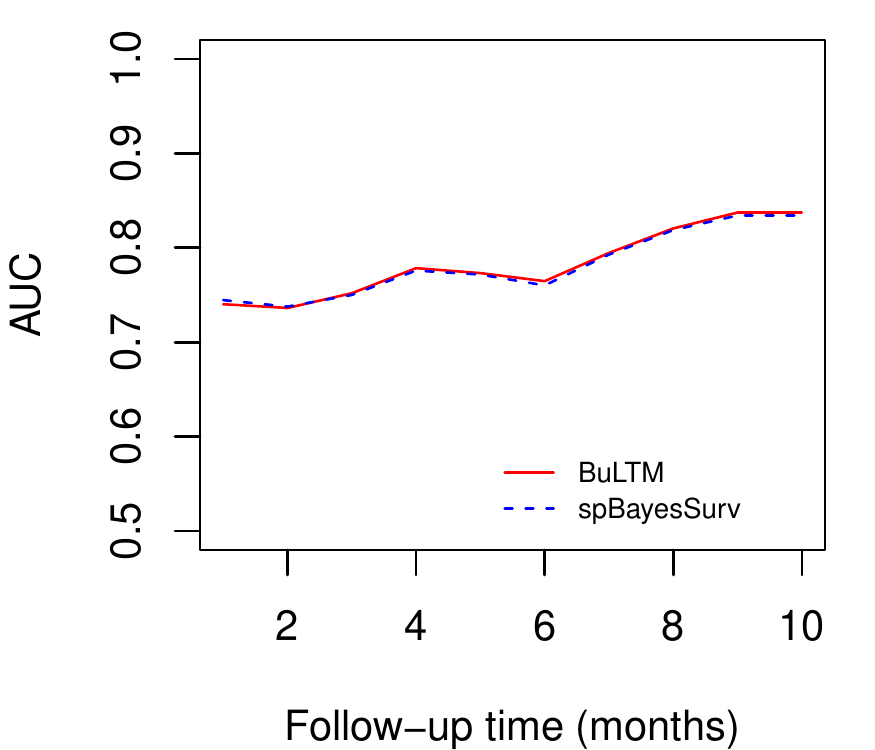}
      \label{HeartAUC_KM}
    \end{minipage}
}
  \subfigure[]{
    \begin{minipage}[t]{0.45\linewidth}
      \centering
      \includegraphics[width=2in]{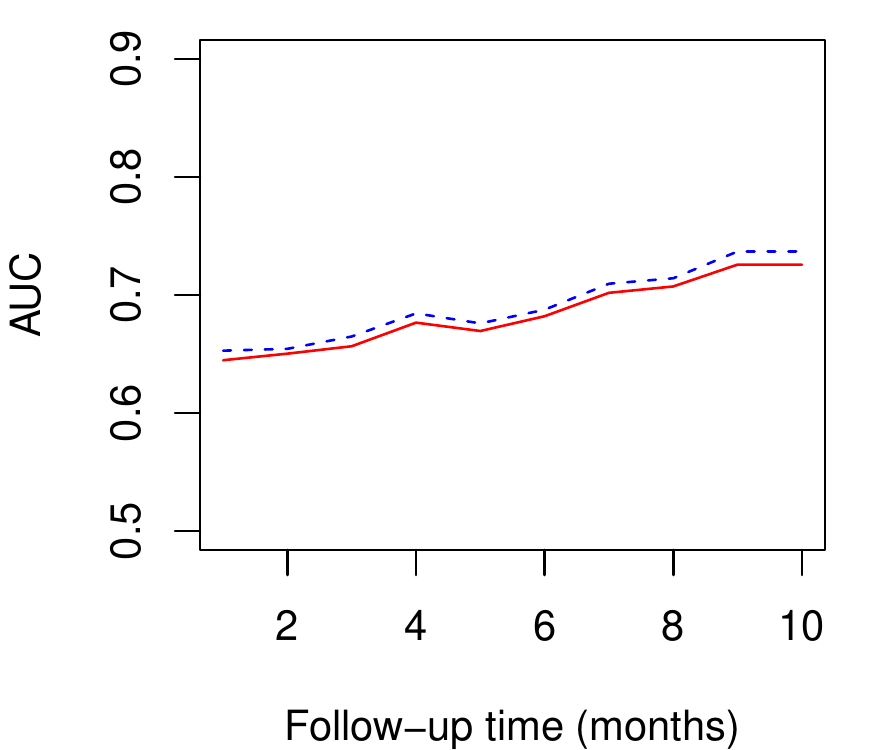}
      \label{HeartAUC_NNE}
    \end{minipage}
}
\vspace{-.3cm}
\caption{\footnotesize  Time dependent survival $\text{AUC}(t)$ computed by estimated relative risks. (a), method ``K-M"; (b), method ``NNE".}
\label{HeartAUC}
\end{figure}

We use the time dependent survival AUC$(t)$ to evaluate estimated relative risks given by the two methods. 
As shown by Figure \ref{HeartAUC}, the estimated relative risks given by the two methods are comparable. 
\begin{table}[!htp]
  \centering
  \footnotesize
  \def~{\hphantom{0}}
  \caption{\label{tabHeart}\footnotesize{Results of estimated $\bbeta$ in the analysis to heart failure clinical records data.} }
  \tabcolsep 8pt
    \begin{tabular}{c|cc|cc}
      \hline
      &  \multicolumn{2}{c}{BuLTM} & \multicolumn{2}{c}{spBayesSurv (PH)} \\
      \hline
      Covariate              & Estimate &     95\%CI      & Estimate &    95\%CI       \\
      $Z_1=$ age     & -0.163   & (-0.433, 0.063) &  -4.670  & (-6.182, -3.135) \\
      $Z_2 =$ anemia &   -0.013  & (-0.036, -0.001) &  -0.412  & (-0.764, -0.066) \\
      $Z_3=$  creatinine phosphokinase & -0.002 & (-0.010, 0.004) & -0.074 & (-0.262, 0.113)\\
      $Z_4=$ diabetes & -0.004 & (-0.020, 0.008) & -0.117 & (-0.476, 0.256) \\ 
      $Z_5 =$ ejection fraction & 0.022 & (0.008, 0.060) & 0.586 & (0.386, 0.785)\\ 
      $Z_6=$ high blood pressure & -0.015 & (-0.042, -0.001) & -0.460 & (-0.807, -0.099)\\
      $Z_7=$ platelets & 0.076 & (-0.033, 0.389) & 1.303 & (-2.836, 5.327) \\
      $Z_8=$ serum creatinine & -0.012 & (-0.033, -0.004) & -0.306 & (-0.421, -0.183) \\
      $Z_9=$ serum sodium & 0.939 & (0.787, 0.997) & 41.347 & (3.248, 74. 256) \\
      $Z_{10} = $ sex & 0.009 & (-0.005, 0.033) & 0.222 & (-0.185, 0.625) \\ 
      $Z_{11} =$ smoking & -0.005 & (-0.024, 0.010) & -0.133 & (-0.542, 0.282)\\
      \hline
    \end{tabular}
\end{table}

\section{Posterior checking}
\begin{figure}[!htp]
    \centering
    \includegraphics[width=0.49\textwidth]{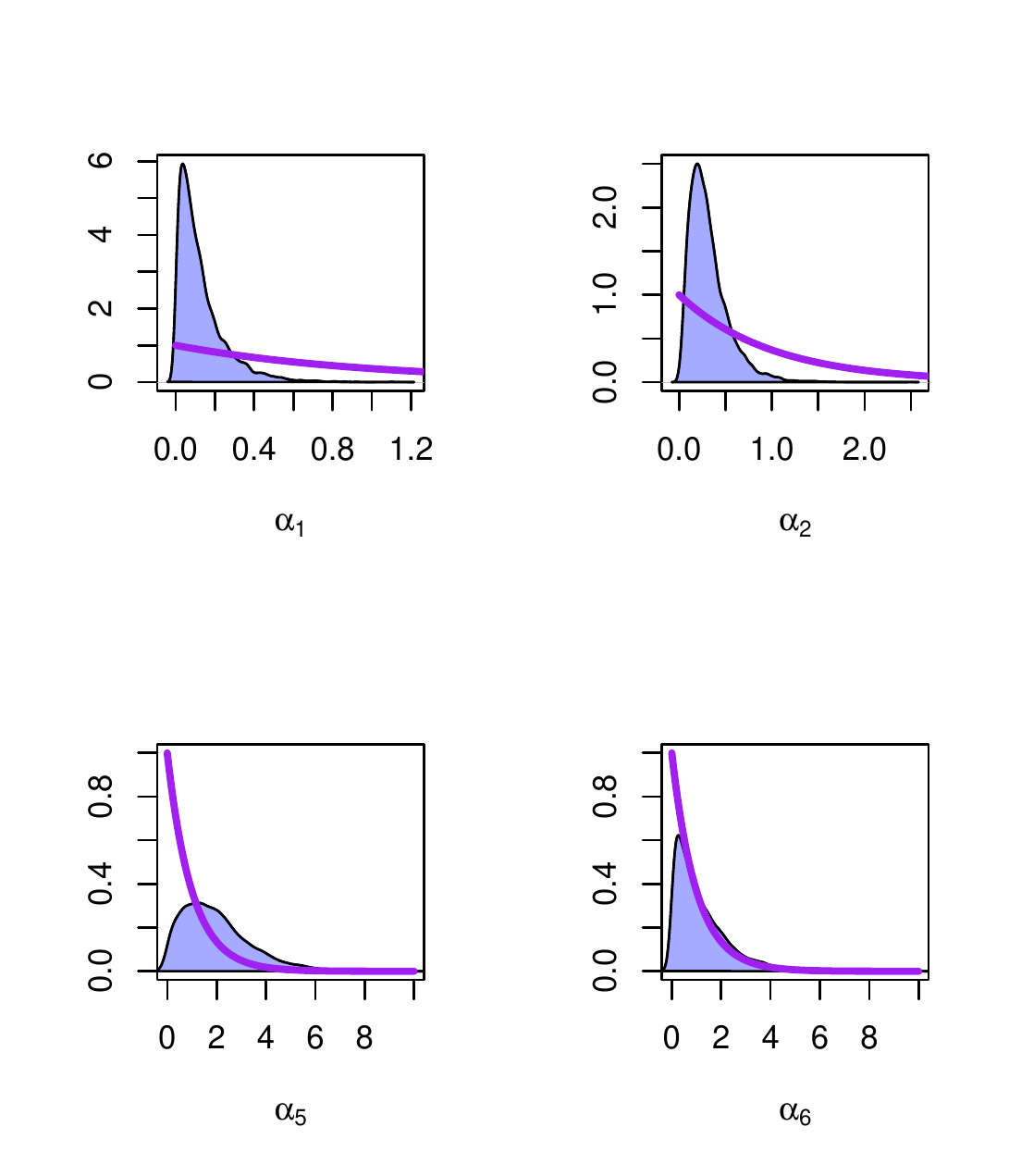}
        \includegraphics[width=0.49\textwidth]{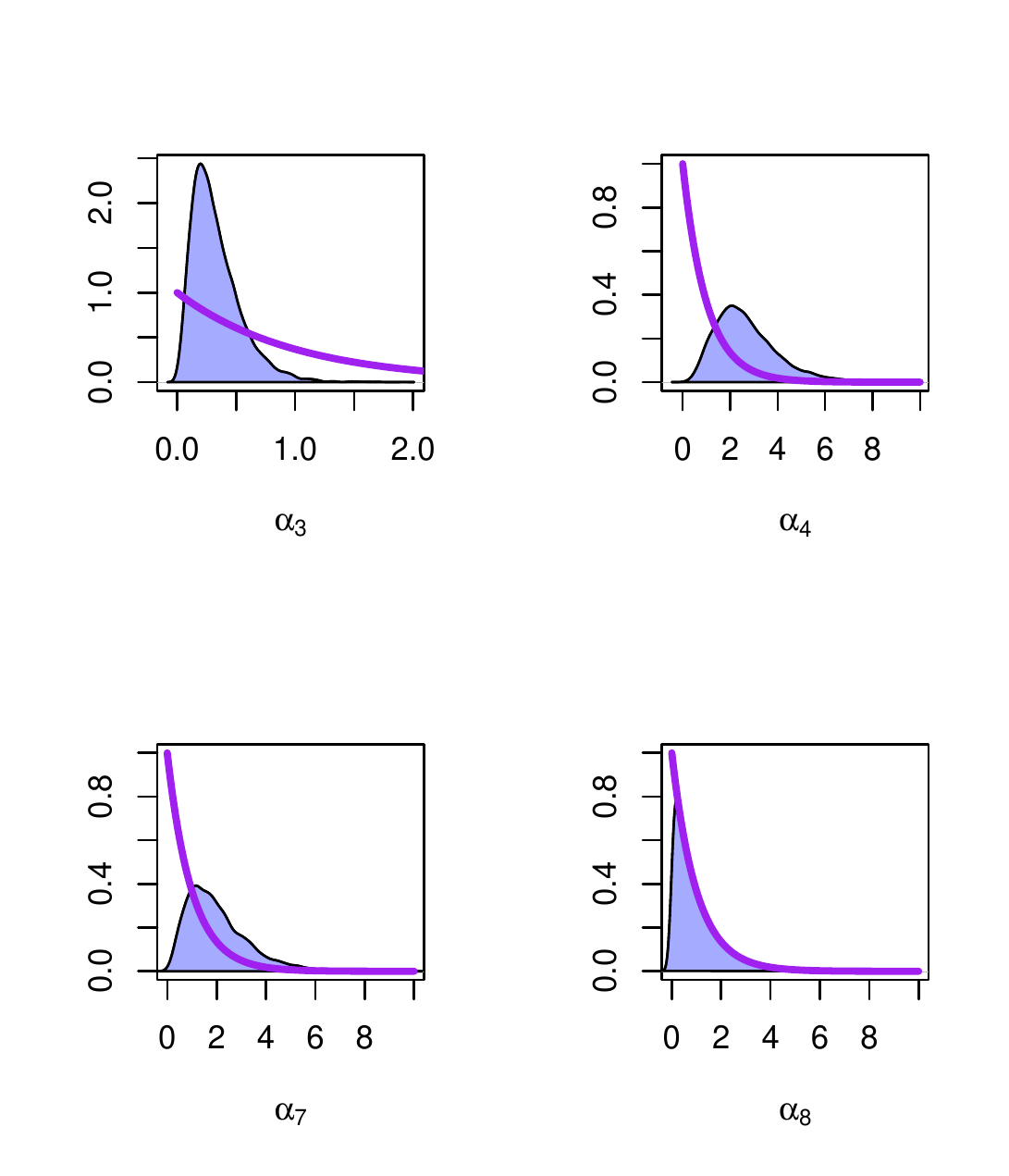}

    \caption{\label{figalpha}\footnotesize{Comparison between the the marginal posterior density  and priors of $\alpha_1, \ldots , \alpha_8$. Shaded region, marginal posterior density; Wide line, prior density of $\exp(1)$.}}
\end{figure}

We assign weakly informative priors for nonparametric components $H$ and $S_\xi$, which are not fully objective priors. 
One may worry whether these priors are so informative that the prior-to-posterior updating is not driven by data. 
We conduct posterior checking on simulation studies and application examples to check the difference between priors and marginal posterior and obtain similar results. 
Here we take our application to veterans lung cancer data set as an example. 
We take $\alpha_j \sim \exp(1)$ for $j=1, \ldots, K$ as weakly informative priors and $p(\bbeta) \propto 1$ as flat priors. 
Figure \ref{figalpha} compares the priors and marginal posterior of the first eight coefficients of I-spline functions. 
For all $\{\alpha_j\}_{j=1}^8$, their variance is controlled by the weakly informative prior, demonstrating the fact that the impact of priors remedies the flat likelihood.
In addition, most of the coefficients in the I-splines prior vary significantly from the prior, evidencing that data drive the prior-to-posterior updating. 

\begin{figure}[!htp]
    \centering
    \includegraphics[width=0.49\textwidth]{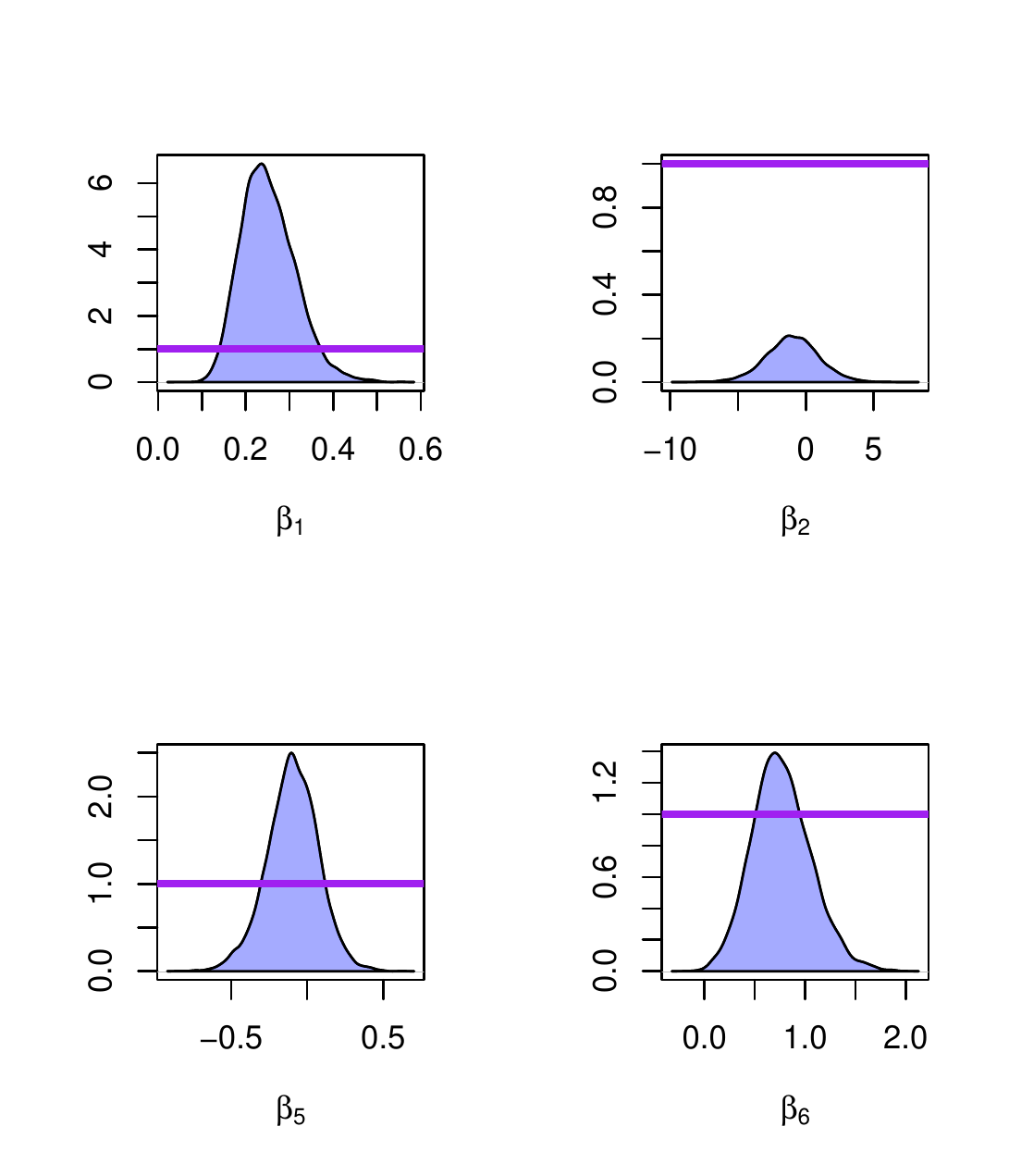}
      \includegraphics[width=0.49\textwidth]{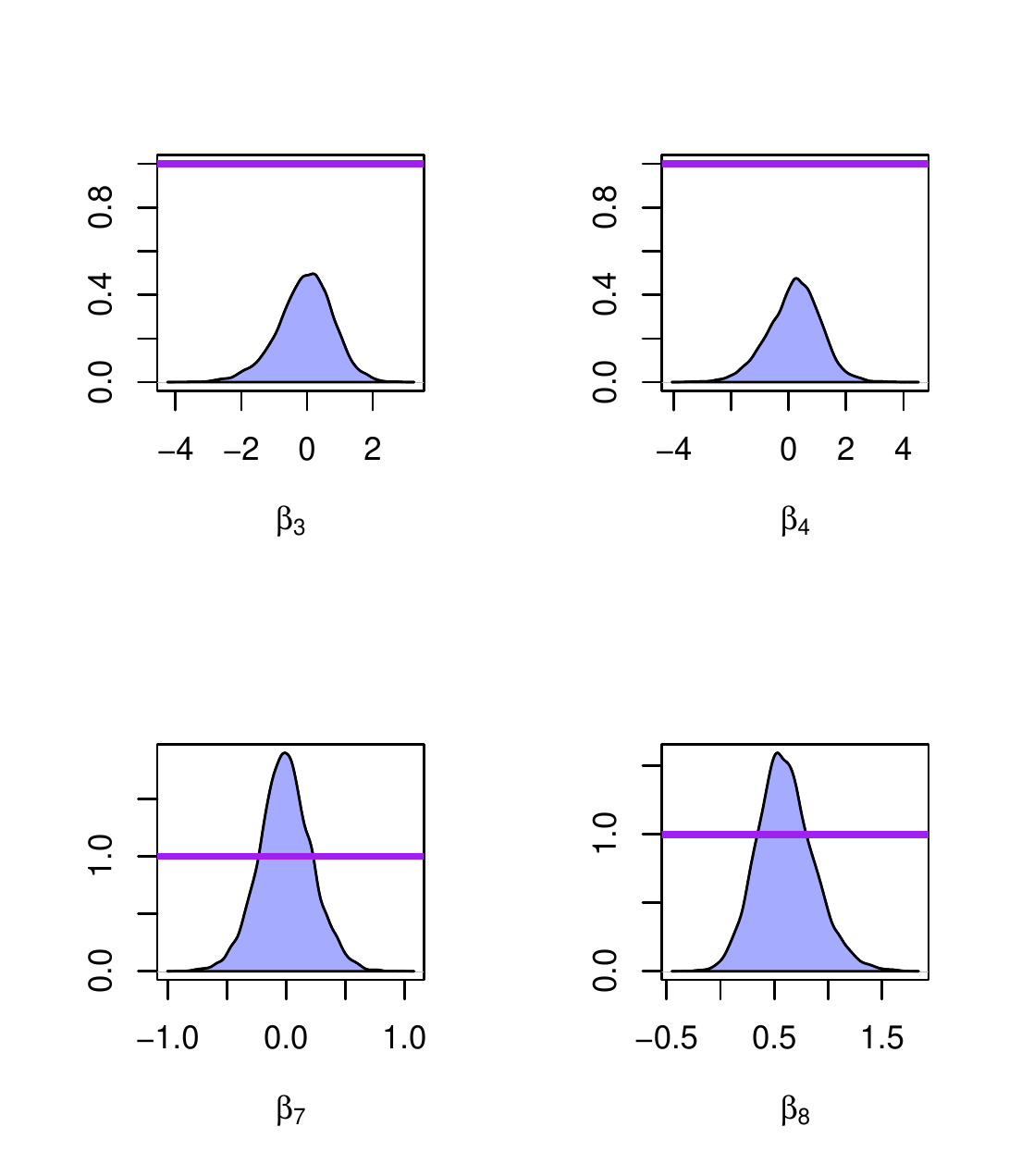}
    \caption{\label{figbeta}\footnotesize{Comparison between the the marginal posterior density of $\beta^*$ without posterior projection and  corresponding priors. The shaded region, posterior density; wide line, flat prior. }}
\end{figure}

Note that comparing the prior and posterior of the fully identified parameter $\bbeta^*$ is meaningless since the projected posterior of $\bbeta$ is certainly different from its prior. 
Therefore, in terms of the parametric component, we compare the priors with the marginal posterior of $\bbeta$, the unconstrained parameter sampled from MCMC.
Fig \ref{figbeta} shows an apparent difference between flat priors and marginal posterior of $\bbeta$, demonstrating that the posterior updating is driven by data. 
An interesting finding is that, even though $\bbeta$ is unidentified, some of the parameters such as $\beta_1$ and $\beta_5$ have low posterior variance and posterior intervals that are short enough. 
This supports the fact that MCMC sampling is workable under unidentified models with weakly informative priors.
Meanwhile, we are aware of the necessity of posterior modification
by checking the marginal posterior of $\bbeta$, since the posterior of $\beta_2$ and $\beta_4$ have heavy-tailed posterior intervals. 

\section{Predictive evaluation metrics}
In this article, we consider two classical metrics to evaluate the predictive capabilities of different survival models, the C index and the integrated Brier score (IBS). 

\noindent{\textbf{C index}}\\
To compute the C-index, we follow the procedure in \cite{Ishwaran2008random}. 
We summarize the procedure as follows:
\begin{enumerate}
    \item[1. ]  Form all possible pairs of survival times over the data.
    \item[2. ] Omit those pairs whose shorter survival time is censored. Omit those tied pairs unless at least one of them is death. Let Permissible denote the total number of permissible pairs.
    \item[3. ] For each untied permissible pair, count 1 if the predicted result is the same as the truth; count 0.5 if the predicted outcomes are tied. 
    For each permissible pair where both are deaths with the same survival time, count 1 if the predicted outcomes are tied; otherwise, count 0.5. 
    For each permissible pair where only one is death and the survival time are tied, count 1 if the death has a worse predicted outcome; otherwise, count 0.5. 
    Let Concordance denote the sum over all permissible pairs.
    \item[4. ] The C index, C, is defined by C = Concordance/Permissible.
\end{enumerate}

\noindent{\textbf{IBS}}\\
The Brier score (BS) is proposed by \cite{graf1999assessment} to evaluate prediction at a certain time point $t$. 
The BS at time $t$ is formulated as
$$
BS(t) = \frac{1}{N} \sum_{i=1}^n \left\{ \frac{S_{T|\Z}(t|\Z_i)]^2}{\hat{G}(T_i)} I(T_i <t , \delta_i = 1) + \frac{[1-S_{T|\Z}(t|\Z_i)]^2}{\hat{G}(T_i)} I(T_i \ge t) \right\}, 
$$
where $\hat{G}(T_i)$ denotes estimated survival probability given by the K-M estimator. 
Then, the IBS is defined as the integral of BS on the interval $(-\infty, \tau)$ for some time $\tau>0$
$$
IBS = \int_{-\infty}^\tau BS(t)dt. 
$$

\newpage
\bibliographystyle{apalike}
\linespread{0.1}
\selectfont
\bibliography{main}

\newcommand{\noop}[1]{}
\begin{thebibliography}{}

\bibitem[Absil and Malick, 2012]{absil2012projection}
Absil, P.-A. and Malick, J. (2012).
\newblock Projection-like retractions on matrix manifolds.
\newblock {\em SIAM Journal on Optimization}, 22(1):135--158.

\bibitem[Ahmad et~al., 2017]{ahmad2017survival}
Ahmad, T., Munir, A., Bhatti, S.~H., Aftab, M., and Raza, M.~A. (2017).
\newblock Survival analysis of heart failure patients: A case study.
\newblock {\em PloS one}, 12(7):e0181001.

\bibitem[Arjas and Gasbarra, 1994]{arjas1994nonparametric}
Arjas, E. and Gasbarra, D. (1994).
\newblock Nonparametric {Bayesian} inference from right censored survival data,
  using the {Gibbs} sampler.
\newblock {\em Statistica Sinica}, 4(2):505--524.

\bibitem[Berchuck et~al., 2021]{berchuck2021bayesian}
Berchuck, S.~I., Janko, M., Medeiros, F.~A., Pan, W., and Mukherjee, S. (2021).
\newblock Bayesian non-parametric factor analysis for longitudinal spatial
  surfaces.
\newblock {\em {Bayesian Analysis}}, TBA:1 -- 30.

\bibitem[Branscum et~al., 2008]{Branscum2008SIM}
Branscum, A.~J., Johnson, W.~O., Hanson, T.~E., and Gardner, I.~A. (2008).
\newblock {Bayesian semiparametric ROC curve estimation and disease diagnosis}.
\newblock {\em Statistics in Medicine}, 27(13):2474--2496.

\bibitem[Burgette et~al., 2021]{Burgette2021BA}
Burgette, L.~F., Puelz, D., and Hahn, P.~R. (2021).
\newblock A symmetric prior for multinomial probit models.
\newblock {\em Bayesian Analysis}, 16(3):1--18.

\bibitem[Cai and Dunson, 2007]{CaiDunson2007JASA}
Cai, B. and Dunson, D.~B. (2007).
\newblock Bayesian multivariate isotonic regression splines: applications to
  carcinogenicity studies.
\newblock {\em Journal of the American Statistical Association},
  102(480):1158--1171.

\bibitem[Carpenter et~al., 2017]{carpenter2017stan}
Carpenter, B., Gelman, A., Hoffman, M.~D., Lee, D., Goodrich, B., Betancourt,
  M., Brubaker, M.~A., Guo, J., Li, P., and Riddell, A. (2017).
\newblock Stan: a probabilistic programming language.
\newblock {\em Journal of Statistical Software}, 76(1):1--32.

\bibitem[Chen, 2002]{chen2002rank}
Chen, S. (2002).
\newblock Rank estimation of transformation models.
\newblock {\em Econometrica}, 70(4):1683--1697.

\bibitem[Cheng et~al., 1995]{cheng1995analysis}
Cheng, S., Wei, L., and Ying, Z. (1995).
\newblock Analysis of transformation models with censored data.
\newblock {\em Biometrika}, 82(4):835--845.

\bibitem[Chiappori et~al., 2015]{chiappori2015nonparametric}
Chiappori, P.-A., Komunjer, I., and Kristensen, D. (2015).
\newblock Nonparametric identification and estimation of transformation models.
\newblock {\em Journal of Econometrics}, 188(1):22--39.

\bibitem[Chicco and Jurman, 2020]{chicco2020machine}
Chicco, D. and Jurman, G. (2020).
\newblock Machine learning can predict survival of patients with heart failure
  from serum creatinine and ejection fraction alone.
\newblock {\em BMC Medical Informatics and Decision Making}, 20(1):1--16.

\bibitem[Colling and Van~Keilegom, 2019]{colling2019estimation}
Colling, B. and Van~Keilegom, I. (2019).
\newblock Estimation of fully nonparametric transformation models.
\newblock {\em Bernoulli}, 25(4B):3762--3795.

\bibitem[de~Castro et~al., 2014]{de2014bayesian}
de~Castro, M., Chen, M.-H., Ibrahim, J.~G., and Klein, J.~P. (2014).
\newblock Bayesian transformation models for multivariate survival data.
\newblock {\em Scandinavian Journal of Statistics}, 41(1):187--199.

\bibitem[Depaoli et~al., 2020]{depaoli2020importance}
Depaoli, S., Winter, S.~D., and Visser, M. (2020).
\newblock The importance of prior sensitivity analysis in bayesian statistics:
  demonstrations using an interactive shiny app.
\newblock {\em Frontiers in Psychology}, 11.

\bibitem[Ding and Nan, 2011]{ding2011sieve}
Ding, Y. and Nan, B. (2011).
\newblock A sieve m-theorem for bundled parameters in semiparametric models,
  with application to the efficient estimation in a linear model for censored
  data.
\newblock {\em The Annals of Statistics}, 39(6):3032--3061.

\bibitem[Dykstra and Laud, 1981]{dykstra1981bayesian}
Dykstra, R. and Laud, P. (1981).
\newblock {A Bayesian nonparametric approach to reliability}.
\newblock {\em The Annals of Statistics}, 9(2):356--367.

\bibitem[Egleston et~al., 2017]{egleston2017latent}
Egleston, B.~L., Uzzo, R.~G., and Wong, Y.-N. (2017).
\newblock Latent class survival models linked by principal stratification to
  investigate heterogenous survival subgroups among individuals with
  early-stage kidney cancer.
\newblock {\em Journal of the American Statistical Association},
  112(518):534--546.

\bibitem[Gelman et~al., 2013]{gelman2013bayesian}
Gelman, A., Carlin, J.~B., Stern, H.~S., Dunson, D.~B., Vehtari, A., and Rubin,
  D.~B. (2013).
\newblock {\em {Bayesian Data Analysis, Third Edition}}.
\newblock CRC press.

\bibitem[G{\o}rgens and Horowitz, 1999]{gorgensHorowitz1999JOE}
G{\o}rgens, T. and Horowitz, J.~L. (1999).
\newblock Semiparametric estimation of a censored regression model with an
  unknown transformation of the dependent variable.
\newblock {\em Journal of Econometrics}, 90(2):155--191.

\bibitem[Graf et~al., 1999]{graf1999assessment}
Graf, E., Schmoor, C., Sauerbrei, W., and Schumacher, M. (1999).
\newblock Assessment and comparison of prognostic classification schemes for
  survival data.
\newblock {\em Statistics in medicine}, 18(17-18):2529--2545.

\bibitem[H{\"a}rdle and Stoker, 1989]{HardleStoker1989jasa}
H{\"a}rdle, W. and Stoker, T.~M. (1989).
\newblock Investigating smooth multiple regression by the method of average
  derivatives.
\newblock {\em Journal of the American statistical Association},
  84(408):986--995.

\bibitem[H{\"a}rdle et~al., 2004]{hardle2004nonparametric}
H{\"a}rdle, W.~K., M{\"u}ller, M., Sperlich, S., and Werwatz, A. (2004).
\newblock {\em Nonparametric and semiparametric models}.
\newblock Springer Science \& Business Media.

\bibitem[Harrell et~al., 1982]{harrell1982evaluating}
Harrell, F.~E., Califf, R.~M., Pryor, D.~B., Lee, K.~L., and Rosati, R.~A.
  (1982).
\newblock Evaluating the yield of medical tests.
\newblock {\em Jama}, 247(18):2543--2546.

\bibitem[Heagerty et~al., 2000]{heagerty2000time}
Heagerty, P.~J., Lumley, T., and Pepe, M.~S. (2000).
\newblock Time-dependent roc curves for censored survival data and a diagnostic
  marker.
\newblock {\em Biometrics}, 56(2):337--344.

\bibitem[Hjort, 1990]{hjort1990nonparametric}
Hjort, N.~L. (1990).
\newblock Nonparametric bayes estimators based on beta processes in models for
  life history data.
\newblock {\em The Annals of Statistics}, 18(3):1259--1294.

\bibitem[Hoffman and Gelman, 2014]{hoffman2014no}
Hoffman, M.~D. and Gelman, A. (2014).
\newblock The {No-U-Turn} sampler: adaptively setting path lengths in
  {Hamiltonian Monte Carlo}.
\newblock {\em Journal of Machine Learning Research}, 15(1):1593--1623.

\bibitem[Horowitz, 1996]{horowitz1996semiparametric}
Horowitz, J.~L. (1996).
\newblock Semiparametric estimation of a regression model with an unknown
  transformation of the dependent variable.
\newblock {\em Econometrica}, 64(1):103--137.

\bibitem[Ishwaran and James, 2002]{ishwaran2002approximate}
Ishwaran, H. and James, L.~F. (2002).
\newblock Approximate dirichlet process computing in finite normal mixtures:
  smoothing and prior information.
\newblock {\em Journal of Computational and Graphical statistics},
  11(3):508--532.

\bibitem[Ishwaran et~al., 2008]{Ishwaran2008random}
Ishwaran, H., Kogalur, U.~B., Blackstone, E.~H., and Lauer, M.~S. (2008).
\newblock Random survival forests.
\newblock {\em The Annals of Applied Statistics}, 2(3):841--860.

\bibitem[Jin et~al., 2003]{jin2003rank}
Jin, Z., Lin, D., Wei, L., and Ying, Z. (2003).
\newblock Rank-based inference for the accelerated failure time model.
\newblock {\em Biometrika}, 90(2):341--353.

\bibitem[Kalbfleisch, 1978]{kalbfleisch1978non}
Kalbfleisch, J.~D. (1978).
\newblock {Non-parametric Bayesian analysis of survival time data}.
\newblock {\em Journal of the Royal Statistical Society: Series B (Statistical
  Methodology)}, 40(2):214--221.

\bibitem[Kottas, 2006]{kottas2006nonparametric}
Kottas, A. (2006).
\newblock {Nonparametric Bayesian survival analysis using mixtures of Weibull
  distributions}.
\newblock {\em Journal of Statistical Planning and Inference}, 136(3):578--596.

\bibitem[Lenk and Choi, 2017]{lenk2017bayesian}
Lenk, P.~J. and Choi, T. (2017).
\newblock Bayesian analysis of shape-restricted functions using {Gaussian}
  process priors.
\newblock {\em Statistica Sinica}, 27(1):43--69.

\bibitem[Lin and Dunson, 2014]{LinDunson2014Biometrika}
Lin, L. and Dunson, D.~B. (2014).
\newblock Bayesian monotone regression using {Gaussian} process projection.
\newblock {\em Biometrika}, 101(2):303--317.

\bibitem[Lin et~al., 2017]{lin2017robust}
Lin, Y., Luo, Y., Xie, S., and Chen, K. (2017).
\newblock Robust rank estimation for transformation models with random effects.
\newblock {\em Biometrika}, 104(4):971--986.

\bibitem[Linton et~al., 2008]{linton2008estimation}
Linton, O., Sperlich, S., Van~Keilegom, I., et~al. (2008).
\newblock Estimation of a semiparametric transformation model.
\newblock {\em The Annals of Statistics}, 36(2):686--718.

\bibitem[Lo, 1984]{lo1984class}
Lo, A.~Y. (1984).
\newblock {On a class of Bayesian nonparametric estimates: I. density
  estimates}.
\newblock {\em The Annals of Statistics}, 12(1):351--357.

\bibitem[Lu and Ying, 2004]{lu2004semiparametric}
Lu, W. and Ying, Z. (2004).
\newblock On semiparametric transformation cure models.
\newblock {\em Biometrika}, 91(2):331--343.

\bibitem[Mallick and Walker, 2003]{mallick2003bayesian}
Mallick, B.~K. and Walker, S. (2003).
\newblock {A Bayesian semiparametric transformation model incorporating
  frailties}.
\newblock {\em Journal of Statistical Planning and Inference},
  112(1-2):159--174.

\bibitem[McCulloch and Rossi, 1994]{MccullochRossi1994JOE}
McCulloch, R. and Rossi, P.~E. (1994).
\newblock An exact likelihood analysis of the multinomial probit model.
\newblock {\em Journal of Econometrics}, 64(1-2):207--240.

\bibitem[McElreath, 2020]{mcelreath2020rethinking}
McElreath, R. (2020).
\newblock {\em Statistical rethinking: A Bayesian course with examples in R and
  Stan}.
\newblock Chapman and Hall/CRC.

\bibitem[McKeague and Tighiouart, 2000]{Mckeague2000Biometrics}
McKeague, I.~W. and Tighiouart, M. (2000).
\newblock Bayesian estimators for conditional hazard functions.
\newblock {\em Biometrics}, 56(4):1007--1015.

\bibitem[M{\"u}ller et~al., 2015]{Muller2015}
M{\"u}ller, P., Quintana, F.~A., Jara, A., and Hanson, T. (2015).
\newblock {\em {Bayesian Nonparametric Data Analysis}}.
\newblock Springer.

\bibitem[Neelon and Dunson, 2004]{neelon2004bayesian}
Neelon, B. and Dunson, D.~B. (2004).
\newblock Bayesian isotonic regression and trend analysis.
\newblock {\em Biometrics}, 60(2):398--406.

\bibitem[Nieto-Barajas et~al., 2012]{nieto2012time}
Nieto-Barajas, L.~E., M{\"u}ller, P., Ji, Y., Lu, Y., and Mills, G.~B. (2012).
\newblock {A time-series DDP for functional proteomics profiles}.
\newblock {\em Biometrics}, 68(3):859--868.

\bibitem[Ohlssen et~al., 2007]{ohlssen2007flexible}
Ohlssen, D.~I., Sharples, L.~D., and Spiegelhalter, D.~J. (2007).
\newblock {Flexible random-effects models using Bayesian semi-parametric
  models: applications to institutional comparisons}.
\newblock {\em Statistics in Medicine}, 26(9):2088--2112.

\bibitem[Park et~al., 2005]{Park2005JCGS}
Park, C.~G., Vannucci, M., and Hart, J.~D. (2005).
\newblock Bayesian methods for wavelet series in single-index models.
\newblock {\em Journal of Computational and Graphical Statistics},
  14(4):770--794.

\bibitem[Perperoglou et~al., 2019]{perperoglou2019review}
Perperoglou, A., Sauerbrei, W., Abrahamowicz, M., and Schmid, M. (2019).
\newblock {A review of spline function procedures in R}.
\newblock {\em BMC Medical Research Methodology}, 19(1):1--16.

\bibitem[Phadia, 2015]{phadia2015prior}
Phadia, E.~G. (2015).
\newblock {\em Prior Processes and Their Applications}.
\newblock Springer.

\bibitem[Ramsay, 1988]{ramsay1988monotone}
Ramsay, J.~O. (1988).
\newblock Monotone regression splines in action.
\newblock {\em Statistical Science}, 3(4):425--441.

\bibitem[Riihim{\"a}ki and Vehtari, 2010]{riihimaki2010gaussian}
Riihim{\"a}ki, J. and Vehtari, A. (2010).
\newblock Gaussian processes with monotonicity information.
\newblock In {\em Proceedings of the thirteenth international conference on
  artificial intelligence and statistics}, pages 645--652. JMLR Workshop and
  Conference Proceedings.

\bibitem[Rodriguez et~al., 2008]{rodriguez2008nested}
Rodriguez, A., Dunson, D.~B., and Gelfand, A.~E. (2008).
\newblock {The nested Dirichlet process}.
\newblock {\em Journal of the American statistical Association},
  103(483):1131--1154.

\bibitem[Scheike, 2006]{scheike2006flexible}
Scheike, T.~H. (2006).
\newblock A flexible semiparametric transformation model for survival data.
\newblock {\em Lifetime Data Analysis}, 12(4):461--480.

\bibitem[Sen et~al., 2022]{sen2022constrained}
Sen, D., Patra, S., and Dunson, D. (2022).
\newblock Constrained inference through posterior projections.
\newblock {\em arXiv}.

\bibitem[Sethuraman, 1994]{sethuraman1994constructive}
Sethuraman, J. (1994).
\newblock {A constructive definition of Dirichlet priors}.
\newblock {\em Statistica Sinica}, 4(2):639--650.

\bibitem[Shi et~al., 2019]{shi2019low}
Shi, Y., Martens, M., Banerjee, A., Laud, P., et~al. (2019).
\newblock {Low information omnibus {(LIO)} priors for Dirichlet process mixture
  models}.
\newblock {\em Bayesian Analysis}, 14(3):677--702.

\bibitem[Shively et~al., 2009]{shively2009bayesian}
Shively, T.~S., Sager, T.~W., and Walker, S.~G. (2009).
\newblock {A Bayesian approach to non-parametric monotone function estimation}.
\newblock {\em Journal of the Royal Statistical Society: Series B (Statistical
  Methodology)}, 71(1):159--175.

\bibitem[Solomon, 1984]{solomon1984effect}
Solomon, P.~J. (1984).
\newblock Effect of misspecification of regression models in the analysis of
  survival data.
\newblock {\em Biometrika}, 71(2):291--298.

\bibitem[Song et~al., 2007]{song2007semiparametric}
Song, X., Ma, S., Huang, J., and Zhou, X.-H. (2007).
\newblock A semiparametric approach for the nonparametric transformation
  survival model with multiple covariates.
\newblock {\em Biostatistics}, 8(2):197--211.

\bibitem[Therneau, 2022]{Terry2022}
Therneau, T.~M. (2022).
\newblock {\em A Package for Survival Analysis in R}.
\newblock R package version 3.4-0.

\bibitem[van~de Schoot et~al., 2021]{van2021bayesian}
van~de Schoot, R., Depaoli, S., King, R., Kramer, B., M{\"a}rtens, K., Tadesse,
  M.~G., Vannucci, M., Gelman, A., Veen, D., Willemsen, J., et~al. (2021).
\newblock Bayesian statistics and modelling.
\newblock {\em Nature Reviews Methods Primers}, 1(1):1--26.

\bibitem[Vehtari et~al., 2021]{vehtari2021rank}
Vehtari, A., Gelman, A., Simpson, D., Carpenter, B., and B{\"u}rkner, P.-C.
  (2021).
\newblock {Rank-normalization, folding, and localization: An improved R for
  assessing convergence of MCMC}.
\newblock {\em Bayesian Analysis}, 16(2):1--28.

\bibitem[Wang and Dunson, 2011a]{wang2011Biometrika}
Wang, L. and Dunson, D.~B. (2011a).
\newblock Bayesian isotonic density regression.
\newblock {\em Biometrika}, 98(3):537--551.

\bibitem[Wang and Dunson, 2011b]{wang2011semiparametric}
Wang, L. and Dunson, D.~B. (2011b).
\newblock {Semiparametric Bayes' proportional odds models for current status
  data with underreporting}.
\newblock {\em Biometrics}, 67(3):1111--1118.

\bibitem[Wang and Berger, 2016]{wang2016estimating}
Wang, X. and Berger, J.~O. (2016).
\newblock Estimating shape constrained functions using {Gaussian} processes.
\newblock {\em SIAM/ASA Journal on Uncertainty Quantification}, 4(1):1--25.

\bibitem[Ye and Duan, 1997]{ye1997nonparametric}
Ye, J. and Duan, N. (1997).
\newblock Nonparametric $ n^{-1/2} $-consistent estimation for the general
  transformation models.
\newblock {\em The Annals of Statistics}, 25(6):2682--2717.

\bibitem[Zeng and Lin, 2006]{zeng2006efficient}
Zeng, D. and Lin, D. (2006).
\newblock Efficient estimation of semiparametric transformation models for
  counting processes.
\newblock {\em Biometrika}, 93(3):627--640.

\bibitem[Zeng and Lin, 2007]{zeng2007efficient}
Zeng, D. and Lin, D. (2007).
\newblock Efficient estimation for the accelerated failure time model.
\newblock {\em Journal of the American Statistical Association},
  102(480):1387--1396.

\bibitem[Zhao et~al., 2009]{zhao2009mixtures}
Zhao, L., Hanson, T.~E., and Carlin, B.~P. (2009).
\newblock Mixtures of polya trees for flexible spatial frailty survival
  modelling.
\newblock {\em Biometrika}, 96(2):263--276.

\bibitem[Zhou et~al., 2022]{Zhou2022RpkgSurvMetriocs}
Zhou, H., Cheng, X., Wang, S., Zou, Y., and Wang, H. (2022).
\newblock {\em SurvMetrics: Predictive Evaluation Metrics in Survival
  Analysis}.
\newblock R package version 0.5.0.

\bibitem[Zhou and Hanson, 2018]{zhou2018unified}
Zhou, H. and Hanson, T. (2018).
\newblock {A unified framework for fitting Bayesian semiparametric models to
  arbitrarily censored survival data, including spatially referenced data}.
\newblock {\em Journal of the American Statistical Association},
  113(522):571--581.

\bibitem[Zhou et~al., 2020]{zhou2020spbayessurv}
Zhou, H., Hanson, T., and Zhang, J. (2020).
\newblock spbayessurv: Fitting bayesian spatial survival models using r.
\newblock {\em Journal of Statistical Software}, 92:1--33.

\end{thebibliography}
\end{document}